\documentclass[preprint]{aa}
\newcommand{\kms}{km~s$^{-1}$}
\usepackage{natbib}
\usepackage{graphicx}
\usepackage{capt-of}
\usepackage[varg]{txfonts}
\usepackage{longtable}
\authorrunning{J.~K.~J{\o}rgensen et al.}

\begin{document}

\title{The ALMA Protostellar Interferometric Line Survey (PILS)}
\subtitle{First results from an unbiased submillimeter wavelength line survey of the
  Class~0 protostellar binary IRAS~16293-2422 with ALMA}

\author{J.~K. J{\o}rgensen\inst{1}, M.~H.~D. van der Wiel\inst{1},
  A.~Coutens\inst{2}, J.~M.~Lykke\inst{1}, H.~S.~P. M\"{u}ller\inst{3}, E.~F.~van Dishoeck\inst{4,5}, H.~Calcutt\inst{1}, P.~Bjerkeli\inst{1,6}, T.~L. Bourke\inst{7}, M.~N. Drozdovskaya\inst{4}, C. Favre\inst{8}, E.~C. Fayolle\inst{9}, R.~T. Garrod\inst{10}, S.~K.~Jacobsen\inst{1}, K.~I. \"{O}berg\inst{9}, M.~V. Persson\inst{4} \and S.~F. Wampfler\inst{11}} \institute{Centre for Star and Planet Formation, Niels Bohr Institute \& Natural History Museum of Denmark, University of
  Copenhagen, {\O}ster Voldgade 5--7, DK-1350 Copenhagen {K}., Denmark
  \and Department of Physics and Astronomy, University College London, Gower St., London, WC1E 6BT, UK
  \and I. Physikalisches Institut, Universit\"{a}t zu K\"{o}ln, Z\"{u}lpicher Str. 77, 50937 K\"{o}ln, Germany
  \and Leiden Observatory, Leiden University, PO Box 9513, NL-2300 RA Leiden, The Netherlands
  \and Max-Planck Institut f\"{u}r Extraterrestrische Physik (MPE), Giessenbachstr. 1, 85748 Garching, Germany
  \and Department of Earth and Space Sciences, Chalmers University of Technology, Onsala Space Observatory, 439 92 Onsala, Sweden
  \and SKA Organization, Jodrell Bank Observatory, Lower Withington, Macclesfield, Cheshire SK11 9DL, UK
  \and Universit\'e Grenoble Alpes and CNRS, IPAG, F-38000 Grenoble, France
  \and Harvard-Smithsonian Center for Astrophysics, 60 Garden Street, Cambridge, MA 02138, USA
  \and Departments of Chemistry and Astronomy, University of Virginia, Charlottesville, VA 22904, USA
  \and Center for Space and Habitability (CSH), University of Bern, Sidlerstrasse 5, CH-3012 Bern, Switzerland
}

\abstract 
{The inner regions of the envelopes surrounding young protostars are characterised by a complex chemistry, with prebiotic molecules present on the scales where protoplanetary disks eventually may form. The Atacama Large Millimeter/submillimeter Array (ALMA) provides an unprecedented view of these regions zooming in on Solar System scales of nearby protostars and mapping the emission from rare species.}
{The goal is to introduce a systematic survey, ``Protostellar Interferometric Line Survey (PILS)'', of the chemical complexity of one of the nearby astrochemical templates, the Class~0 protostellar binary IRAS~16293$-$2422, using ALMA, to understand the origin of the complex molecules formed in its vicinity. In addition to presenting the overall survey, the analysis in this paper focuses on new results for the prebiotic molecule glycolaldehyde, its isomers and rarer isotopologues and other related molecules.}
{An unbiased spectral survey of IRAS~16293$-$2422 covering the full frequency range from 329 to 363~GHz (0.8~mm) has been obtained with ALMA, in addition to a few targeted observations at 3.0 and 1.3~mm. The data consist of full maps of the protostellar binary system with an angular resolution of 0.5$''$ (60~AU diameter), a spectral resolution of 0.2~\kms\ and a sensitivity of 4--5~mJy~beam$^{-1}$~km~s$^{-1}$ -- approximately two orders of magnitude better than any previous studies. } 
{More than 10,000 features are detected toward one component in the protostellar binary, corresponding to an average line density of approximately one line per 3~\kms. Glycolaldehyde, its isomers, methyl formate and acetic acid, and its reduced alcohol, ethylene glycol, are clearly detected and their emission well-modeled with an excitation temperature of 300~K. For ethylene glycol both lowest state conformers, $aGg'$ and $gGg'$, are detected, the latter for the first time in the ISM. The abundance of glycolaldehyde is comparable to or slightly larger than that of ethylene glycol. In comparison to the Galactic Center these two species are over-abundant relative to methanol, possibly an indication of formation of the species at low temperatures in CO-rich ices during the infall of the material toward the central protostar. Both $^{13}$C and deuterated isotopologues of glycolaldehyde are detected, also for the first time ever in the ISM. For the deuterated species a D/H ratio of $\approx$5\% is found with no differences between the deuteration in the different functional groups of glycolaldehyde, in contrast to previous estimates for methanol and recent suggestions of significant equilibration between water and -OH functional groups at high temperatures. Measurements of the $^{13}$C-species lead to a $^{12}$C:$^{13}$C ratio of $\approx$30, lower than the typical ISM value. This low ratio may reflect an enhancement of $^{13}$CO in the ice due to either ion-molecule reactions in the gas before freeze-out or differences in the temperatures where $^{12}$CO and $^{13}$CO ices sublimate.}
{The results reinforce the importance of low temperature grain surface chemistry for the formation of prebiotic molecules seen here in the gas after sublimation of the entire ice mantle. Systematic surveys of the molecules thought to be chemically related, as well as the accurate measurements of their isotopic composition, hold strong promises for understanding the origin of prebiotic molecules in the earliest stages of young stars.}

\offprints{Jes K.\,J{\o}rgensen, \email{jeskj@nbi.ku.dk}}
\maketitle

\section{Introduction}\label{introduction}
Understanding how, when and where complex organic and potentially prebiotic molecules are formed is a fundamental goal of astrochemistry and an integral part of origins of life studies. The recent images from the Atacama Large Millimeter/submillimeter Array (ALMA) of a potentially planet-forming disk around a young star with an age of only 0.5--1~Myr, HL Tau, \citep{hltau} has highlighted the importance of the physics and chemistry of the early protostellar stages: how do stars evolve during their earliest evolutionary stages and in particular, to what degree does the chemistry reflect this early evolution relative to, e.g., the conditions in the environment from which the stars are forming? Already during its first years ALMA has demonstrated enormous potential for addressing these issues with its high angular resolution and sensitivity making it possible to zoom in on solar system scales of young stars and map the chemical complexity in their environments \citep[e.g.,][]{pineda12,jorgensen12,jorgensen13,persson13,codella14,sakai14,friesen14,lindberg14alma,oya14,murillo15,podio15,belloche16,muller16}.

A particular focus of ALMA observations is the search for complex molecules in regions of low-mass star formation. Over the last decade it has become clear that the chemical complexity toward the innermost envelopes of solar-type protostars can rival that of more massive hot cores \citep[see, e.g., review by][]{herbst09}. The presence of complex molecules is not solely attributed to such regions, however, but also found toward cold prestellar cores \citep[e.g.,][]{oberg10,bacmann12,vastel14} and toward outflow driven shocks \citep[e.g.,][]{arce08,sugimura11,mendoza14}. The big questions that remain include \emph{(a)} what degree of molecular complexity can arise during the protostellar stages, \emph{(b)} how exactly do complex organics form, \emph{(c)} what the roles are of grain-surface/ice-mantle vs. gas-phase reactions at low and high temperatures for specific molecules, and \emph{(d)} what the importance is of external conditions (e.g., cosmic ray induced ionization, UV radiation) and the physical environment (e.g., temperature).

Many of these questions can potentially be adressed through systematic surveys with ALMA: with its high angular resolution we can zoom in on the smallest scales of young stars making it possible to unambiguously identify the emitting regions for different molecules.  The advantage of studying the hot inner regions of the envelopes around protostars is that the ices there are fully sublimated and all the molecules are present in the gas-phase. ALMA's high sensitivity and spectral resolution allows for identification of faint lines of rare species and also observations of more quiescent sources (e.g., of lower masses) for which line confusion is reached at a much deeper level than for sources of higher masses. One particularly interesting source in this context is the well-studied protostellar binary, IRAS~16293$-$2422, the first low-mass protostar for which complex organic molecules \citep{vandishoeck95,cazaux03} as well as prebiotic species \citep{jorgensen12}, were identified -- the latter already during ALMA science verification.

This paper presents an overview of an unbiased survey, \emph{Protostellar Interferometric Line Survey (PILS)}\footnote{http://youngstars.nbi.dk/PILS}, of IRAS~16293$-$2422 with ALMA covering a wide frequency window from 329 to 363~GHz at 0.5\arcsec\ angular resolution (60~AU diameter) as well as selected other frequencies around 1.3~mm (230~GHz) and 3.0~mm (100~GHz). The paper provides a first overview of the observations and data and presents new results concerning the presence of glycolaldehyde and related species, as well as the first detections of its rarer $^{13}$C and deuterated isotopologues. The paper is laid out as follows: Sect.~\ref{overview} presents a detailed review of studies of the physics and chemistry of IRAS~16293$-$2422 as background for this and subsequent PILS papers and Sect.~\ref{observations} presents an overview of the details of the observations and reduction. Sect.~\ref{results} presents the overall features of the datasets including the continuum emission at the three different wavelengths and information about the line emission, while Sect.~\ref{analysis} focuses on the analysis of the emission from glycolaldehyde and related molecules (Sect.~\ref{glycolaldehydeisomers}) and its isotopologues (Sect.~\ref{isotopologues}) with particular emphasis on the constraints on formation scenarios for these species. Section~\ref{summary} summarises the main findings of the paper.

\section{An overview of IRAS~16293$-$2422}\label{overview}
IRAS~16293$-$2422 (hereafter IRAS16293) is a deeply embedded young stellar object located in the L1689 region in the eastern part of the $\rho$ Ophiuchus cloud complex studied extensively through larger scale infrared and submillimeter continuum and line maps \citep[e.g.,][]{tachihara00,nutter06,young06,jorgensen08,padgett08}. The traditionally quoted distance for $\rho$~Oph was $\approx$160~pc \citep{whittet75}. However, most recent distance estimates for the bulk of the $\rho$~Oph cloud complex place it nearer at approximately 120~pc based on extinction measurements \citep{knude98,lombardi08} as well as VLBI parallax measurements \citep{loinard08}. One dedicated measurement of the parallax of water masers toward IRAS~16293$-$2422 place it at a much larger distance than the rest of the cloud ($\approx$178~pc), but \cite{rivera15} argue that this estimate may be hampered by the weak and highly variable nature of water masers and quote new measurements that place IRAS~16293$-$2422 at the nearer distance. We therefore adopt a distance of 120~pc throughout this paper. For this distance, the observed luminosity of the source is 21$\pm 5$~L$_\odot$ estimated by pure integration of its SED with data from the mid-infrared from \emph{Spitzer} \citep{iras16293letter}, far-infrared from ISO-LWS \citep{correia04} and \emph{Herschel}/SPIRE \citep{makiwa14}, and submillimeter \citep{schoeier02}. To put the new ALMA PILS results in context, the following presents an overview of physical and chemical studies of this source to date.

\subsection{Physics}
Much of our current picture for star formation goes back to the 1980's: at the time infrared observations (in particular, the full survey of the infrared sky by the Infrared Astronomical Satellite) gave rise to identification and empirical classification of young stellar objects \citep{lada84} and combined with theoretical work on the collapse of dense cores and the formation of disks \citep[e.g.,][]{terebey84,adams86} led to the generally accepted picture of the evolution of young stellar objects \citep{shu87}. One of the particular goals of these studies was to identify the youngest protostars and IRAS~16293$-$2422 with its very red colors in that context became a particularly interesting target for immediate follow-up studies using ground-based millimeter wavelength single-dish telescopes and small interferometers. The focus of these early line studies was to address whether the circumstellar material was predominantly characterised by infall toward the protostar \citep{walker86} or rather by rotation in a disk-like structure \citep{mundy86,menten87}.

\cite{andre93} introduced the term ``Class 0'' protostar for young stellar objects thought to have accreted less than half of their final mass and still deeply embedded in their protostellar envelopes. \citeauthor{andre93} noted that IRAS~16293$-$2422 was a candidate of such a source. Its very strong submillimeter emission made it unlikely that it was simply an edge-on Class~I object, but instead a very young, still deeply-embedded, protostar. The distribution of the dust around the source was recognized early on as being consistent with originating in a power-law density profile envelope \citep{walker90} following $\rho \propto r^{-p}$ with $p\approx 1.5-2.5$ as theoretically predicted for a singular-isothermal sphere or free-falling envelope. \cite{ceccarelli00model}, \cite{schoeier02} and others have modeled the lower resolution single-dish observations of IRAS~16293$-$2422 and found that such envelope profiles reproduce far-infrared/(sub)millimeter line and continuum observations, and that such models thereby could be taken as a useful reference for, e.g., description of the chemistry at larger scales ($\gtrsim$1000~AU). Mapping of the protostellar envelope with single-dish telescopes demonstrated that the material could be well-characterised by infall on a few thousand AU scales \citep{zhou95,narayanan98}, although more quantitative statements are complicated due to the underlying physical structure that appears at higher angular resolution as multiple separated velocity components \citep[e.g.,][]{hotcorepaper}.

On smaller scales IRAS~16293$-$2422 was the first protostar identified as a binary, being resolved into two separate components, \emph{IRAS16293A} and \emph{IRAS16293B}, at radio \citep{wootten89} and millimeter \citep{mundy92} wavelengths with a separation of $\approx$5.1\arcsec\ or 620~AU \citep{looney00,chandler05}. Each of the two components show compact millimeter continuum emission on 100~AU scales, signs of two protostars each surrounded by a compact disk-like structure embedded within a larger circumbinary envelope \citep{looney00,schoeier04}. The early observations demonstrated that the continuum fluxes toward the two components have different dependencies with wavelength, with IRAS16293A showing a flattened continuum spectrum at wavelengths longer than 3~mm, while IRAS16293B has a continuum spectrum that is well described by a single power-law, $F_\nu \propto \nu^\alpha$, with $\alpha = $~2.0--2.5 from cm through submm wavelengths \citep{mundy92,hotcorepaper,chandler05}. A plausible explanation for this difference is that the emission from IRAS16293A is a combination of dust continuum radiation with shock-ionised emission at longer wavelengths \citep{chandler05}, while IRAS16293B is dominated by optically thick dust continuum emission. IRAS16293A is in fact resolved into multiple components at longer wavelengths \citep{wootten89,chandler05,loinard07,pech10} and possibly itself a tight binary within a distance of about 1\arcsec\ (120~AU).  The early claims of rotation in a ``disk'' encompassing IRAS16293A and IRAS16293B was likely a result of a difference in the LSR velocities of the two sources \citep[3.1 vs. 2.7~\kms;][]{iras16293sma} for which the physical origin is still not fully understood. IRAS16293A does show a small velocity gradient in the NE-SW direction close to the location of the continuum peak that could be attributed to rotation in a disk located in this direction \citep{pineda12,girart14}, although the velocity profile of this component is only consistent with Keplerian rotation if the dynamical mass of IRAS16293A is very small ($\sim$0.1~$M_\odot$; \citealt{favre14eSMA}). Also, the relative contributions to the overall luminosity of IRAS~16293$-$2422 from the different sources remain unclear.

The binarity is also reflected in the complex outflow morphologies observed toward the system. The outflow structure has been studied in different tracers and at very different angular resolutions from its first detections \citep{fukui86,wootten87}. It was recognised early on as being quadrupolar in nature \citep{walker88,mizuno90} with one collimated pair of lobes in the NE-SW direction and one less collimated (and less well-aligned set of lobes) in the E-W direction. This complicated outflow morphology has led to some discussion concerning the nature of IRAS16293A and IRAS16293B: it has long been accepted that IRAS16293A is of protostellar nature and the NE-SW outflow driven by this source \citep[e.g.,][]{castets01,stark04}. However, the lack of clear outflow structures associated with IRAS16293B itself has made its nature more ambiguous: \cite{stark04} suggested that IRAS16293B was in fact a more evolved T~Tauri star responsible for the ``fossil'' outflow in the E-W direction but higher resolution CO images from the SMA showed that a compact outflow in the E-W direction originates close to IRAS16293A, possibly the ``current'' manifestation of the larger scale, less collimated, E-W flow \citep{yeh08}. Also, direct detections of infall toward the source \citep{chandler05,pineda12,jorgensen12} suggest that it is in fact in an early evolutionary stage. Based on ALMA CO 6-5 images, \cite{loinard13} argued that blue-shifted emission seen to the southeast of IRAS16293B (see also \citealt{yeh08} and \citealt{iras16293sma}) was the manifestation of a very young (mono-polar) outflow from IRAS16293B -- but \cite{kristensen13} using the same data argued that it rather was a bow-shock associated with a (new) NW-SE outflow driven from IRAS16293A. 

Finally, larger scale maps reveal a companion core to that associated with IRAS~16293$-$2422 itself, IRAS16293E, offset by about 1.5\arcmin. This core was first recognized as a prominent NH$_3$ core \citep{mizuno90} and is almost as bright in submillimeter continuum as the core hosting IRAS~16293$-$2422 itself \citep[e.g.,][]{nutter06}. \cite{castets01} mapped the region in H$_2$CO, N$_2$H$^+$ and other tracers using single-dish telescopes and suggested that the source was in fact also a Class~0 protostar driving a separate outflow in the northwest/southeast direction. However, \cite{stark04} instead argued that this was the reflection of the interaction between the outflow driven by IRAS~16293$-$2422 with the dense IRAS16293E core -- itself therefore likely a dense prestellar core. In support of this interpretation, \cite{jorgensen08} did not find any signs of an embedded protostar toward IRAS16293E through \emph{Spitzer} mid-infrared observations, but rather extended 4.5~$\mu$m emission (molecular hydrogen) originating toward IRAS~16293$-$2422 and extending to and around IRAS16293E.

\subsection{Chemistry}
 
\subsubsection{Single-dish observations}

IRAS16293 has a particularly rich molecular line spectrum for a low-mass protostar, as has been recognized since the late 1980s.  Given the limited sensitivity of early interferometers, chemical studies were mainly performed with single dish telescopes, especially the then newly available Caltech Submillimeter Observatory (CSO) and James Clerk Maxwell Telescope (JCMT) equipped with sensitive submillimeter receivers and located on Mauna Kea, a high and dry site with easy access to this southern source \citep{vandishoeckppiii}. Individual line settings in the 230 and 345 GHz windows revealed 265 lines belonging to 24 different molecules \citep{blake94,vandishoeck95}.  Because of its rich spectrum, IRAS16293 was quickly dubbed the low-mass counterpart of Orion-KL. Through rotational diagrams, non-LTE excitation calculations and line profile analyses, at least three different physical and chemical components were identified: a compact, turbulent, warm ($>$80 K), and dense ($\sim 10^7$ cm$^{-3}$) region rich in Si- and S-bearing molecules as well as complex organic molecules such as CH$_3$OH and CH$_3$CN; the quiescent circumbinary envelope best traced in common molecules like CS, HCO$^+$ and H$_2$CO; and a colder outer envelope and surrounding cloud core seen in radicals such as CN, C$_2$H and C$_3$H$_2$ with very narrow lines.

These early data also revealed that IRAS16293 has very high abundances of deuterated molecules, with C$_2$D/C$_2$H and HDCO/H$_2$CO values $>$0.1. This extreme fractionation is thought to be a result of gas-grain chemistry, benefitting from much longer timescales at very low temperatures and high densities compared with high mass sources \citep{vandishoeck95}.  Subsequent deeper observations revealed that doubly- and even triply-deuterated molecules are common toward IRAS16293 and the nearby cold clump IRAS16293E: D$_2$CO \citep{ceccarelli98d2co,loinard00}, ND$_2$H \citep{loinard01,lis06,gerin06}, ND$_3$ \citep{roueff05}, D$_2$O \citep{butner07,vastel10}, CHD$_2$OH \citep{parise02} and CD$_3$OH \citep{parise04}. The extreme deuteration of D$_2$CO over scales of a few thousand AU was considered to be a smoking gun of active grain surface chemistry \citep{ceccarelli01d2co}. Around the same time, the detection of H$_2$D$^+$ in protostellar sources including IRAS16293 \citep{stark99,stark04} and pre-stellar cores \citep{caselli03} pointed toward enhanced abundances of gaseous H$_2$D$^+$, D$_2$H$^+$ and even D$_3^+$ in cold cores assisted by heavy freeze-out of CO, which in turn drive high deuteration fractions of other gas-phase species \citep{roberts00,roberts03}. D$_2$H$^+$ was indeed first detected toward IRAS16293E by \cite{vastel04}.

Another astrochemical milestone came from highly sensitive IRAM 30m spectra revealing a much larger variety of complex organic molecules than found in early data, including HCOOCH$_3$, CH$_3$OCH$_3$ and C$_2$H$_5$CN \citep{cazaux03}. Subsequent full spectral scans at 3, 2, 1 and 0.9~mm with the IRAM 30m and JCMT as part of the TIMASSS survey \citep{caux11} detected thousands of lines belonging to simple and complex species \citep{jaber14}, including the first detection of an amide, NH$_2$CHO (formamide), in a low-mass protostar \citep{kahane13}. An early search for the simplest amino acid, glycine, was unsuccessful \citep{ceccarelli00glycine}. IRAS16293 has also been surveyed with the HIFI instrument on the {\it Herschel} Space Observatory in various bands between 460~GHz and 2~THz \citep{ceccarelli10,hilyblant10,bacmann10,bottinelli14} as well as the GREAT instrument on SOFIA \citep{parise12}, revealing various (deuterated) hydrides, water lines (see below) and high-excitation lines of heavier molecules.

Water is clearly abundant near IRAS16293, revealed originally by maser emission at radio wavelengths \citep{wootten89,furuya01} and subsequently by thermal emission using the {\it Infrared Space
  Observatory} (ISO) \citep{ceccarelli98h2o,ceccarelli99} and most recently with the {\it Herschel} Space Observatory \citep{ceccarelli10,coutens12}. {\it Herschel} observed lines over a large energy range as well as from minor isotopologues, H$_2^{18}$O and H$_2^{17}$O, which are less dominated by the strong outflow emission than H$_2^{16}$O. Several lines of deuterated water, HDO \citep{coutens12} and D$_2$O \citep{vastel10,coutens13} were detected as well.

Determination of abundances from this wealth of data has been hindered by the realization that there are gradients in temperature, density and chemistry throughout the source.  Early analyses hinted at a jump in abundances of organic molecules in the inner envelope by up to two orders of magnitude taking into account the beam dilution of the inner warm region in the large observing beam \citep{vandishoeck95}. Quantitative abundance determinations with varying temperature and density profiles using jump abundance models were subsequently introduced when non-LTE excitation and radiative transfer codes became available \citep{vandertak00,ceccarelli00model,schoeier02,ceccarelli03,doty04}. The jump in abundance was typically put at dust temperatures around 100 K when water ice sublimates, together with any species embedded in the ice \citep{fraser01}, although H$_2$CO was recognized to return to the gas at lower temperatures around 50 K \citep{ceccarelli01d2co,schoeier04} and CO at even lower temperatures around 25 K \citep{jorgensen02}. This inner region with temperatures above 100 K is also called the hot core or hot corino. The realization of varying abundances also pointed to another conundrum, namely that optically thin C$^{18}$O emission could not be used as a tracer of the H$_2$ column density since the emission arises from a much larger area than that of the organic molecules. This is one of the main reasons for the large discrepancies in derived abundances that are scattered throughout the literature \citep[see][for further details]{herbst09}. For molecules with a sufficient number of observed lines originating from a range of upper energy levels, inner and outer envelope abundances can be determined within the context of the adopted physical model \citep{ceccarelli00h2co,schoeier02,maret04,maret05,hotcoresample,coutens12}. In the systematic search for complex organic molecules in the TIMASSS survey, it was found that a number of species, including ketene, acetaldehyde, formamide, dimethyl ether and methyl formate, show emission not just from the inner hot core but also from the cold extended envelope probed by the single-dish observations \citep{jaber14}.

The early data also triggered the debate of the importance of thermal sublimation of ices in the inner hot core region versus shocks associated with the outflows. The broad line profiles of Si- and S-bearing molecules as well as some complex organic molecules like CH$_3$OH and CH$_3$CN hinted at the importance of sputtering along the outflow walls, much like seen for SiO and CH$_3$OH at outflow spots away from the protostar in this and other sources \citep[e.g.,][]{bachiller97,hirano01,garay02,n1333i2art}. These lines are clearly broader than can be explained by an infalling envelope.  A particularly important molecule in the debate of the importance of shocks versus quiescent hot core emission is water itself, with both origins being argued in the literature \citep{ceccarelli00model,nisini02}. New data on optically thin isotopologues as well as spatially resolved interferometry by \cite{persson13} and such as presented here provide evidence for both scenarios.

\subsubsection{Interferometric observations}

Early interferometer observations revealed elongated C$^{18}$O and CS emission coincident with the dust distributions on $\sim$800~AU scales, with NH$_3$ arising from a larger (8000~AU diameter) region \citep{mundy90,walker90}. Also, very strong SO emission was found centered on IRAS16293A but not on IRAS16293B, providing early evidence for chemical differentiation between the two sources \citep{mundy92}.

Once interferometers grew from 3 to 6 telescopes and the Submillimeter Array (SMA) was inaugurated, IRAS16293 became a prime target for spatially resolved astrochemical studies. The first arcsecond resolution images by \cite{bottinelli04iras16293} and \cite{kuan04} demonstrated that the emission from complex organic molecules peak at the locations of the components of the protostellar binary on scales argued to be consistent with thermal evaporation. Both studies showed some evidence for differences between the intensities of lines of different species toward the two sources, but with the emission toward IRAS16293A generally being brighter. A systematic interferometric study \citep{bisschop08,iras16293sma} demonstrated the association of some nitrogen-containing organic molecules (HNCO, CH$_3$CN) primarily with IRAS16293A while a number of oxygen-containing species are present with comparable intensities toward both sources and one, acetaldehyde (CH$_3$CHO), almost exclusively associated with IRAS16293B. When normalised to the column densities of methanol, these differences translate into relative abundance differences of up to an order of magnitude between different species toward the two sources \citep{bisschop08}. Subarcsecond resolution observations by \cite{chandler05} indicated that the emission from molecules characterised by high rotation temperatures peak toward one of the shock positions in the vicinity of IRAS16293A rather than the protostar itself. The presence of shocks on small scales toward IRAS16293B is demonstrated by high angular resolution images of SiO \citep{iras16293sma}, with some evidence for the importance of shocks on other species (including water on larger scales). Recently, \cite{oya16} analyzed the kinematics around source A and found evidence for chemical differentiation between OCS on the one hand and CH$_3$OH and CH$_3$OCHO on the other with the latter emission more compact, possibly associated with weak shocks at the centrifugal barrier of the disk forming around IRAS16293A.

Despite the natural focus of the interferometric studies on the warm, compact gas, the observations also demonstrated the importance of the colder parts of the protostellar environments. For example, observations of lower excited H$_2$CO \citep{schoeier04} and HNCO \citep{bisschop08} transitions show that significant abundance (and optical depth) variations are present in the colder gas in the circumbinary envelope. The images also demonstrated spatial differences between species such as C$^{18}$O, DCO$^+$ and N$_2$D$^+$ in the colder gas that can be explained through relatively simple gas-phase chemistry \citep{iras16293sma}.

The chemical richness (and differentiations) of the gas toward IRAS16293, seen in the 6--8 element interferometric data, made it an obvious target for science verification observations with ALMA at 220~GHz or 1.4~mm \citep{pineda12,jorgensen12} and at 690~GHz or 0.43~mm \citep{baryshev15}. These early observations were already more than an order of magnitude more sensitive than the previous interferometric data and, as mentioned above, showed the first discovery of a prebiotic molecule, glycolaldehyde, toward a solar-type protostar \citep{jorgensen12}. Additional science verification observations at 690~GHz (0.4~mm) produced images of one line of the H$_2^{18}$O water isotopologue. Together with observations of both H$_2^{18}$O and HDO from the Submillimeter Array, these measurements constrained the HDO/H$_2$O abundance ratio to $9\times 10^{-4}$ \citep{persson13}, much lower than the value in the colder parts of the envelope \citep[e.g.,][]{coutens12} and more in line with the ratios for Earth's oceans and Solar System comets.

\section{Observations}\label{observations}
IRAS~16293$-$2422 was observed as part of the ``Protostellar Interferometric Line Survey (PILS)'' program (PI: Jes K. J{\o}rgensen). The survey consists of an unbiased spectral survey covering a significant part of ALMA's Band~7 (wavelengths of approximately 0.8~mm) in ALMA's Cycle~2 (project-id: 2013.1.00278.S) as well as selected windows in ALMA's Bands~3 ($\approx$3~mm, or 100~GHz) and 6 ($\approx$1.3~mm, or 230~GHz) obtained in Cycle~1 (project-id: 2012.1.00712.S). The following sections describe each of these datasets\footnote{The data are available through the ALMA archive at https://almascience.eso.org/aq.} and the reduction process.

\subsection{Band~7 data: 329--363~GHz unbiased line survey}
\subsubsection{Data, calibration and imaging}
The Band~7 part of the survey covers the frequency range from 329.147~GHz to 362.896~GHz in full. Data were obtained from both the array of 12~m dishes (typically 36--41~antennas in the array at the time of observations) and the ``Atacama Compact Array (ACA)'', or ``Morita Array'', of 7~m dishes (typically 8--10~antennas). Table~\ref{obslog7} presents a full log of the Band~7 observations. The pointing center was in both cases set to be a location between the two components of the binary system at $\alpha_{\rm J2000}=$16$^{\text{h}}$32$^{\text{m}}$22\fs72; $\delta_{\rm J2000}=-$24\degr28\arcmin34\farcs3. In total 18~spectral settings were observed: each setting covers a bandwidth of 1875~MHz (over four different spectral windows, 468.75~MHz wide). To limit the data rate the data were downsampled by a factor of two relative to the native spectral resolution of the array, resulting in a spectral resolution of 0.244~MHz ($\approx 0.2$~km~s$^{-1}$) over 1920 channels for each spectral window. Each setting was observed with approximately 13~minutes integration on source (execution blocks of approximately 40~minutes including calibrations) for the 12~m array and double that for the ACA.

The calibration and imaging proceeded according to the standard recipes in CASA. Titan, or in a few instances Ceres, were used as flux calibrators while the quasar J1517$-$2422 ($\approx$1.2~Jy) was used as bandpass calibrator. Observations of the science target were interspersed with observations of the closeby quasar J1625-2527 ($\approx$0.5~Jy; 1.8\degr) that was used as phase calibrator. During the reduction, data of bad quality were flagged, partly automatically by the CASA tasks and partly through manual inspection. A phase-only self-calibration was performed on the continuum datasets and applied to the full data cubes before combining the 12~m array and ACA datasets and performing the final imaging. The resulting spectral line data cubes have a noise RMS for the combined datasets of about 7--10~mJy~beam$^{-1}$~channel$^{-1}$, which translates into a uniform sensitivity of approximately 4--5~mJy~beam$^{-1}$~km~s$^{-1}$ with beam sizes ranging from $\approx$0.4--0.7$''$ depending on the exact configuration at the date of observation (Table~\ref{obslog7}). For ease of comparison accross the different spectral windows and extraction of spectra, a combined dataset with a circular restoring beam of 0.5\arcsec\ was produced as well.

\subsubsection{Continuum subtraction}
For analysis and, in particular, the imaging of individual lines it is desirable to have continuum subtracted products. The standard procedure is to define one or more spectral regions with ``line-free'' channels to be used for continuum estimates and subtraction either in the $(u,v)$- or image-plane. In this case, however, because of the high sensitivity (and dynamic range) the resulting image cubes are strongly line confused in large parts of the spectral ranges even toward IRAS16293B with the narrowest lines. Also, some velocity gradients are seen, which makes a general definition of ``line-free'' regions impossible. Finally, fitting each spatial pixel by hand is not practical because of the large number of individual spectra (each datacube representing more than 10,000 spectra and spectral coverage of more than 140,000 independent channels).

\begin{figure}
\resizebox{\hsize}{!}{\includegraphics{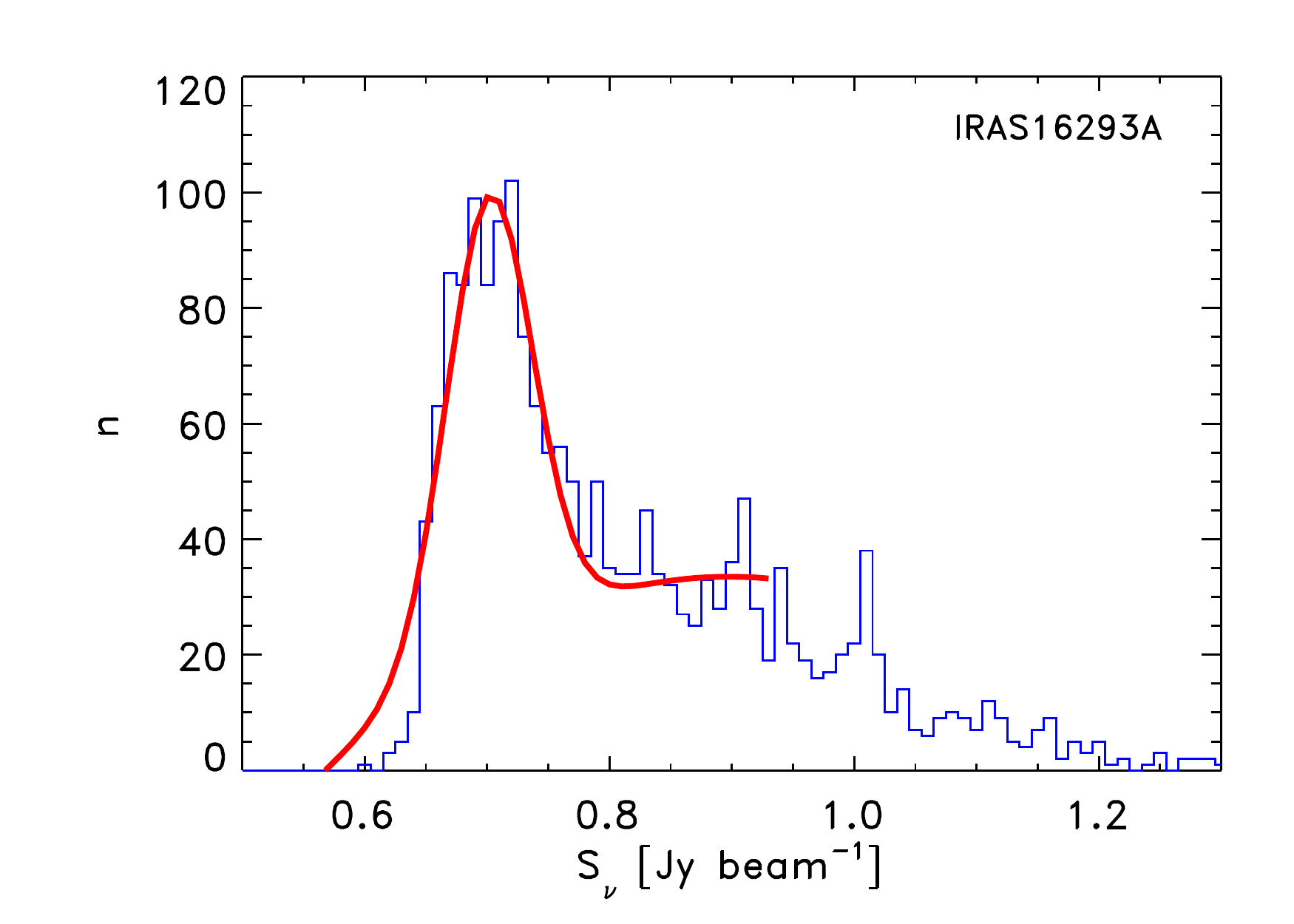}}
\resizebox{\hsize}{!}{\includegraphics{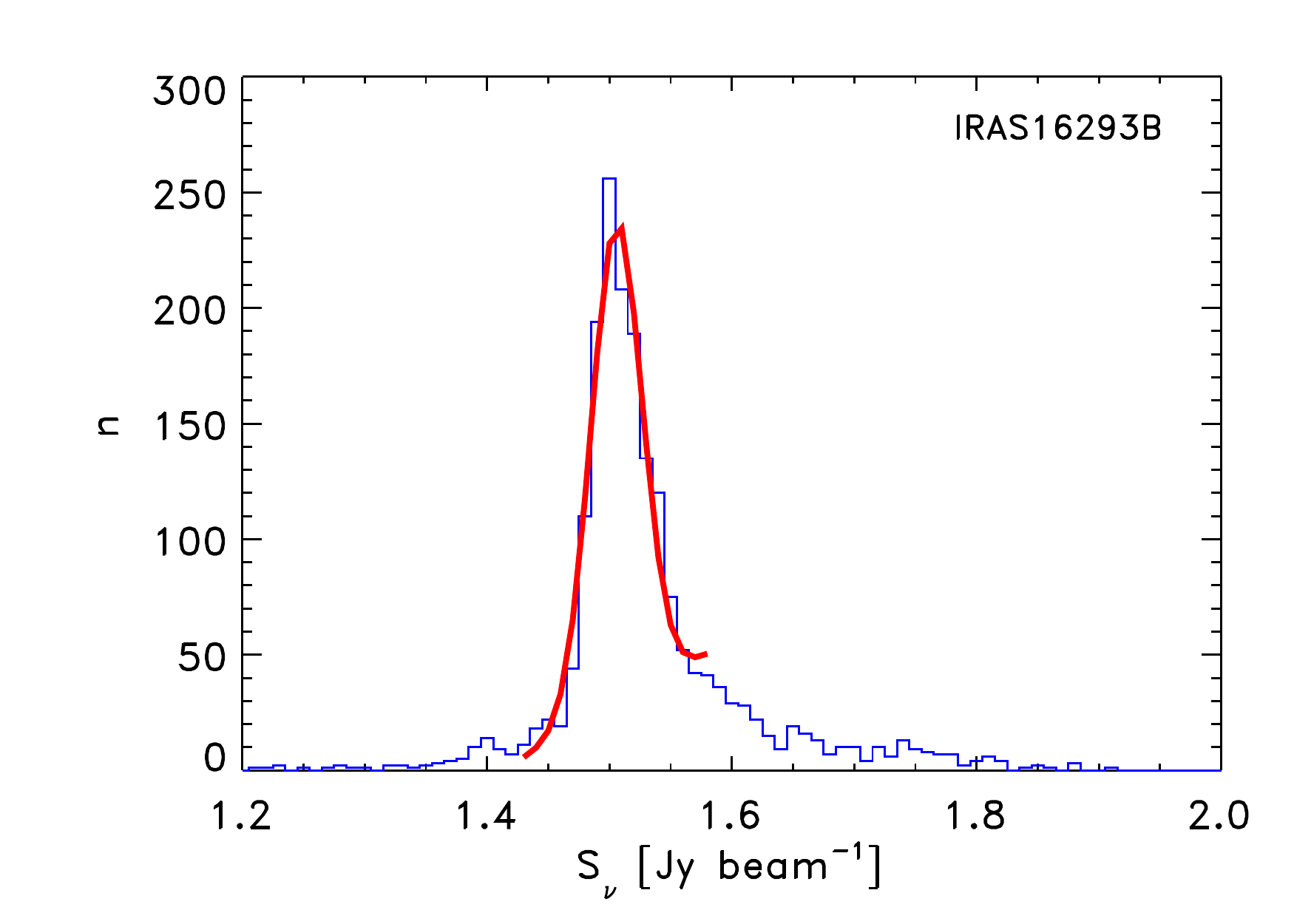}}
\caption{Methodology for continuum subtraction illustrated with data from two representative pixels toward IRAS16293A (\emph{upper}) and IRAS16293B (\emph{lower}): shown are the flux distributions for the two pixels (histogram) with the resulting fit overlaid (red line).}\label{continuum_sub}
\end{figure}

To circumvent these issues and treat the data homogeneously, the continuum subtraction is done in a statistical/iterative manner: for each pixel in a datacube the density distribution of flux-values is created (Fig.~\ref{continuum_sub}). For a pixel with only a contribution from continuum emission (or noise) but no line emission, this distribution will represent a symmetric Gaussian centered at the continuum level with a width corresponding to the channel-by-channel RMS noise, $\sigma$. For pixels with both continuum and line emission this Gaussian will be modified by an exponential tail toward higher values, with a number of datapoints at lower flux values in case of absorption (in particular, toward IRAS16293B). In either of these cases, however, the continuum level can still be recognised as the leading edge in the flux distribution (Fig.~\ref{continuum_sub}) with the overall distribution appearing as a log-normal or skewed Gaussian distribution. 

The continuum is then determined in two automated steps: first, the peak of the distribution is determined over the entire range of the flux values with a simple symmetric Gaussian fit resulting in an estimate of the centroid $F$ and its width $\Delta F$. In case of pixels with no or little line emission the width is small, but for line contaminated regions the distribution becomes broader and the exact location of the peak more uncertain. Subsequently, a skewed Gaussian is fitted to the part of the distribution within $F\pm 3\Delta F$ and the new centroid (now not necessarily symmetric) is recorded as the continuum level that can then be used for continuum subtraction for that particular pixel. This method does not fit the low flux tail corresponding to absorption lines, but generally these channels do not contribute significantly to the overall distribution. Experiments with the observed cubes and synthetic models suggest that the continuum level can be estimated in this manner to be accurate to within 2$\sigma$.  

\begin{figure}
\resizebox{\hsize}{!}{\includegraphics{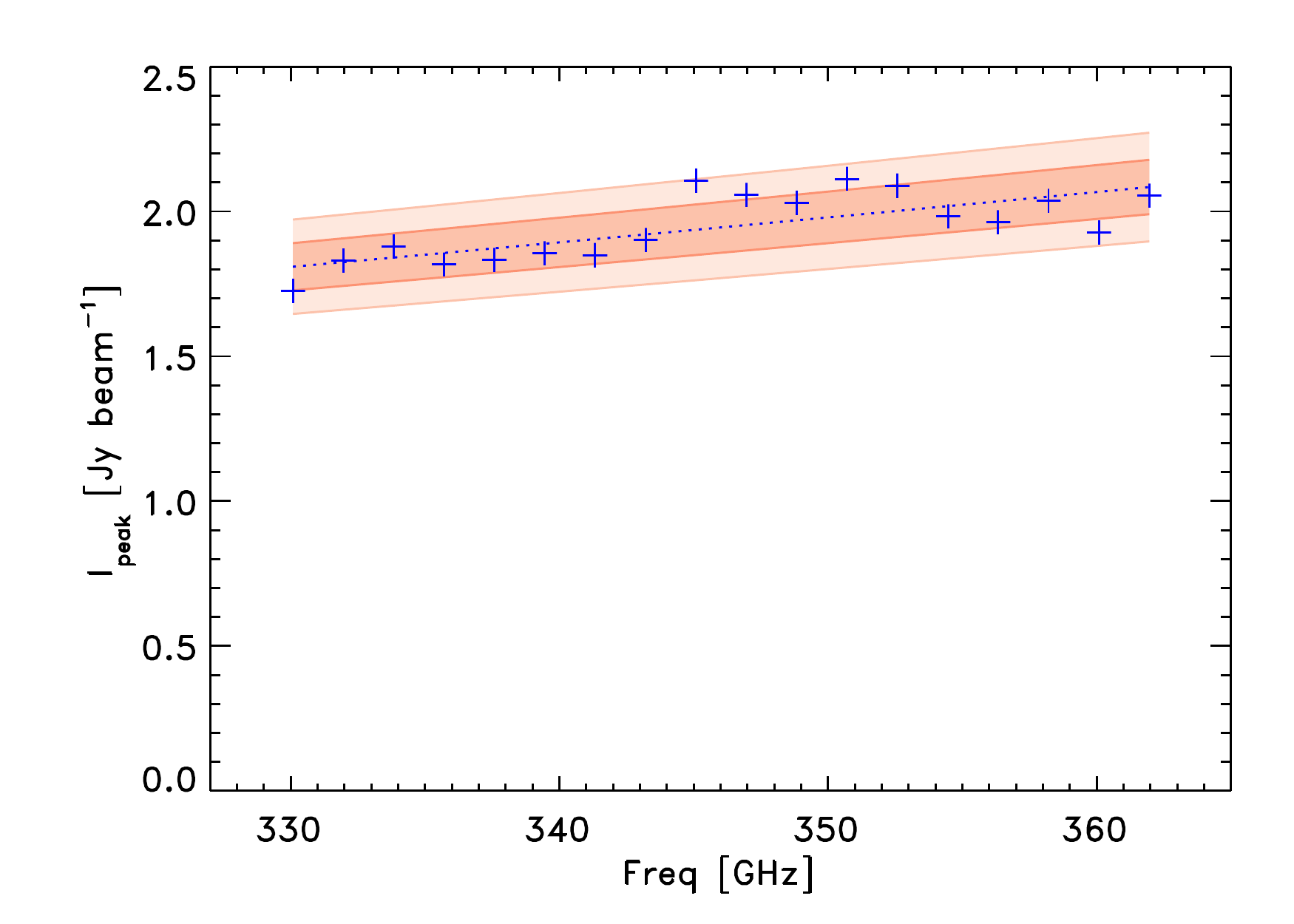}}
\caption{Continuum peak flux (in a 0.5$''$ beam toward IRAS16293B for the 18 different spectral setups (plus-signs) as function of frequency. A slight increase in flux as function of frequency is seen that can be approximated by a power-law $F_\nu \propto \nu^{2}$. The shaded areas correspond to this dependency $\pm 5$\% (darker color) and $\pm 10$\% (lighter color). About 70\% of the measurements are within the $\pm 5$\% region with all of the measurements within the $\pm 10$\% area.}\label{continuum_slope}
\end{figure}

Figure~\ref{continuum_slope} compares the continuum peak flux toward IRAS16293B for each of the 18 spectral set\-ups as a function of frequency. As expected a slight slope is seen with flux increasing as function of frequency consistent with a power-law $F_\nu \propto \nu^\alpha$ with $\alpha \approx 2$ as expected for optically thick dust continuum emission. This plot also provides a direct estimate of the calibration accuracy: the dark and light shaded areas correspond to respectively $\pm 5$\% and $\pm 10$\% around the continuum power-law. Thirteen out of the eighteen measurements (or 72\%) are within $\pm 5$\% while all of the measurements lie within the $\pm 10$\% ranges. Assuming that the scatter is entirely due to errors in the calibration, this suggests that the (flux) calibration accuracy is better than 5\% for the ALMA observations.

\begin{table*}
\caption{Log of Band~7 observations.}\label{obslog7}
\begin{tabular}{llllll} \hline\hline
Setting & Frequency [GHz] & \multicolumn{2}{c}{Observing date}      & Number of antennas$^{a}$ & Beam$^b$ \\
& & \multicolumn{1}{c}{12~m array}  & \multicolumn{1}{c}{ACA}         & & \\ \hline
a       & 329.150--331.025 & 2015-May-16 & 2014-Jun-08 &   41 / 10          & 0.49$''$$\times$0.37$''$ ($+$77\degr)\\
b       & 331.025--332.900 & 2015-May-17 & 2014-Jun-14 &   36 / 9           & 0.49$''$$\times$0.40$''$ ($+$87\degr)\\
c       & 332.900--334.775 & 2015-May-17 & 2014-Jun-17 &   36 / 8           & 0.55$''$$\times$0.41$''$ ($-$87\degr)\\
d       & 334.775--336.650 & 2015-May-17 & 2014-Jun-08 &   36 / 10          & 0.43$''$$\times$0.39$''$ ($+$72\degr)\\
e       & 336.650--338.525 & 2015-May-17 & 2015-Apr-04 &   36 / 9           & 0.45$''$$\times$0.39$''$ ($+$79\degr)\\
f       & 338.525--340.400 & 2015-May-21 & 2014-Jun-30 &   36 / 10          & 0.44$''$$\times$0.40$''$ ($+$37\degr)\\
g       & 340.400--342.275 & 2015-May-21 & 2014-Jun-14 &   36 / 9           & 0.42$''$$\times$0.38$''$ ($+$56\degr)\\
h       & 342.275--344.150 & 2015-May-21 & 2014-Jun-11 &   36 / 10          & 0.43$''$$\times$0.37$''$ ($+$56\degr)\\
i       & 344.150--346.025 & 2015-Apr-05 & 2014-Jun-29 &   39 / 11          & 0.87$''$$\times$0.60$''$ ($+$86\degr)\\
j       & 346.025--347.900 & 2015-Apr-05 & 2014-Jun-12 &   40 / 9           & 0.81$''$$\times$0.57$''$ ($+$82\degr)\\
k       & 347.900--349.775 & 2015-Apr-04 & 2014-Jun-04 &   37 / 10          & 0.82$''$$\times$0.57$''$ ($+$85\degr)\\
l       & 349.775--351.650 & 2015-May-17 & 2014-Jun-10 &   36 / 8           & 0.65$''$$\times$0.39$''$ ($-$81\degr)\\
m       & 351.650--353.525 & 2015-Apr-03 & 2014-Jun-10 &   38 / 8           & 0.79$''$$\times$0.58$''$ ($+$79\degr)\\
n       & 353.525--355.400 & 2015-May-19 & 2014-Dec-29 &   37 / 11          & 0.44$''$$\times$0.34$''$ ($-$82\degr)\\
o       & 355.400--357.275 & 2015-May-20 & 2014-Jun-17 &   37 / 8           & 0.40$''$$\times$0.37$''$ ($+$45\degr)\\
p       & 357.275--359.150 & 2015-May-20 & 2014-Jun-16 &   37 / 10          & 0.42$''$$\times$0.37$''$ ($+$53\degr)\\
q       & 359.150--361.025 & 2015-May-20 & 2014-Jun-14 &   37 / 9           & 0.44$''$$\times$0.38$''$ ($+$60\degr)\\
r       & 361.025--362.900 & 2015-May-21 & 2014-Jun-08 &   36 / 10          & 0.43$''$$\times$0.38$''$ ($+$75\degr)\\ \hline
\end{tabular}

Notes: $^{a}$Number of antennas available for the main array and ACA, respectively. $^b$Synthesized beam size and position angle for the 12~m and ACA datasets combined.
\end{table*}

\subsection{Band~3 and Band~6 data: observations of specific spectral windows at 100 and 230 GHz}
In addition to the unbiased survey in Band~7, three spectral setups were observed in Band~6 and Band~3 between May and July 2014 targeting the same position. Those setups were optimised to $\emph{(i)}$ observe the emission from a range of glycolaldehyde lines previously measured in the laboratory as an extension of the detection of that molecule \citep{jorgensen12} and $\emph{(ii)}$ map species in the chemical network related to glycolaldehyde formation. Table~\ref{cycle1log} lists windows targeted in these set\-ups as well as the logs of observations. For these settings only the 12-m array was utilised, and any more extended emission will thus be resolved-out. The calibration otherwise proceeded in the same manner as for the Band~7 data.

\begin{table*}
\caption{Log of ALMA Band~3 and Band~6 observations.}\label{cycle1log}
\begin{tabular}{llllll}\hline\hline
      & Window & Frequency [GHz]    & Date              & Number of antennas & Beam$^a$ \\ \hline
B3    & 0      & 92.78--93.02       & 2014-Jul-03       & 31                 & 1.13$''$$\times$0.93$''$ ($+$15\degr) \\
      & 1      & 89.45--89.74       &                   &                    &           \\
      & 2      & 102.48--102.73     &                   &                    &           \\
      & 3      & 103.18--103.42     &                   &                    &           \\ \hline
B6-1  & 0      & 239.40--239.86     & 2014-Jun-14       & 35                 & 0.48$''$$\times$0.43$''$ ($+$64\degr) \\
      & 1      & 240.15--240.61     &                   &                    &           \\
      & 2      & 224.75--225.21     &                   &                    &           \\
      & 3      & 221.76--222.22     &                   &                    &           \\ \hline
B6-2  & 0      & 247.32--247.79     & 2014-May-22       & 35                 & 0.62$''$$\times$0.48$''$ ($+$77\degr) \\ 
      & 1      & 250.28--250.75     &                   &                    &           \\
      & 2      & 231.03--231.51     &                   &                    &           \\
      & 3      & 232.18--232.65     &                   &                    &           \\ \hline
\end{tabular}

$^a$Synthesized beam size and position angle for the dataset.
\end{table*}

\section{Results}\label{results}
\subsection{Continuum emission}\label{continuum_results}
\begin{figure}
\resizebox{\hsize}{!}{\includegraphics{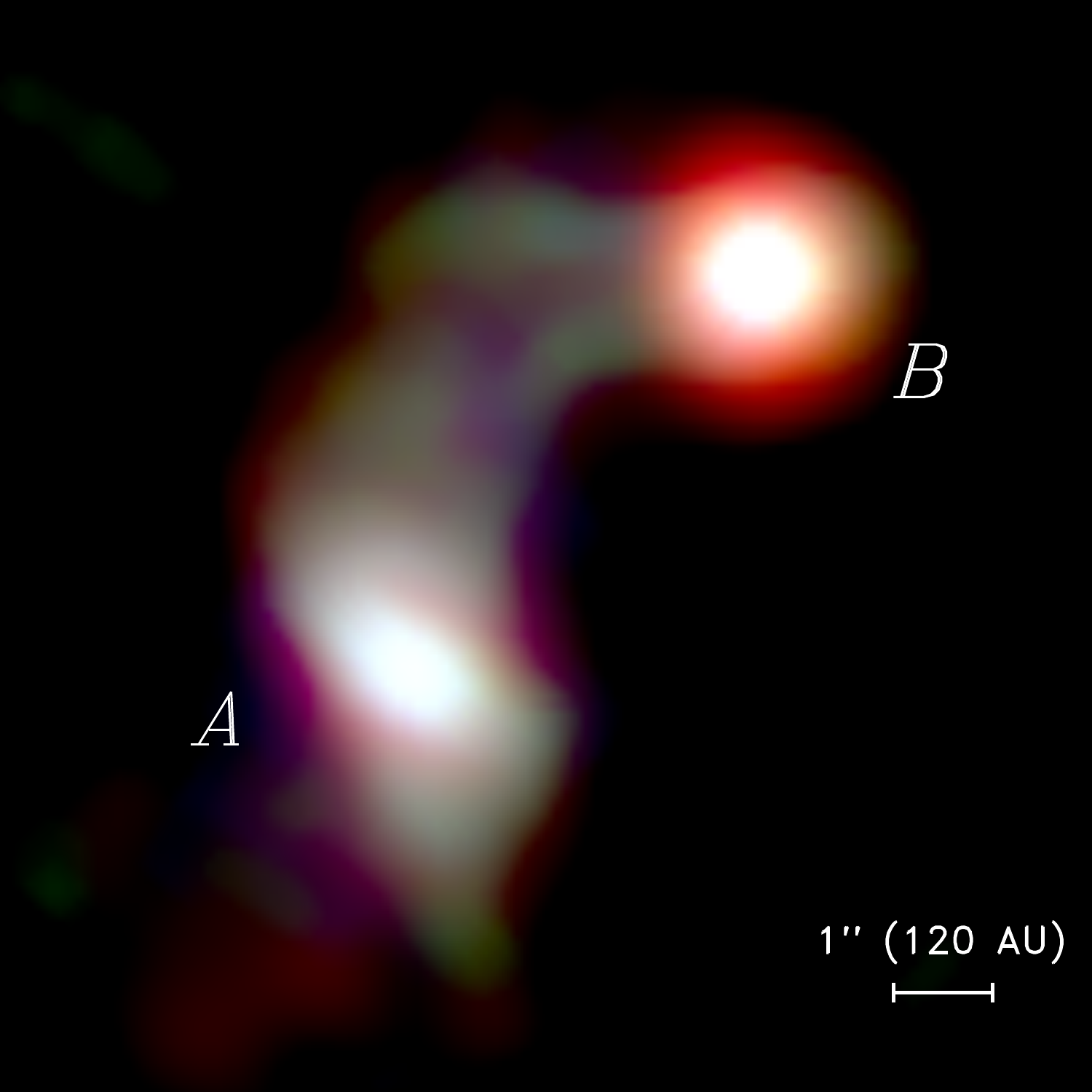}}
\caption{Three-color image showing the continuum at 3.0~mm, 1.3~mm and 0.87~mm (ALMA Bands 3, 6 and 7) in red, green and blue, respectively. Before the combination the 1.3~mm and 0.87~mm images were smoothed to match the resolution of the 3.0~mm data.}\label{threecolor}
\end{figure}

\begin{figure*}
\resizebox{\hsize}{!}{\includegraphics{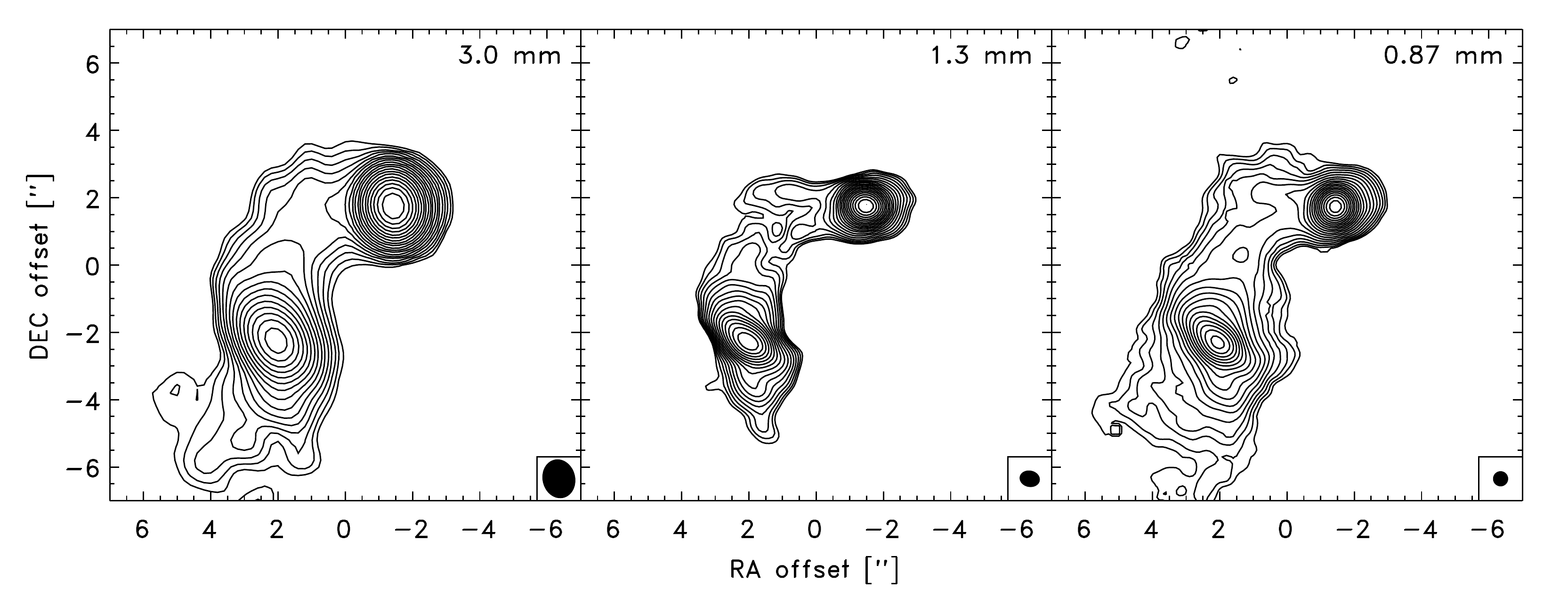}}
\caption{Continuum images at 3.0, 1.3 and 0.87~mm (left, middle and right) at the angular resolution of each dataaset. The 0.87~mm images include both 12~m array and ACA data, while the 3.0 and 1.3~mm images only contain data from the 12~mm array. The contour levels are given as 20 logarithmically divided levels between the 0.5\% and 100\% fof the peak flux at the given wavelength. The RA and DEC offsets are relative to the phase center for the observations.}\label{continuum_comparison}
\end{figure*}

Much can be learned about the structure of the IRAS~16293$-$2422 system by straightforward inspection of the dust continuum maps.
Figure \ref{threecolor} shows a three-color composite of the continuum towards the system in the 3.0~mm, 1.3~mm and 0.87~mm bands, while Fig.~\ref{continuum_comparison} shows the continuum at the three different wavelengths separately at the angular resolution of each individual dataset. The extended emission connecting the two sources, also noted in the science verification data \citep{pineda12}, is clearly seen. It shows a characteristic bend toward the north of IRAS16293A/east of IRAS16293B. East of IRAS16293B two separate stream-lines pointing away from the source are seen. Toward IRAS16293A additional extended continuum emission is observed toward the southwest. This extension coincides with the N$_2$D$^+$ emission picked-up in SMA observations \citep{iras16293sma} and likely reflects cold material with a high column density.

Another very striking feature of the maps is the clear differences in the morphologies and colors of the emission toward the two protostars. IRAS16293A appears clearly elongated in the northeast/southwest direction (with an aspect ratio of 1.9). \cite{chandler05} used continuum maps at 305~GHz from the SMA to study the system: in ``super-resolution'' images (images for which higher weight was given to the longer baselines and subsequently restored with a beam slightly smaller than the usual ``synthesized beams''), they noted a similar extension, but also found that the source broke up into two separate components named ``Aa'' and ``Ab''. These separate components are not seen in our images even though the beam in the ALMA observations (with regular weighting) is comparable to those ``super-resolution'' images. Our images are therefore more in support of the interpretation that the submillimeter continuum emission toward IRAS16293A represents a more edge-on disk system, also supported by the velocity gradient seen in this direction \citep{pineda12,girart14}.

In stark contrast, the emission toward IRAS16293B is very circular with an elliptical aspect ratio $<1.1$. Its colors are also much redder than both the extended emission and that toward IRAS16293A. This is consistent with the suggestion that the emission toward IRAS16293B is optically thick at the 50~AU scales imaged here as well as at higher frequencies \citep[e.g.,][]{zapata13}. The fact that it appears so optically thick at even longer wavelengths tightens the constraints on the minimum amount of dust toward IRAS16293B: assuming typical dust opacities from \cite{ossenkopf94} with $\kappa_\nu=0.0182$~cm$^{2}$~g$^{-1}$ at 850~$\mu$m an optical depth $\tau_\nu > 1$ implies that the column density toward the continuum peak is
\begin{equation}
N({\rm H}_2) = \frac{\tau_\nu}{\mu_{H_2} m_{\rm H}\kappa_\nu} > 1.2 \times 10^{25}~{\rm cm}^{-2}
\end{equation}
and consequently for a distribution over a 50~AU (diameter) projected circular region that the mass is $\gtrsim$~0.01~$M_\odot$. While this may not appear like a significant amount, it is in fact 1--2 orders of magnitude above the mass on similar scales from larger scale envelope models \citep[e.g.,][]{schoeier02} for IRAS~16293$-$2422 extrapolated to these scales. Likewise, if one assumes that the distribution of the material along the line of sight is comparable to the projected extent on the sky the lower limit to the column density translates into to a density $\gtrsim 3\times 10^{10}$~cm$^{-3}$: if this dust indeed is located in a face-on disk-like structure such as implied by the circular distribution and narrow line-widths toward the source \citep[e.g.,][]{iras16293sma}, the density should be expected to be even higher than this lower limit.

\subsection{Line emission}
\begin{figure*}\centering
\resizebox{\hsize}{!}{\includegraphics{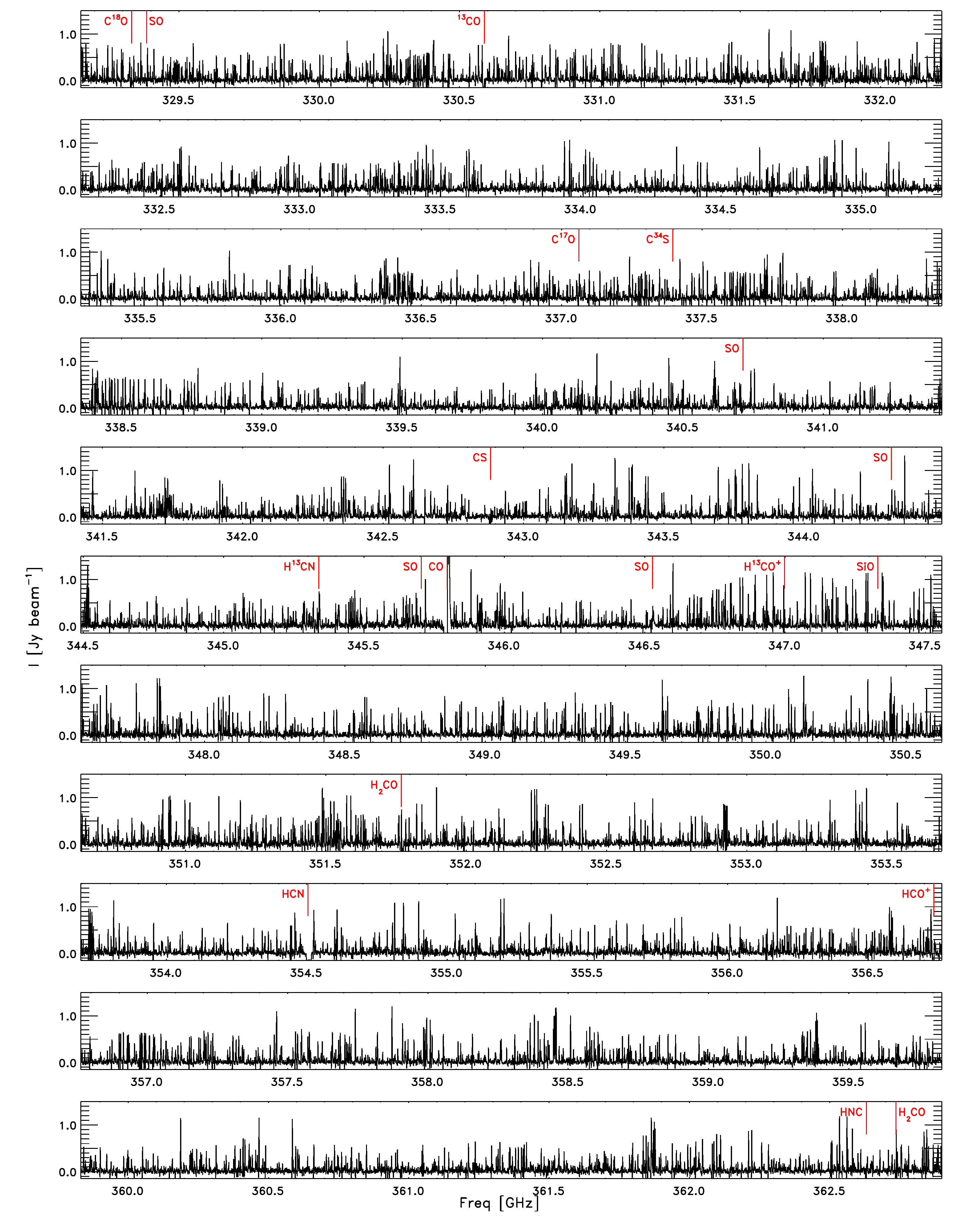}}
\caption{Spectrum in a 0.5$''$ beam toward a position half a beam offset from the continuum peak of IRAS16293B. Frequencies of prominent lines of a few key species are marked by vertical lines.}\label{full_spec}
\end{figure*}

\begin{figure*}
\resizebox{0.2\hsize}{!}{\phantom{xx}}\resizebox{0.4\hsize}{!}{\includegraphics{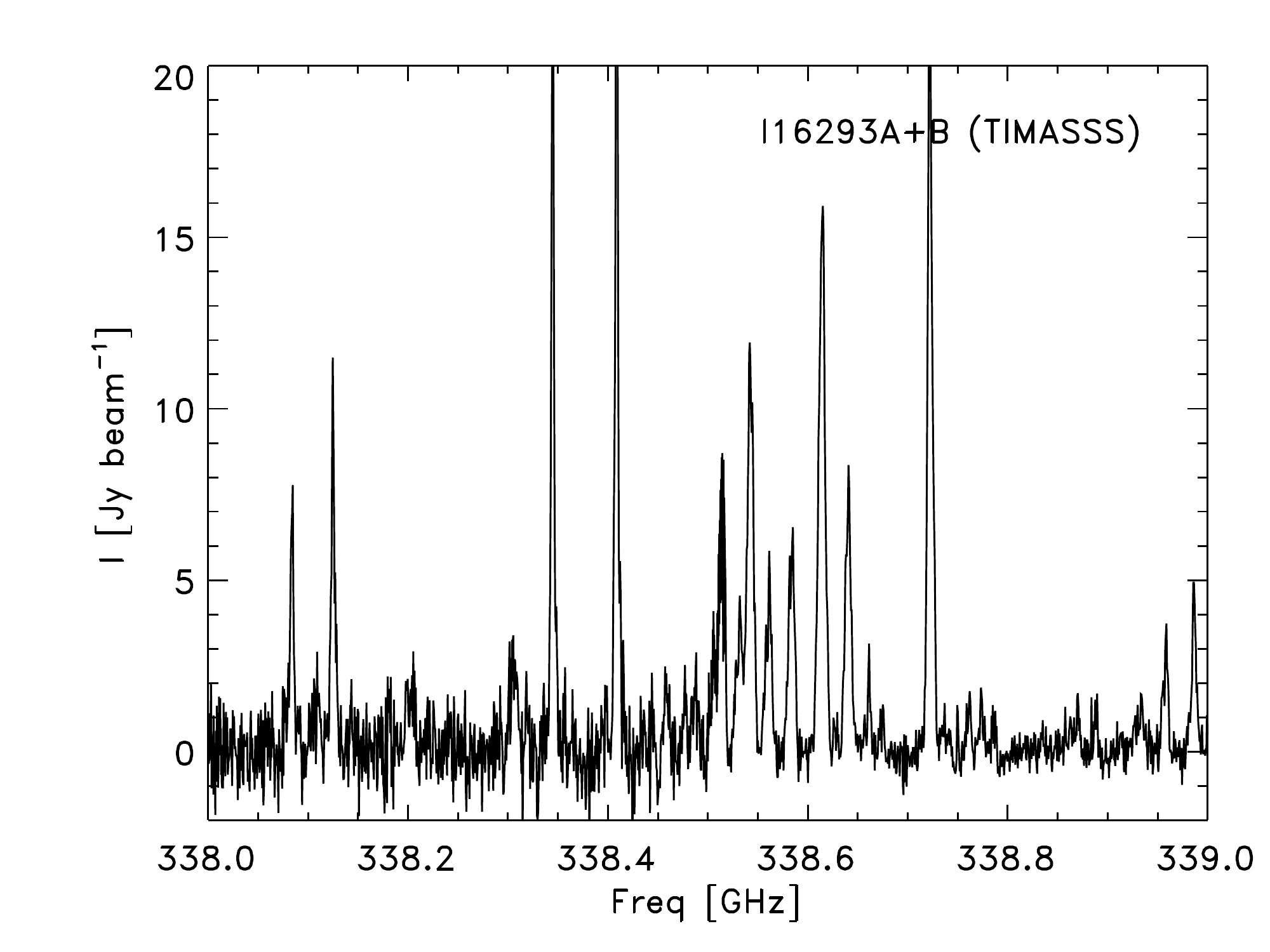}}\resizebox{0.2\hsize}{!}{\phantom{xx}}
\resizebox{0.8\hsize}{!}{\includegraphics{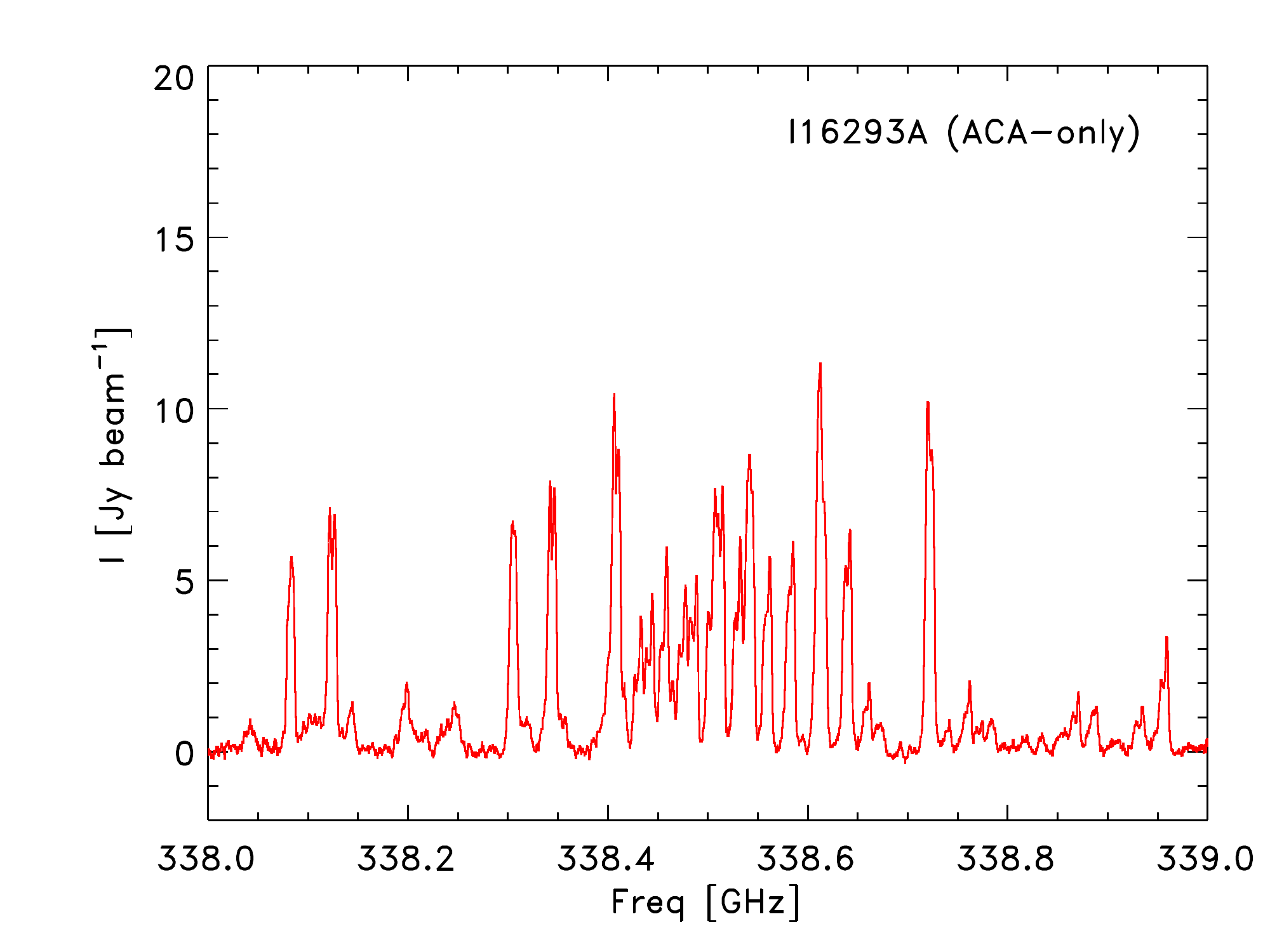}\includegraphics{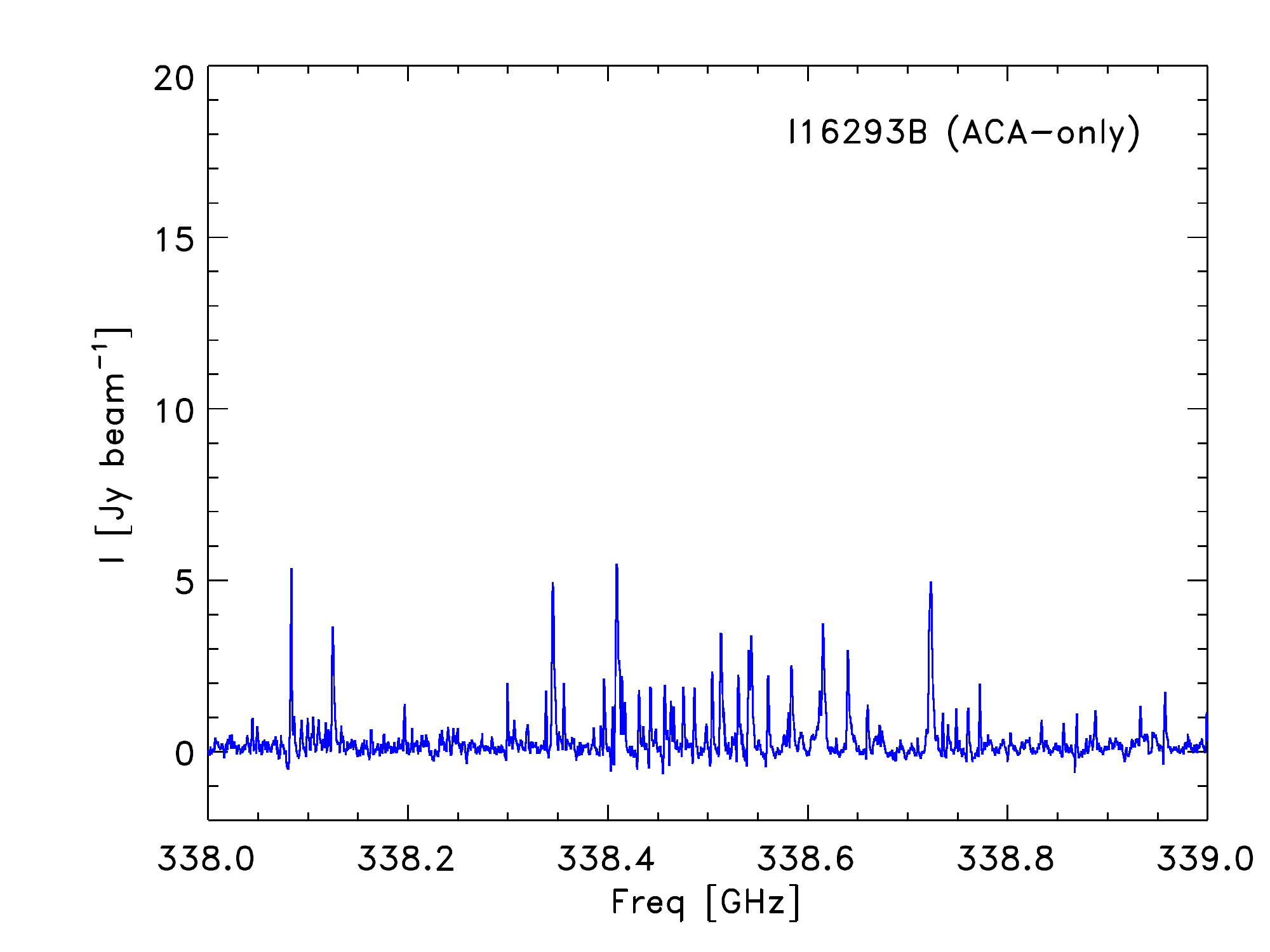}}
\resizebox{0.8\hsize}{!}{\includegraphics{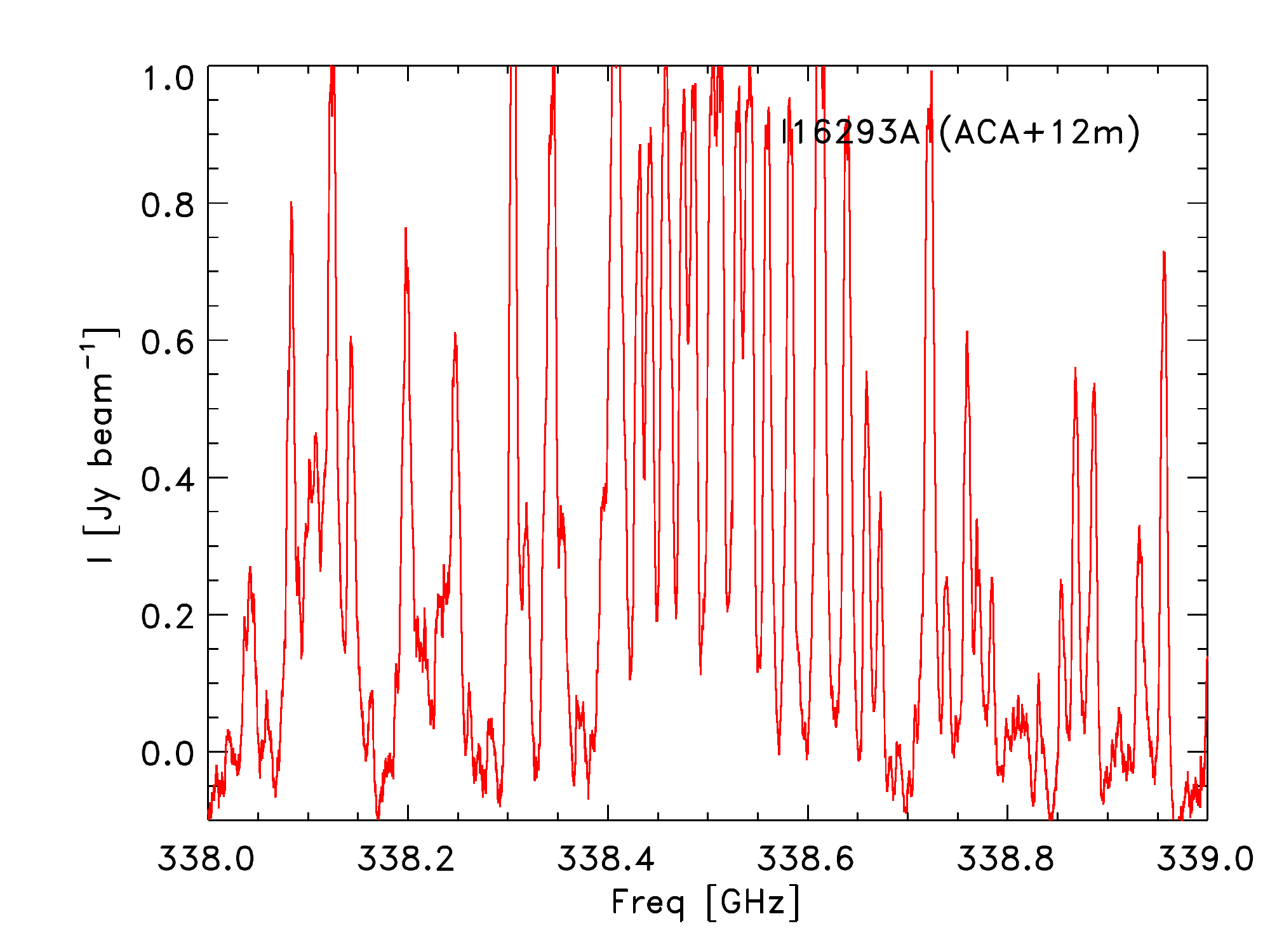}\includegraphics{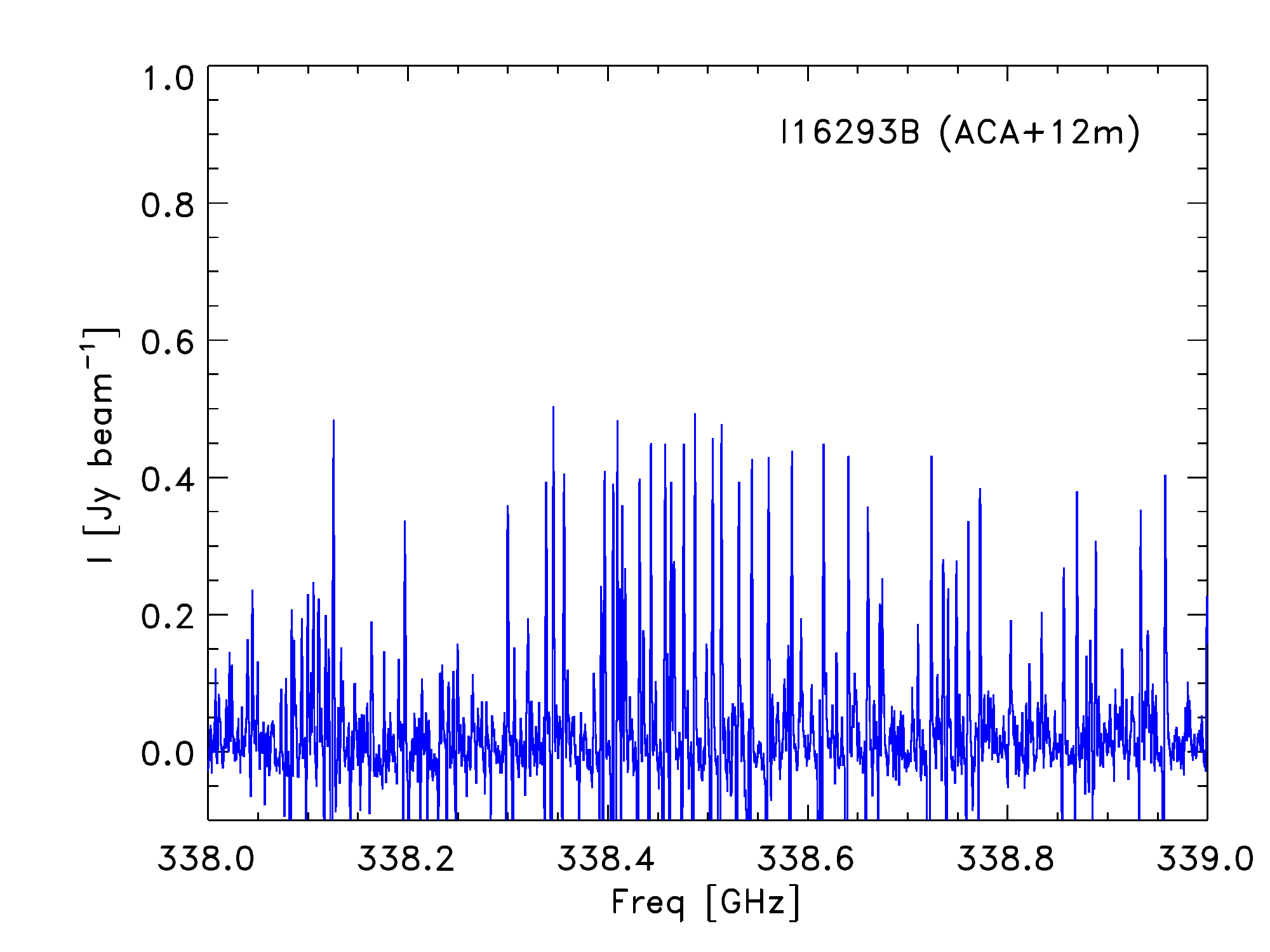}}
\resizebox{0.8\hsize}{!}{\includegraphics{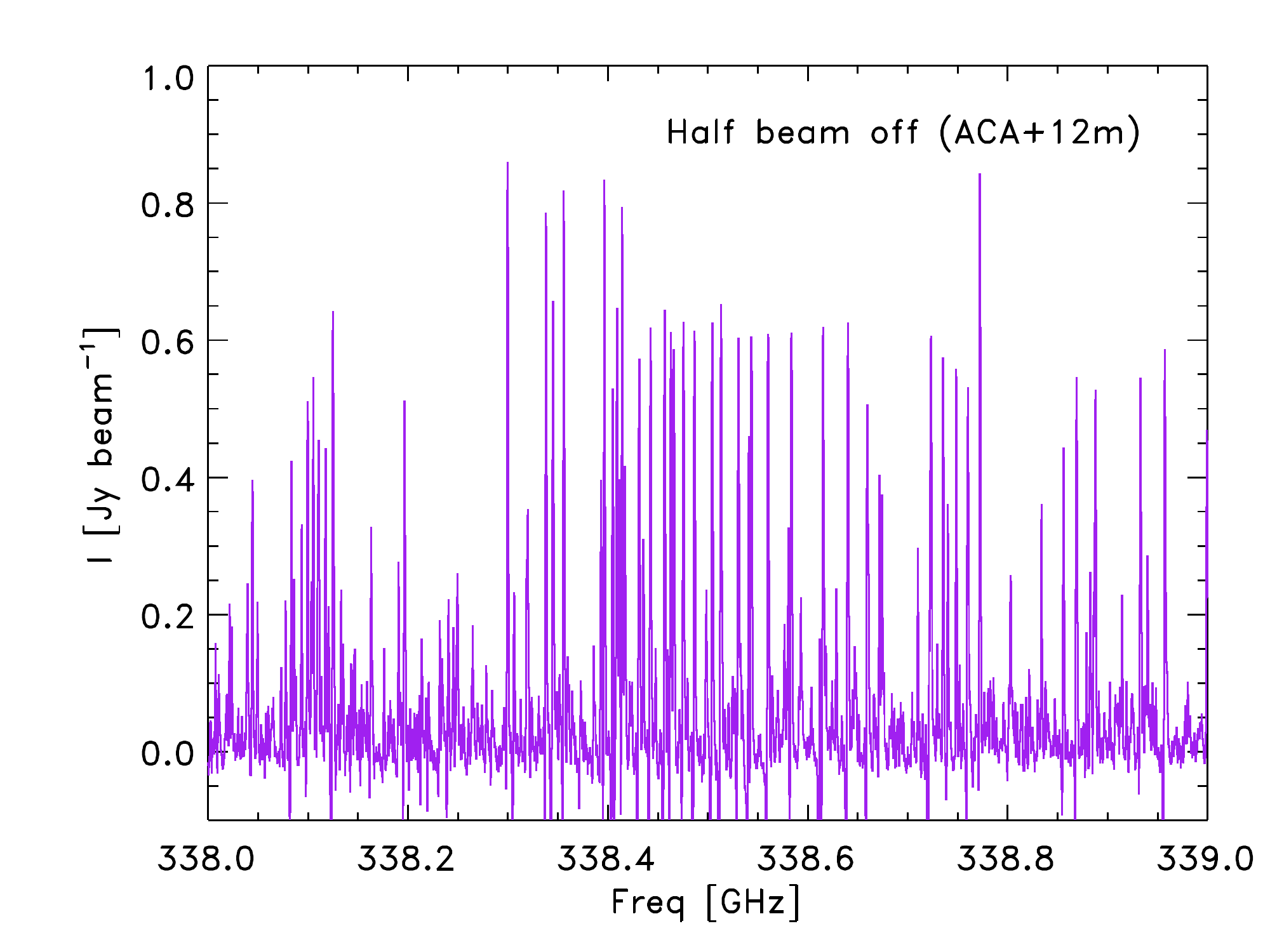}\includegraphics{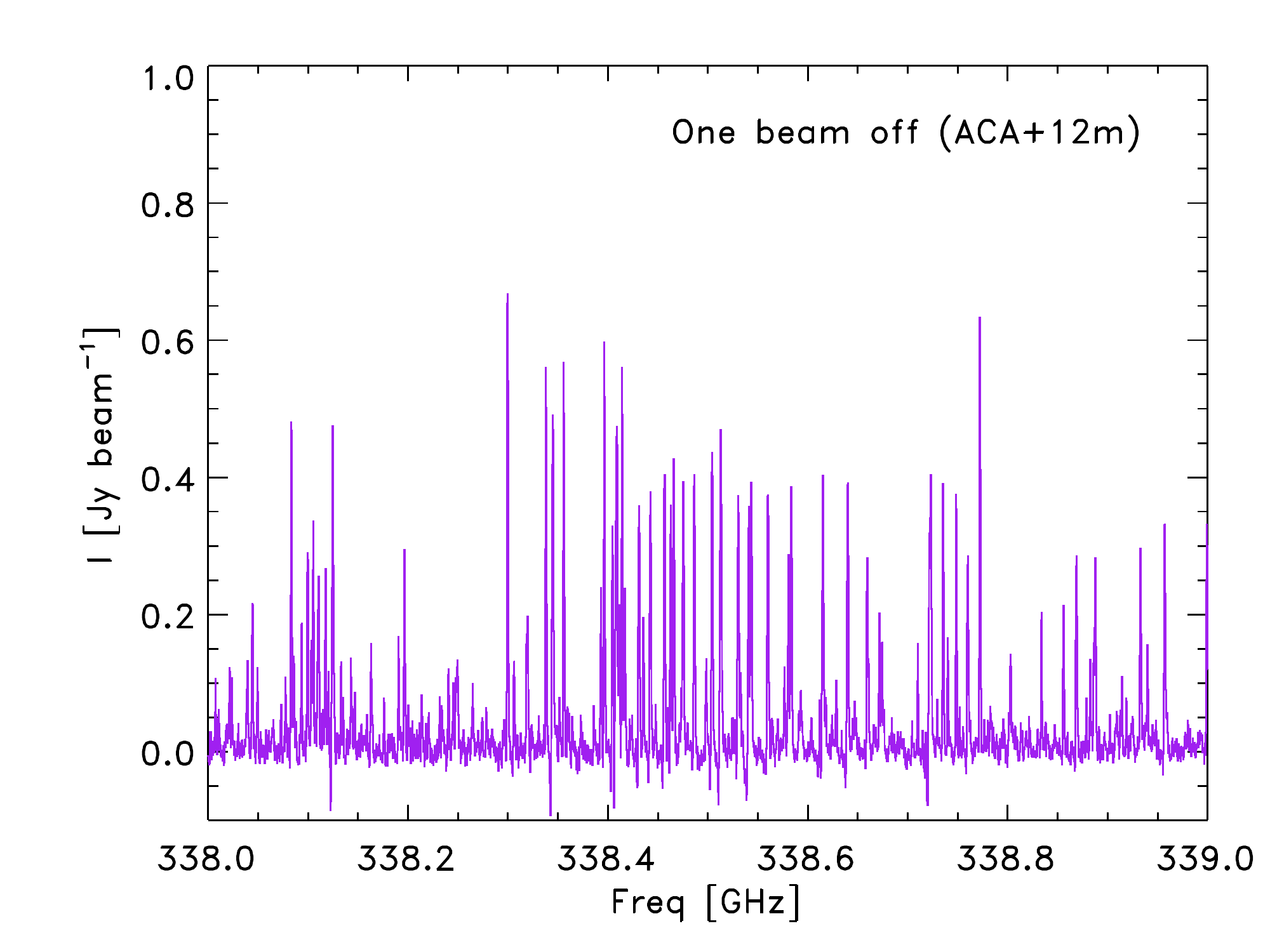}}
\caption{Comparison between the spectra in the 338~GHz window around the prominent CH$_3$OH branch from the single-dish observations from the TIMASSS survey \citep{caux11} (top row), the ACA-only data toward IRAS16293A and IRAS16293B (second row; left and right, respectively), the full data toward IRAS16293A and IRAS16293B (third row; left and right, respectively) and positions offset by a half and full beam southwest of the IRAS16293B continuum position (bottom row). Note that the scale on the Y-axis in the bottom two rows differ from the top rows.}\label{spectra_comparison}
\end{figure*}

The incredible line-richness of IRAS~16293$-$2422 makes it a natural template source for astrochemical studies. Figure~\ref{full_spec} shows the full spectrum from the Band~7 data toward a position offset by 0.25$''$ (a half beam) from IRAS16293B that is used for the analysis in this paper, while the figures in Appendix~\ref{Band36spectra} show the spectra from the Band~3 and 6 data. Figure~\ref{spectra_comparison} compares the observed spectra in the 338--339~GHz spectral range (including the main CH$_3$OH $7_k-6_k$ branch) from JCMT observations taken as part of the larger single-dish TIMASSS survey \citep{caux11} to spectra towards the two continuum peaks in the ACA-only and full ALMA datasets. Also shown are spectra toward positions a half and a full beam separated from IRAS16293B, respectively.

The difference in line widths between IRAS16293A and IRAS16293B noted previously \citep[e.g.,][]{bottinelli04iras16293,iras16293sma} is clearly seen in Fig.~\ref{spectra_comparison} with the lines toward IRAS16293A about a factor of 5 broader than the narrow $\approx$~1~\kms\ (FWHM) lines toward IRAS16293B. This difference is consistent with the interpretation above that IRAS16293A is an edge-on system and IRAS16293B face-on. The very narrow line widths for IRAS16293B make it such an ideal source for line identifications compared to, e.g., the Galactic Center and typical high-mass star formation regions with widths of 5--10~\kms. Furthermore, the strong variations in the peak strengths of the different lines between all seven panels is a clear indication that the excitation conditions and/or chemistry change significantly over the studied scales.

The single-dish spectrum with its beam of approximately 14$''$ naturally encompasses both IRAS16293A and IRAS16293B: the line widths in this particular spectrum are generally closer to IRAS16293A suggesting that the data are more strongly weighted toward this source. Again, this is consistent with the emission being more extended, and thus dominating in the single-dish beam, even though the peak strengths are not significantly different. Some of the brighter lines in the single-dish spectrum are in fact stronger than in the ACA data as one should expect but there are also counter-examples. This likely indicates that the single-dish observations were not targeted exactly between the two sources, and thus that for slightly extended transitions some of the flux is not picked-up in full by the single-dish spectra. The point source RMS noise level in the single-dish data is 0.15--0.3~Jy~beam$^{-1}$~km~s$^{-1}$, whereas the ACA-only data already is about a factor of 5 more sensitive. The combined dataset is an additional order of magnitude more sensitive with a RMS level of 5~mJy~beam$^{-1}$~km~s$^{-1}$. Furthermore the ALMA observation produced fully sampled maps with a higher angular resolution.

\begin{figure*}
\resizebox{\hsize}{!}{\includegraphics{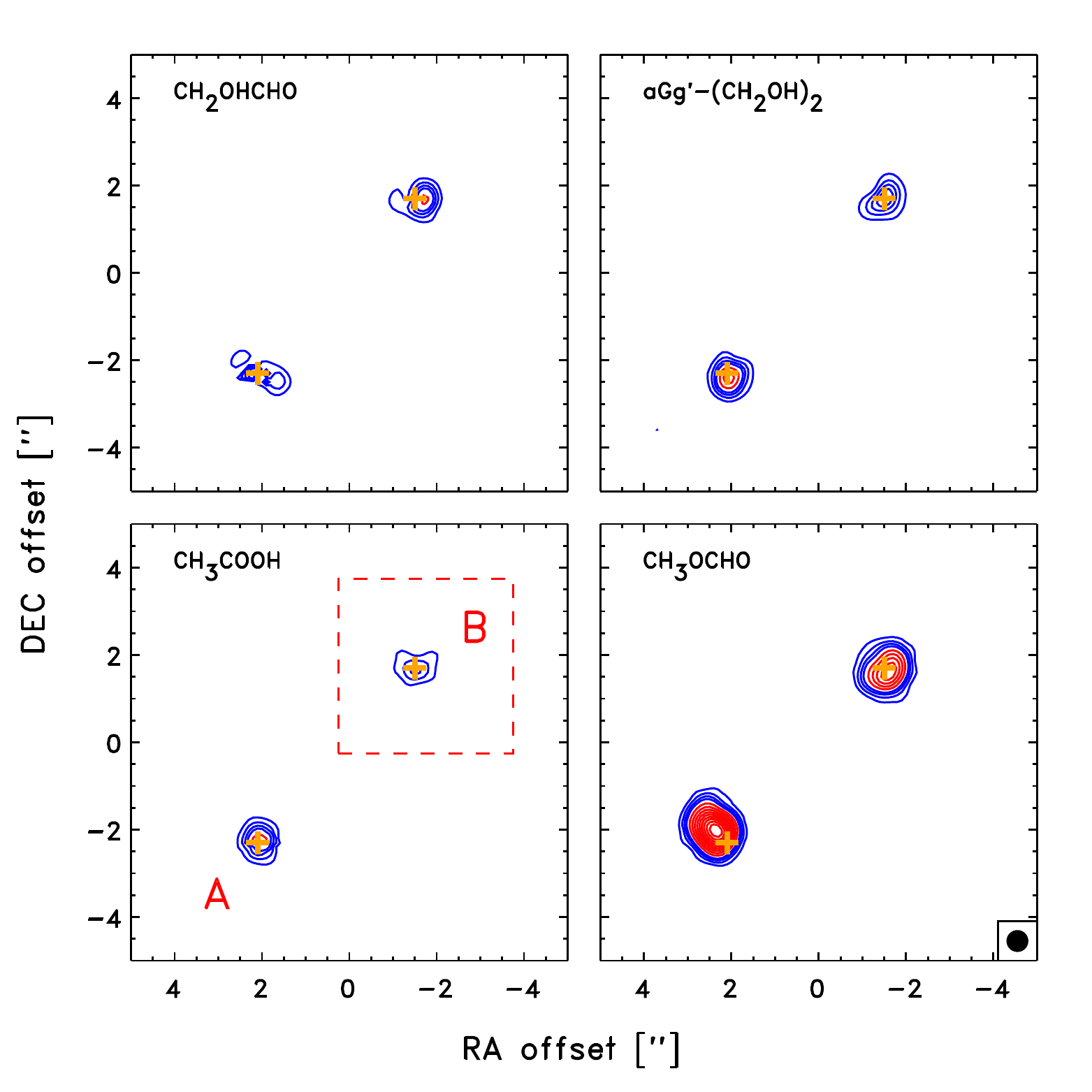}\includegraphics{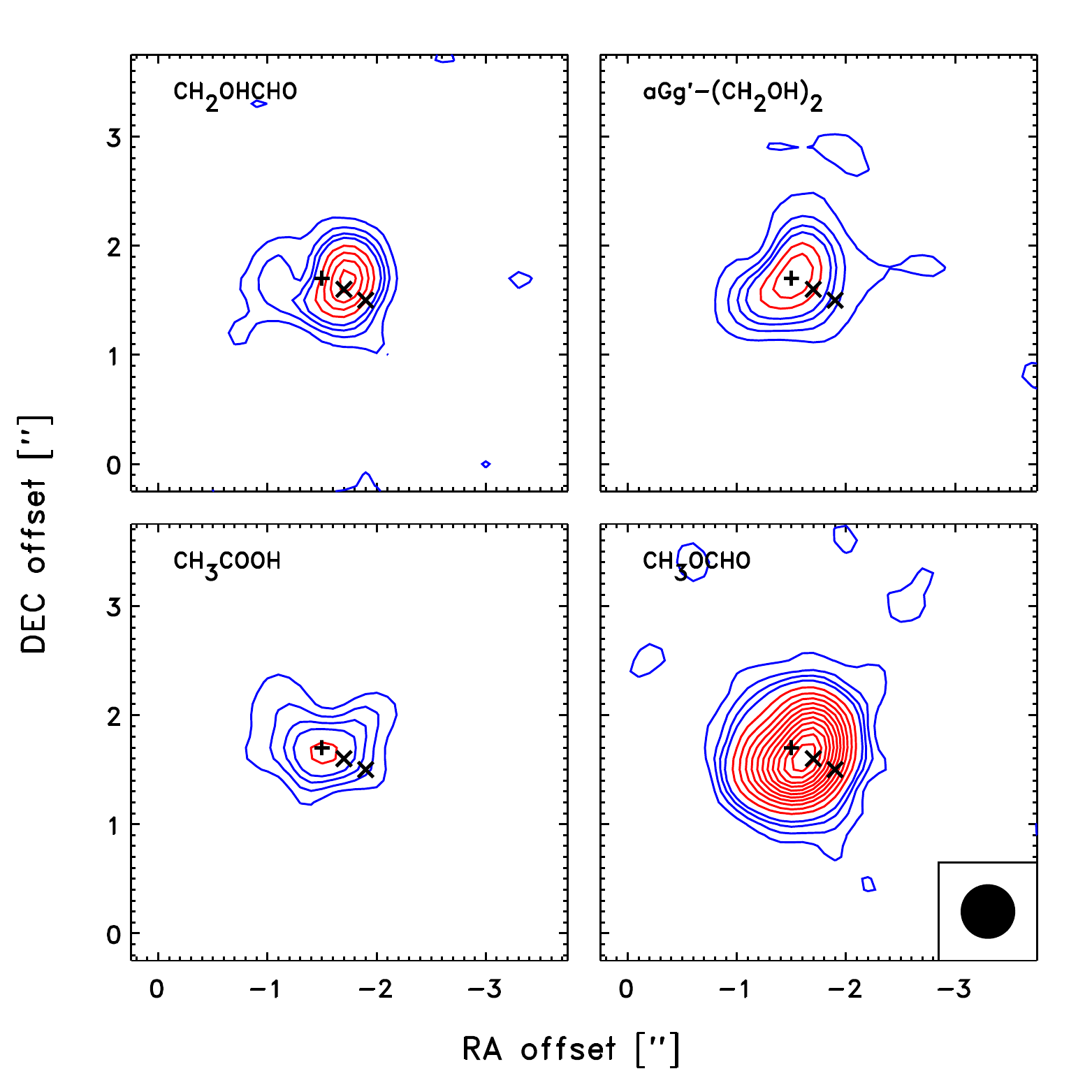}}
\caption{Maps of representative transitions of glycolaldehyde and related species. The four panels to the left show the entire IRAS~16293$-$2422 system while the four panels to the right zoom-in on IRAS16293B (indicated as the red dashed square in the lower left panel). The mapped species are glycolaldehyde (362.406~GHz, $E_{\rm up}$=266~K; \emph{upper left}), ethylene glycol (348.550~GHz, $E_{\rm up}$=329~K; \emph{upper right}), acetic acid (352.872~GHz, $E_{\rm up}$=275~K; \emph{lower left}) and methyl formate (360.467~GHz, $E_{\rm up}$=324~K; \emph{lower right}). In each map the contours represent the integrated emission over $\pm 0.5$~\kms\ in steps of 15~mJy~km~s$^{-1}$ (approximately 3$\sigma$). In each panel the plus-sign marks the location of the continuum sources associated with IRAS16293A and IRAS16293B (labeled by the A and B in the lower left panel) -- while the crosses in the right panels represent the positions offset by half a beam and a full beam from IRAS16293B. These are the positions used for line identification/modeling.}\label{maps}
\end{figure*}

While comparing the ACA-only and combined ALMA data, a very striking feature is the emergence of a number of absorption lines. This is one of the key arguments for not considering the continuum peak position of IRAS16293B for line identification and modeling. Instead we extract the spectra at positions half a beam and one beam southeast of the continuum peak (see Fig.~\ref{maps} as well as higher angular resolution Band~9 data of \citealt{baryshev15}; their Fig.~13). As seen in Fig.~\ref{spectra_comparison} this shift means that most lines become pure emission lines: the peak of line emission is located at a position half a beam offset from the continuum peak, but some of the brighter lines still show self-absorption there. At the position separated one beam from the continuum peak, very few absorption features remain, but the fluxes of lines that are not strongly optically thick is also weakened by a factor $\approx$1.8, likely reflecting the drop in (column) density moving away from the center: in the band~6 data, the continuum flux drops by factors $\approx$2.2 and 4.7 by going from the peak position to the positions a half and full beam offset, respectively, reflecting the drop in column density and, to a smaller degree, temperature. Which of the positions are preferable for the analysis depends on the species considered. It should be emphasised that for species with similar distributions, one can either compare column densities derived at the same positions, or apply the correction factor above, for statements about relative abundances.

Over the full 33.7~GHz spectral range approximately 10,000 lines are seen toward the position half a beam away from the continuum peak, i.e., on average 1~line per 3.4~MHz. Toward this position the average line flux per channel is 60~mJy~beam$^{-1}$ whereas the continuum flux is 1.1~Jy~beam$^{-1}$, i.e., approximately a 5\% contamination of the continuum by line emission. The line density is about a factor 10 higher than what was found by \cite{caux11} in the part of the TIMASSS single-dish survey covering the same frequencies. This improvement is in part due to the higher sensitivity of the ALMA, but also reflects the narrower lines toward IRAS16293B in the spatially resolved ALMA data making it easier to separate individual features. Compared to single-dish studies of IRAS16293 \citep[e.g.,][]{blake94,vandishoeck95,caux11} it is noteworthy that the lines from the most common species do not stand out prominently close to the continuum peaks, partly because of the high optical depth for the transtions of the most common isotopologues causing those to be in absorption against the continuum. This is, for example, seen for species such as CO, HCO$^+$, CS and HCN. For some of their rarer isotopologues the emission lines are seen, but their line brightness at the scales of the interferometric data are not particularly high compared to the lines from the complex organics. This is in contrast to single-dish observations that also pick up the more extended emission of the more common species and lower excited transitions. 

While assigning all of the lines seen in Fig.~\ref{full_spec} is beyond the scope of this initial paper, a statistical argument can be made for the nature of these 10,000 lines. Based on previous estimates of the column densities of complex organic molecules toward IRAS16293B we calculate synthetic spectra for the most commonly organics seen toward hot cores and hot corinos: formaldehyde, methanol, methyl cyanide, isocyanic acid, ethanol, acetaldehyde, methyl formate, dimethyl ether and ketene. The model predicts that approximately 25\% of the transitions in Fig.~\ref{full_spec} are from the main isotopologues of these species. In addition, some regular non-organic species have lines in the frequency range but the bulk of the remaining species can likely be attributed to isotopologues of the complex organics ($^{13}$C, $^{18}$O, $^{17}$O, $^{15}$N, D and all possible combinations) appearing at the sensitivity of the ALMA observations \citep[][and Sect.~5 of this paper]{coutens16} as well as the ``next level'' of complex organics, i.e., species with three carbon atoms \citep[e.g.,][]{lykke16}.

\section{Analysis}\label{analysis}
In this section we present an analysis of the ALMA data focusing, in particular, on the simple sugar-like molecule, glycolaldehyde, previously detected in the Science Verification data \citep{jorgensen12} and related species. Section~\ref{excitation} analyses the assumption of LTE at the densities representative for the scales traced by the ALMA data and presents an estimate of the column density for methanol useful as a future reference for abundance measurements. Section~\ref{glycolaldehydeisomers} analyses the emission from glycolaldehyde as well as related species, ethylene glycol (the reduced alcohol of glycolaldehyde) and acetic acid (the third isomer in the glycolaldehyde/methyl formate group), while Sect.~\ref{isotopologues} presents detections of the $^{13}$C and deuterated isotopologues of glycolaldehyde.

\subsection{Excitation analysis at observed scales}\label{excitation}
The dust continuum emission from the material on the scales probed by ALMA provides important constraints on the physical structure -- and thus the molecular excitation. Ideally the full non-LTE radiative transfer problem needs to be solved to use the line observations for, e.g., constraints on column densities or abundances. However, for most of the complex organic molecules dominating the spectra, no collisional rate coefficients are available and such calculations are therefore not possible. However, the high densities implied from the optical thickness of the dust continuum emission (Sect.~\ref{continuum_results}) means the calculations under the assumption of LTE are likely to be sufficient.

To test the validity of the LTE assumption, we use the escape probability code \emph{Radex} \citep{vandertak07radex} to solve the radiative transfer problem for the largest complex molecule methanol (CH$_3$OH), for which collision rates are available from the Leiden Atomic and Molecular Database \citep[LAMDA][]{lamda} based on calculations by \cite{rabli10}. For a range of densities, $n({\rm H}_2)$, and kinetic temperatures, $T_{\rm kin}$, we calculate the excitation temperatures, $T_{{\rm ex},i}$, for each of the CH$_3$OH transitions (both A- and E-type) in the 329 to 363~GHz range of the survey. We then estimate the root mean square deviation between each of these excitation temperatures and the kinetic temperature:
\begin{equation}
T_{\rm rms} = \sqrt{\frac{1}{N}\sum_i(T_{{\rm ex},i}-T_{\rm kin})^2}
\end{equation}
where $N$ is the total number of transitions in the band: $N=82$ in the frequency range from 329 to 363~GHz. $T_{\rm rms}$ thereby provides a measure of whether LTE is a good approximation in general for the specific value of $n({\rm H}_2)$ and $T_{\rm kin}$: if LTE is a good approximation for a given transition, then $T_{{\rm ex},i} \approx T_{\rm kin}$. A low $T_{\rm rms}$ would therefore signify that most transitions are well reproduced in LTE.

\begin{figure}
\resizebox{\hsize}{!}{\includegraphics{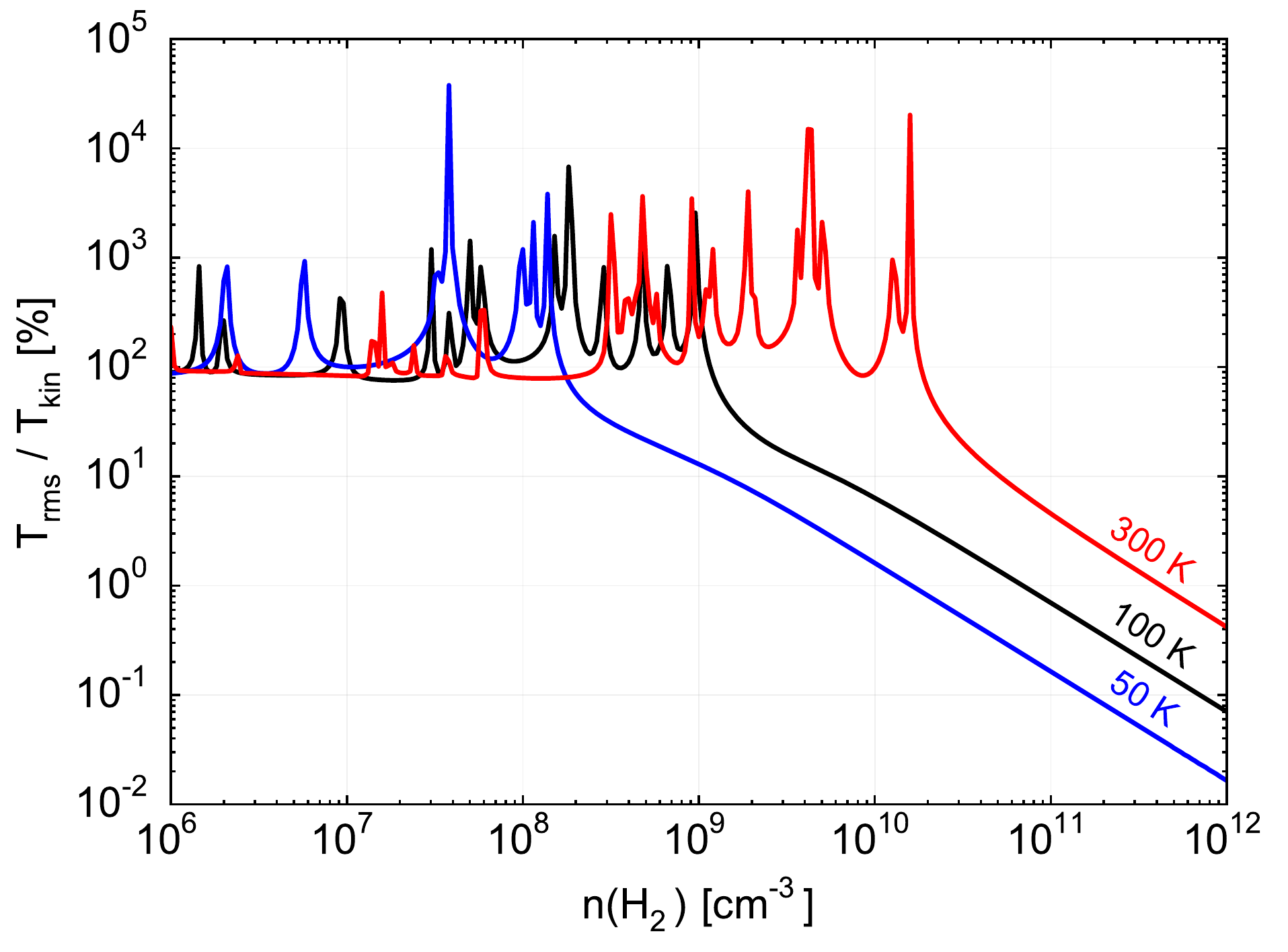}}
\caption{Excitation of CH$_3$OH transitions: the lines show the rms deviation $T_{\rm rms}$ for the excitation temperatures, $T_{\rm ex}$, relative to the kinetic temperature from calculations with the Radex escape probability code \citep{vandertak07radex} for all CH$_3$OH transition in the 329 to 363~GHz band as function of H$_2$ density. The three different lines show results for kinetic temperatures of 50, 100 and 300~K in blue, black and red, respectively.}\label{ch3oh_excitation}
\end{figure}

Figure~\ref{ch3oh_excitation} shows the result of these excitation calculations for CH$_3$OH for models with $T_{\rm kin}$ of 50, 100 and 300~K, respectively, with densities varying from $10^6-10^{12}$~cm$^{-3}$. At low densities significant fluctuations are seen due to transitions masing at specific densities. Still, at those densities even the brighter transitions are subthermally excited, leading to the very large rms deviation between the individual excitation temperatures and the adopted value for $T_{\rm kin}$. However, for each set of models $T_{\rm rms}/T_{\rm kin}$ clearly decreases at densities higher than $10^9-10^{10}$~cm$^{-3}$. For densities higher than $3 \times 10^{10}$~cm$^{-3}$, the lower limit according to the continuum calculations (Sect.~\ref{continuum_results}), the deviations are less than 15\% for these three kinetic temperatures -- a confirmation that the excitation is close to being in LTE. In conclusion, with the lower limit to the density at the scales of the ALMA observations implied by the continuum optical depth, LTE is a good approximation for CH$_3$OH and presumably for most other molecules.

Although it is often assumed that the dust continuum emission is optically thin at submillimeter wavelengths and thus a tracer of column density, this is not the case for IRAS16293B and the column density quoted in Sect.~\ref{continuum_results} is \emph{at best} a lower limit to the actual column density of material at these scales. Abundance estimates are therefore much better expressed relative to a chemically related molecule. To supply such a reference we fitted the observed lines from CH$_3$$^{18}$OH in the Band~7 data with synthetic spectra (Fig.~\ref{18methanol_spectra1}). As indicated in Fig.~\ref{full_spec} and the figures in Appendix~\ref{Band36spectra}, the Band~7 and 6 data are approaching the line confusion level at specific frequencies. For this reason it is not practical to identify individual lines and create rotation diagrams based on those. Instead, the approach taken for this analysis, as well as for other species in Bands~6 and 7, is to calculate synthetic spectra of the targeted molecules and directly compare those to the extracted spectra. This model also contains a reference model for other identified species and thereby makes it possible to address whether a given line is a genuine identification. Following the arguments above, the synthetic spectra are computed under the assumption that the excitation obeys LTE. The parameters needed to produce a synthetic spectrum are then the column densities and excitation temperature of the particular molecule, the extent of the emission as well as the line width and LSR velocity. We adopt a source size of 0.5\arcsec, justified, e.g., by the maps shown in Fig.~\ref{maps}: the maps of the emission from the complex organics in general show a centrally condensed component toward IRAS16293B, which is marginally resolved with an extent of 0.6--0.9$''$ consistent with a deconvolved extent of approximately 0.5$''$ (see also \citealt{coutens16} and \citealt{lykke16}). It should be noted that since the main purpose is to establish a reference scale for comparison between relative abundances of different complex organics present in the same gas based on optically thin transitions, the exact value of the intrinsic source size is less critical. In this analysis we keep the line width and LSR velocity fixed, which appears to reproduce all lines well at the position half a beam from the continuum peak. 

Since we aim to use CH$_3$$^{18}$OH lines to provide a reference column density for comparison to other species we adopt here an excitation temperature of 300~K, which appears to be required from modeling of isotopologues of formamide (NH$_2$CHO) isocyanic acid (HNCO) as well as glycolaldehyde and ethylene glycol \citep[][and the following section]{jorgensen12}. As seen in Fig.~\ref{18methanol_spectra1} most of the transitions are in fact well-modeled with this excitation temperature. Two counter-examples are the lowest excitation transitions ($E_{\rm up}$ of 16 and 35~K, respectively) for which colder material may contribute to the observed line intensities. Also, a few other examples of ``worse'' fits are seen where the CH$_3^{18}$OH is blended with other species, e.g., the transition at 343.135~GHz ($E_{\rm up}$ = 204~K) that is located between two bright methyl formate transitions (at 343.134 and 343.136~GHz) or the 345.858~GHz and 361.052~GHz transitions ($E_{\rm up}$ = 326 and 338~K) that are overlapping with lines of ethylene glycol and formamide, respectively (the latter in fact strongly self-absorbed; see also \citealt{coutens16}). Still, besides these cases, it appears that there is no systematic over- or underestimate of either the lowest or highest excitation transitions, an indication that the adopted excitation temperature is representative for the \emph{bulk} of the gas at this position. The fact that all of the lines are well-fit with one LSR velocity and width indicates that this single warm component dominates the emission at the scales of the ALMA beam and is relatively homogeneous in terms of its physical conditions.

With the excitation temperature of 300~K, the derived column density for CH$_3$$^{18}$OH is $4\times 10^{16}$~cm$^{-2}$ within a 0.5$''$ beam toward the position offset by half a beam from the continuum peak, corresponding to a column density of the main CH$_3$OH isotopologue of $2\times 10^{19}$~cm$^{-2}$ with a standard ISM $^{16}$O/$^{18}$O ratio of 560 \citep{wilson94} or an \emph{upper limit} to its abundance of $3\times 10^{-6}$ with respect to H$_2$ using the lower limit to the total column density from the continuum emission. Again, we emphasize that \emph{even as a limit} the abundance with respect to H$_2$ is highly uncertain and dependent on the exact relation between the physical components traced by the line and continuum emission. However, for molecules distributed similarly to CH$_3$OH and thought to be chemically related, the abundance reference relative to CH$_3$OH is useful.

\subsection{Glycolaldehyde, ethylene glycol and acetic acid}\label{glycolaldehydeisomers}
Glycolaldehyde (CH$_2$OHCHO) is interesting from an astrobiological point of view as it is a simple sugar-like molecule and under Earth-like conditions is related to the formation of ribose. Since its first detection toward the Galactic Center (\citealt{hollis00}; see also \citealt{hollis01,hollis04,halfen06,requenatorres08}) it has been detected in a number of other places in the ISM, both associated with formation of stars of high \citep{beltran09,calcutt14}, intermediate \citep{fuente14} and low mass \citep{jorgensen12,coutens15,taquet15}. Recently, glycolaldehyde has also been detected in comet C/2014 Q2 (Lovejoy) \citep{biver15}. Ethylene glycol is closely related to glycolaldehyde being its reduced alcohol version. It has been seen toward most of the regions where glycolaldehyde is detected -- as well as a few more high-mass star forming regions \citep{lykke15}, the Orion-KL nebula \citep{brouillet15} and toward a few comets where it appears to be more abundant than glycolaldehyde \citep[e.g.,][]{crovisier04,biver14}. Two other species worth mentioning are the two isomers of glycolaldehyde, methyl formate (CH$_3$OCHO) and acetic acid (CH$_3$COOH). The former is relatively common in high-mass star forming regions \citep[e.g.,][]{brown75,ellder80,blake87,macdonald96,gibb00,bisschop07} and has also been detected toward low-mass protostars \citep[e.g.,][]{cazaux03,bottinelli04n1333i4a,iras2sma,sakai06} and even prestellar cores \citep[e.g.][]{oberg10,bacmann12}, while acetic acid is less abundant \citep{mehringer97,remijan02} and tentatively detected toward low-mass protostars \citep{cazaux03,shiao10}. Studies of the relative abundances of these species may provide interesting insights into their formation and possibly also provide the link between the physical conditions in star forming regions and the solar system. For example, significant differences are seen between the relative abundances of ethylene glycol and glycolaldehyde between different regions of high- and low-mass star forming regions and comets \citep[e.g.,][]{lykke15} with low glycolaldehyde abundances relative to ethylene glycol seen in some star forming regions and comets but the Galactic Center shows comparable abundances of the two species \citep{hollis02,belloche13}.

\subsubsection{Spectroscopic data and vibrational corrections}
The spectroscopic data used in the analysis here have different origins: the data for glycolaldehyde are based on laboratory measurements by \cite{butler01}, \cite{widicus05} and \cite{carroll10} including the $\varv=0-3$ forms and are provided by the JPL catalog \citep{jpl}. The JPL catalog also supplies data for methyl formate based on measurements by \cite{ilyushin09}. The data for acetic acid were taken from the Spectral Line Atlas of Interstellar Molecules (SLAIM) that is available through the Splatalogue\footnote{http://www.splatalogue.net} interface (F.~J. Lovas, private communication, \citealt{remijan07}) with the partition function given by \cite{mehringer97}. Ethylene glycol has a more complex structure than the other species: it is a triple rotator with coupled rotation possible around the C-C and two C-O bonds. This means that ethylene glycol can be characterised by a total of ten different conformers distributed over the two main groups (gauche and anti), characterised by the arrangements of the two OH groups (the gauche forms more energetically favorable). Of these, the lowest state $aGg'$ conformer is the one that has been detected in star forming regions. The next lowest $gGg'$ conformer lies about 290~K higher than the $aGg'$ conformer \citep{muller04}: it has been searched for in the Orion hot core but not detected with an upper limit of 0.2 times the column density of the $aGg'$ conformer \citep{brouillet15}. In the ALMA Science Verification data, a single line of this conformer was tentatively detected \citep{jorgensen12}, but no lines of the main conformer were covered in the observed spectral range. The spectroscopy of the two conformers is provided by the Cologne Database for Molecular Spectroscopy \citep[CDMS;][]{cdms1,cdms2} based on measurements by \cite{christen03} and \cite{muller04} for the $aGg'$ and $gGg'$ conformers, respectively.

The high temperatures on the scales of the ALMA observations \citep[$\sim$300~K][and below]{jorgensen12} adds complications to the direct use of the tabulated spectroscopic data and partition functions. In the case of an organic molecule with several low-lying vibrational modes, such as glycolaldehyde, the high temperatures mean that several vibrational states will be populated besides the vibrational ground state. The infrared spectrum of gas-phase glycolaldehyde was measured quite recently \citep{johnson13}. There are three fundamental modes below 520~K vibrational energy. These fundamentals have been considered in the calculation of the partition function for the JPL catalogue entry. They increase the partition function by a factor of $\sim$1.9 at 300~K. Partition function values in the CDMS entries refer to the ground state only. Accounting for associated overtone and combination states in the harmonic oscillator approximation and lowering the $\nu _{17}$ mode because of its pronounced anharmonicity yields a vibrational factor of 2.59, which increases to 2.80 upon consideration of the next three modes below 1250~K. All other modes are above 1550~K and contribute combined less than 2\% to the vibrational factor at 300~K, which is less than the uncertainty of the procedure. Thus, for correct estimates of the column densities of the main isotopologue of glycolaldehyde the main species needs to be increased by a factor 1.47 compared to estimates derived using the partition function given in the JPL catalog.

The gas-phase infrared spectrum of ethylene glycol was published several years ago 
\citep{buckley67}. Comparison with quantum-chemical calculations \citep{howard05} 
suggests that all seven features measured below 1500~K are all fundamentals and that they may 
constitute all fundamentals below 1500~K. These modes together yield a vibrational factor of 4.02. 
The contribution by the higher lying \textit{gGg'} conformer should increase the total column density 
additionally by a factor of $\sim$1.38 at 300~K \citep{muller04}.

\subsubsection{Line identification, temperatures and column densities}
\begin{figure}
\resizebox{1.0\hsize}{!}{\includegraphics{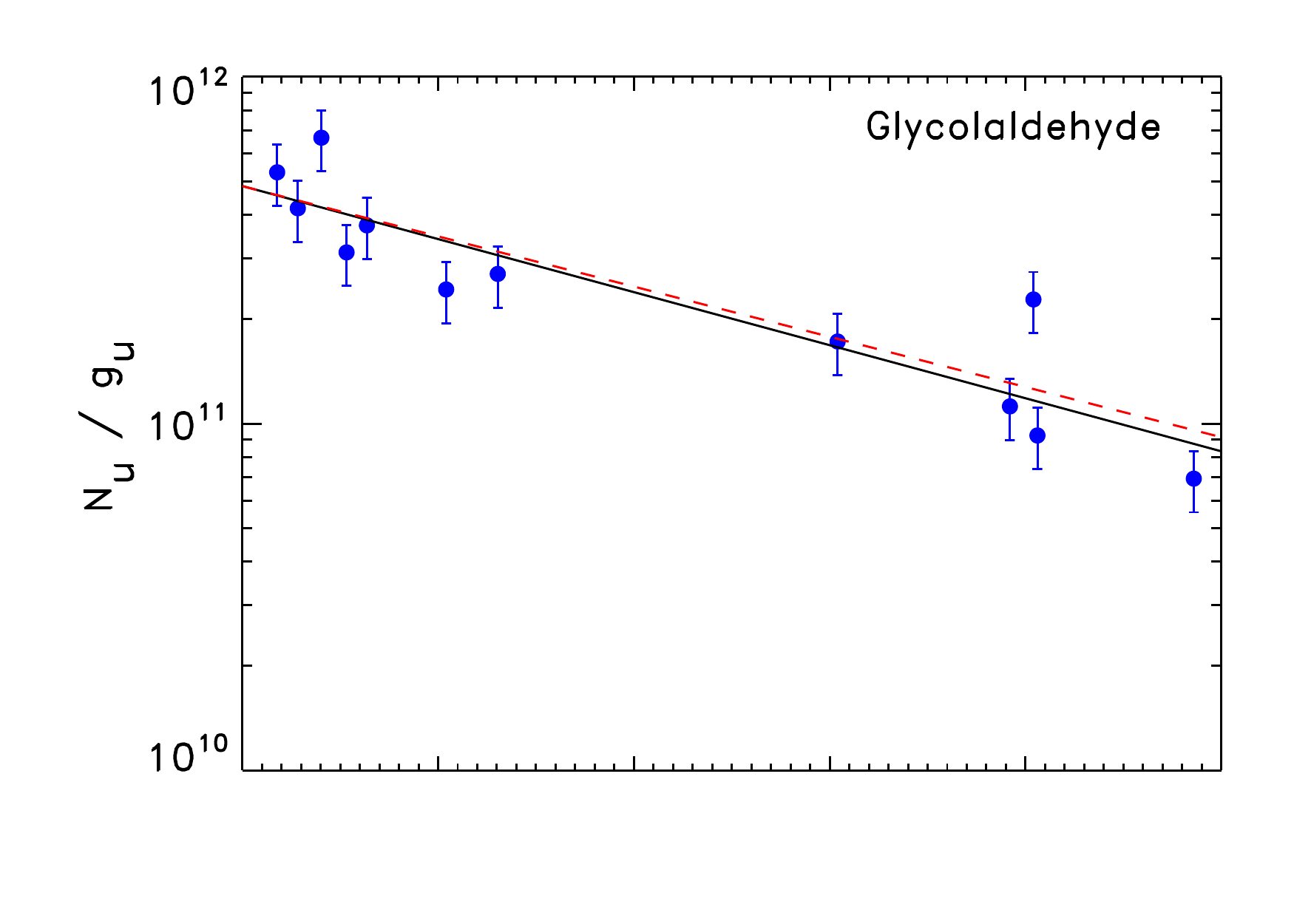}}\vspace{-1.5cm}
\resizebox{1.0\hsize}{!}{\includegraphics{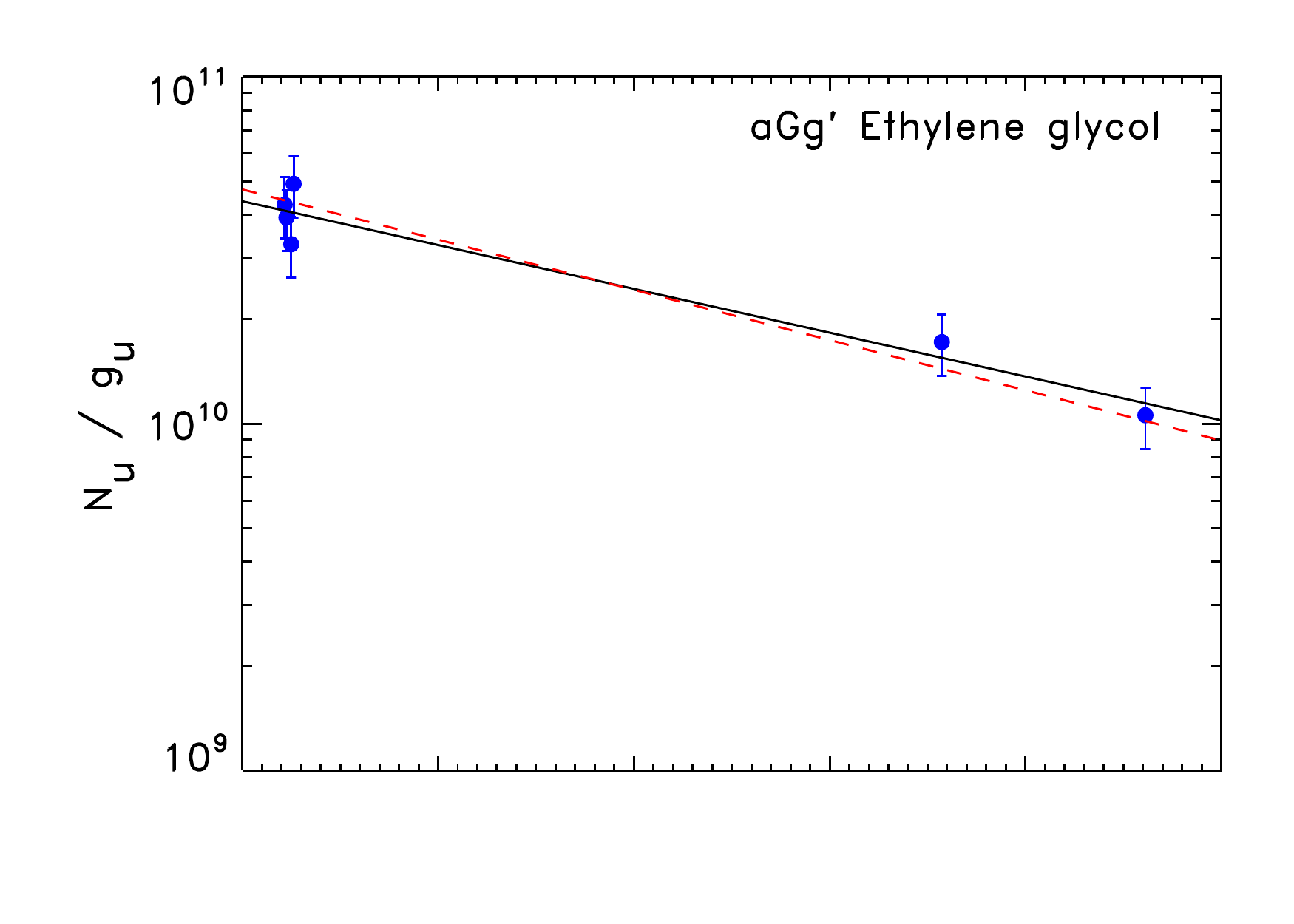}}\vspace{-1.5cm}
\resizebox{1.0\hsize}{!}{\includegraphics{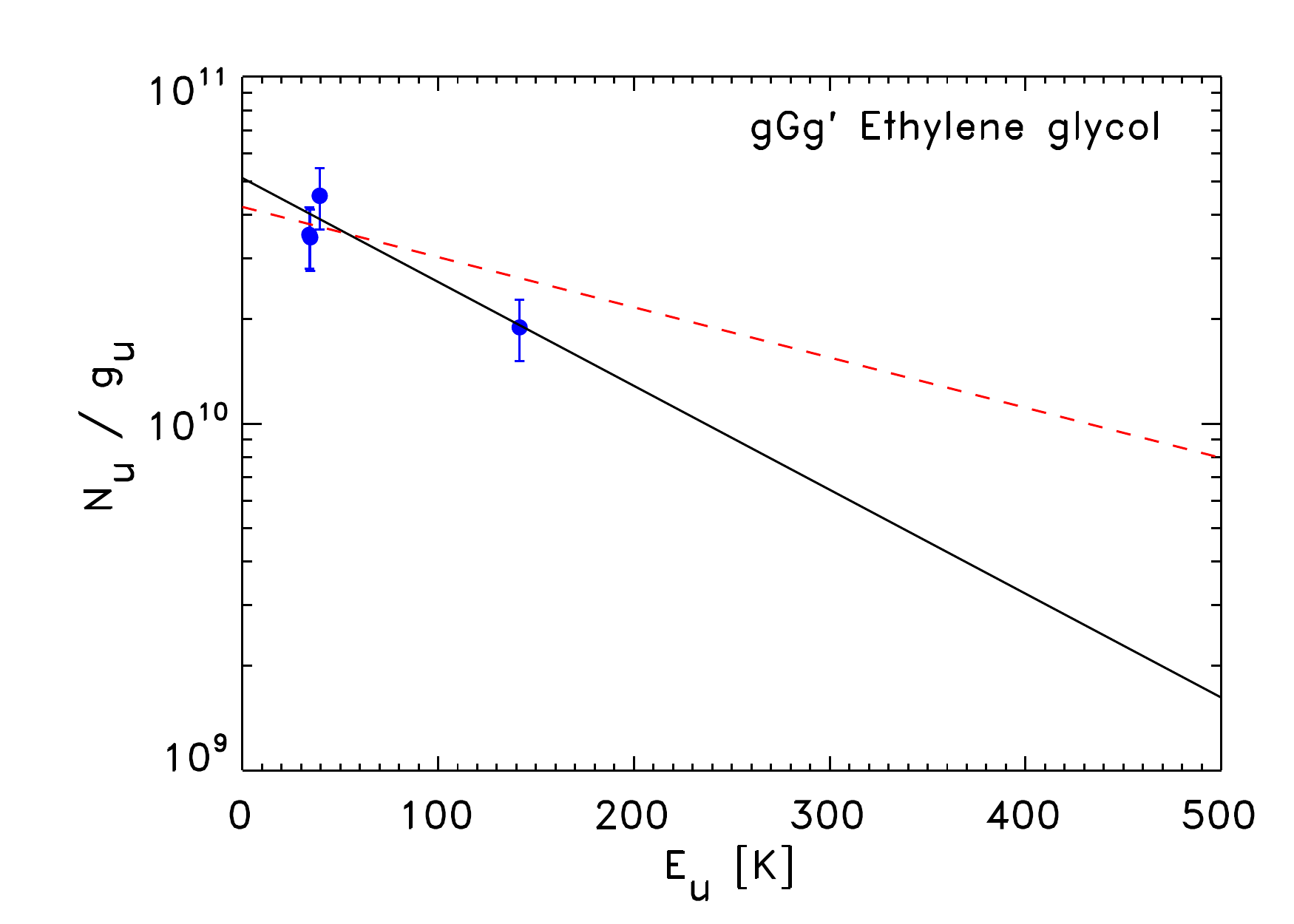}}
\caption{Rotation diagrams for glycolaldehyde and the two conformers ($aGg'$ and $gGg'$) of ethylene glycol based on the observations in Band~3. The solid line indicates the best fit line to the rotation diagram for each species and the dashed red line the best fit for a fixed temperature of 300~K. The error-bars represent 1$\sigma$ uncertainties.}\label{rotationdiagrams}
\end{figure}

To identify the emission from each of the species the lines of glycolaldehyde and ethylene glycol are first identified in Band~3: the spectral windows in the 3~mm observations were selected to cover a number of the low excitation $\varv = 0$ transitions of glycolaldehyde for which the frequencies had previously been measured in the laboratory. As those spectra are much less line confused than those at higher frequencies, it is meaningful to fit the individual lines. Table~\ref{band3measurements} lists the parameters from Gaussian fits to the lines of glycolaldehyde as well as the two conformers of ethylene glycol in the central beam toward IRAS16293B and Fig.~\ref{rotationdiagrams} shows rotation diagrams for these three species: if the emission of a set of lines is in local thermodynamic equilibrium and optically thin, the column density per statistical weight $N_u/g_u$ (directly derived from their measured line strengths) for the lines as function of their energies above the ground-state $E_{\rm up}$ will follow the Boltzmann distribution. The total column density of the molecule and its kinetic temperature can thus be derived from linear fits to plots of $\ln \left(N_u / g_u\right)$ as function of $E_{\rm up}$ where the column density can be derived as the intersect of the fit with the Y-axis and the temperature from the slope of the fit \citep[e.g.,][]{blake87,goldsmith99}. We used the parameters from these rotation diagram fits (Fig.~\ref{rotationdiagrams}) to create synthetic spectra for each of the observed spectral windows to strengthen the discoveries. Figs.~\ref{gcaeg_band3}--\ref{gcaeg_band6-4} compare the model predictions to the full spectral ranges covered in the Band~3 and Band~6 parts of the survey while Figs.~\ref{glycolaldehyde_spectra1}, \ref{Ga-ethyleneglycol_spectra1} and \ref{Gg-ethyleneglycol_spectra1} show the observed and synthetic spectra for the 24 lines each of glycolaldehyde, $aGg'$ and $gGg'$ ethylene glycol, respectively, that are predicted to be the brightest in the Band~7 data while still having optical depths less than 0.2. A strong test of the assignments is that no lines are predicted to show emission at frequencies where none is seen in the spectra. 

As can be observed from these fits, clear detections are seen of glycolaldehyde and the two conformers of ethylene glycol. The detection of ethylene glycol marks the first detection toward IRAS~16293$-$2422 and specifically the detection of the $gGg'$ conformer (tentatively seen in the science verification data; \citealt{jorgensen12}) is the first solid detection of this conformer in the ISM at all. For glycolaldehyde and the lower state conformer of ethylene glycol, the $aGg'$ conformer, good fits are obtained with rotation temperatures of approximately 300~K, as was also found in the analysis of glycolaldehyde and methyl formate in the science verification data \citep{jorgensen12}. The fits for the higher state conformer of ethylene glycol, the $gGg'$ conformer, has a lower rotation temperature, but the few lines for this species cover a rather limited range in $E_{\rm u}$ up to about 140~K and a rotation temperature of 300~K, similar to that of the other conformer and glycolaldehyde, is consistent with the data as well. Furthermore, when comparing to the Band~6 and 7 data, it also appears that all lines are well-reproduced with an excitation temperature of 300~K. In particular, there are no systematics with either low- or high-excitation transitions being generally over- or under-produced, a clear indication that the excitation temperature of 300~K is a good approximation. 

In terms of column densities, several complications arise when comparisons are done across the different wavelengths.  Whereas the 3~mm transitions are all expected to be optically thin, both glycolaldehyde and ethylene glycol have lines at the higher frequencies becoming optically thick. This is partly an excitation effect and partly a result of the smaller beam and makes the column density estimates slightly more uncertain. Because of the differences in beam sizes between the Band~3 and Band~6/7 data one should also be cautious about what beam filling factor to adopt for the analysis. Also, the optically thick continuum affects the observed line emission. Still, if only transitions predicted to have optical depths lower than 0.1 are considered, it seems that the best fit column densities are slightly lower for the two ethylene glycol conformers relative to glycolaldehyde at the higher frequencies in contrast to the case for the pure rotation diagram fits to the Band~3 data where the three species have similar column densities within the uncertainties. A possible explanation for this may be that ethylene glycol is slightly more extended around the protostar itself than glycolaldehyde -- and thus contributing more in the larger beam at 3~mm while also being more obscured by the optically thick continuum in the Band~7 data. Hints of this are seen in the maps of the representative transitions in Fig.~\ref{maps}, but higher resolution imaging at lower frequencies as well as observations of isotopologues of ethylene glycol would be important to test whether line optical thickness or the distribution is more important. No strong lines of acetic acid are covered in the Band~3 data, but it is clearly seen in the Band~6 and 7 datasets (see Figs~\ref{gcaeg_band3}--\ref{gcaeg_band6-4} as well as Fig.~\ref{aceticacid}). To model the emission, the same excitation temperature as derived for the glycolaldehyde and ethylene glycol lines is adopted for the derivation of its column density using similar comparisons between observed and synthetic spectra as for CH$_3^{18}$OH.

\begin{table*}\label{fitsresults}
\caption{Column densities of glycolaldehyde, the two conformers of ethylene glycol and acetic acid.}
\begin{tabular}{llll}\hline\hline
Species                & $T_{\rm rot}$$^{a}$ & $N(T=300\,\mathrm{K})$$^{b}$  & $N_{\rm synt}$$^{c}$          \\ \hline
Glycolaldehyde         & $284\pm 27$~K       & $3.3\times 10^{16}$~cm$^{-2}$ & $6.8\times 10^{16}$~cm$^{-2}$ \\ 
$aGg'$ ethylene glycol & $344\pm 52$~K       & $3.3\times 10^{16}$~cm$^{-2}$ & $1.1\times 10^{17}$~cm$^{-2}$ \\ 
$gGg'$ ethylene glycol & $145\pm 45$~K       & $3.0\times 10^{16}$~cm$^{-2}$ & $1.0\times 10^{17}$~cm$^{-2}$ \\
Acetic acid$^{d}$      & $\ldots$            & $\ldots$                      & $6\times 10^{15}$~cm$^{-2}$   \\
Methanol$^{e}$         & $\ldots$            & $\ldots$                      & $2\times 10^{19}$~cm$^{-2}$   \\ \hline
\end{tabular}

Notes: $^a$Derived rotation temperature from fit to Band~3 data (Fig.~\ref{rotationdiagrams}). $^{b}$Column densities from Band~3 data assuming that $T = 300$~K and that the emission is uniformly extended compared to the beam. $^{c}$Column density from fits of synthetic spectra to the Band~7 data assuming a source size of 0.5\arcsec\ (see also \citealt{lykke16}) and including vibrational corrections. $^d$ From fitting synthetic spectra to the Band~6 and 7 data (Figs~\ref{gcaeg_band3}--\ref{gcaeg_band6-4} as well as Fig.~\ref{aceticacid}) adopting an excitation temperature of 300~K. $^e$Derived by fitting synthetic spectra in Band~7 as discussed in Sect.~\ref{excitation}.
\end{table*}

\subsubsection{Comparison with models}
\begin{figure*}
\resizebox{\hsize}{!}{\includegraphics{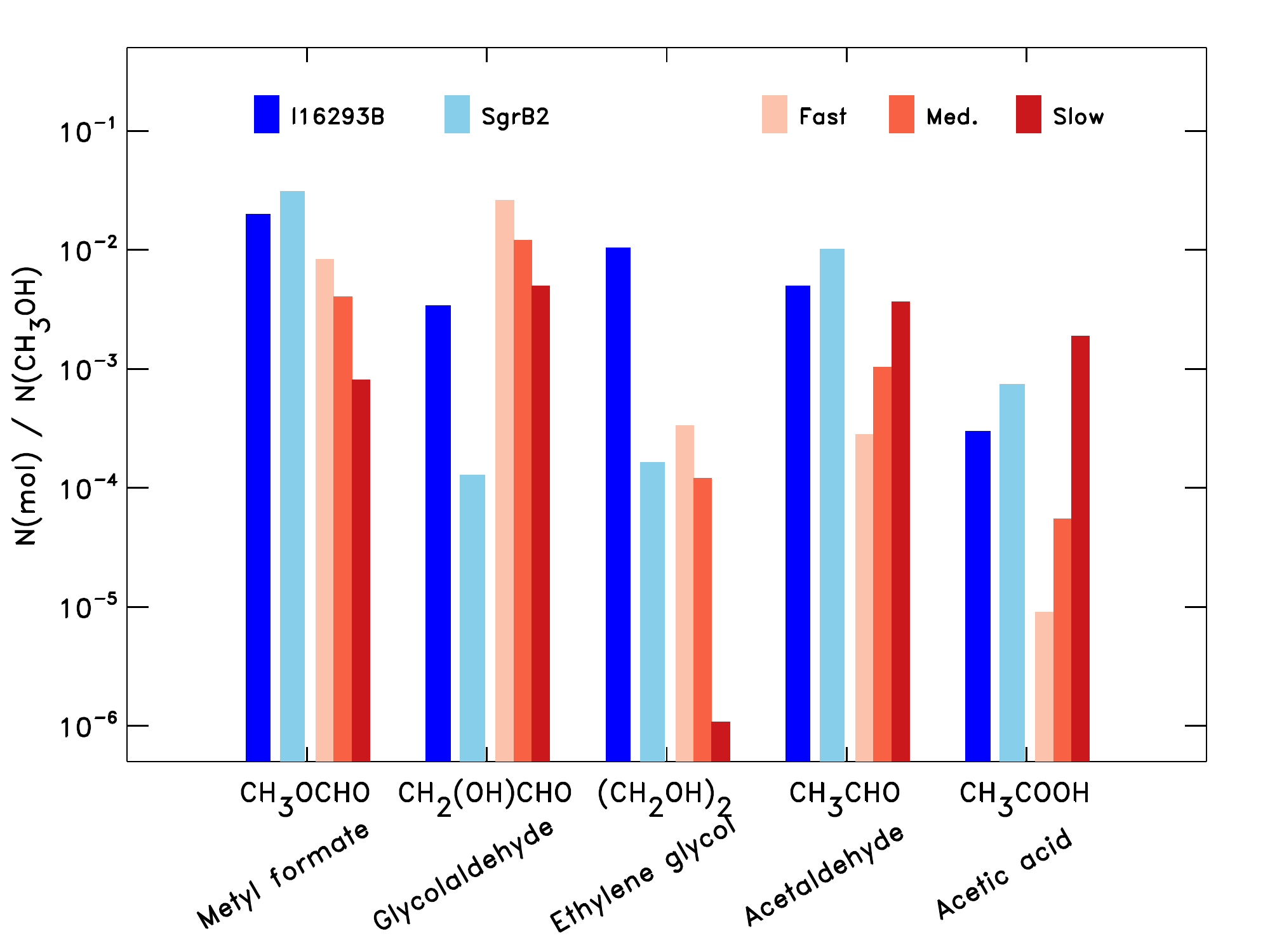}}
\caption{Abundances of methyl formate (CH$_3$OCHO), glycolaldehyde (CH$_2$(OH)CHO), ethylene glycol ((CH$_2$OH)$_2$), acetaldehyde (CH$_3$CHO) and acetic acid (CH$_3$COOH) relative to methanol (CH$_3$OH) toward IRAS16293B (this paper) compared to SgrB2(N) from \cite{belloche13} and to the slow, medium and fast models of \cite{garrod13}.}\label{rob_model}
\end{figure*}

Figure~\ref{rob_model} compares the derived column densities for methyl formate, glycolaldehyde, ethylene glycol, acetaldehyde and acetic acid relative to methanol for IRAS16293B, SgrB2(N) \citep{belloche13} and three versions of the three-phase chemical models of hot cores by \cite{garrod13}. The models of \citeauthor{garrod13} simulate gas-phase, grain surface and mantle chemistry during the formation of high-mass hot cores. The physical models represent the initial collapse followed by gradual warm-up of the gas and grains, the latter with three different rates, ``slow'' with warm-up to 200~K in $1\times 10^6$~years, ``medium'' reaching 200~K in $2\times 10^5$~years and ``fast'' in $5\times 10^4$~years. The model values in Figure~\ref{rob_model} represent the peak fractional abundances for the individual species in these three model instances. The most notable aspect is the good agreement between the abundances relative to methanol for methyl formate, acetaldehyde and acetic acid with those in SgrB2(N), but somewhat low abundances of glycolaldehyde and ethylene glycol. The ethylene glycol abundance taking into account both the $aGg'$ and $gGg'$ conformers is higher by about a factor 3 than that of glycolaldehyde, in line with measurements for comets and the other solar-type protostar, NGC~1333-IRAS2A, where glycolaldehyde and the $aGg'$ conformers have been detected \citep{coutens15}. This is in contrast to what was inferred based on the tentative detection of the $gGg'$ conformer based on the SV data \citep{jorgensen12}. The reason for this discrepancy is that the analysis of the SV data did not take into account the vibrational correction factors for ethylene glycol and only considered one of the conformers. As noted above these corrections are non-negligible at the temperature of 300~K.

Generally both observations and models find abundances for the five species relative to methanol in the few~0.01\%-1\% ranges, however, some significant differences are seen within this range. For example, the models strongly overproduce the glycolaldehyde abundances relative to methyl formate when comparing to the observations toward both IRAS16293B and SgrB2(N). This was also noted by \cite{garrod13} who speculated that it was caused by differences in the formation efficiencies of methyl formate and glycolaldehyde through their primary grain-surface/ice formation routes involving HCO and respectively CH$_3$O and CH$_2$OH for methyl formate and glycolaldehyde. The models of \cite{garrod13} require photodissociation in the ices to lead to those radicals, but recent experiments \citep{fedoseev15,chuang16} show that methyl formate, glycolaldehyde and ethylene glycol can also be formed in ices at low temperatures through H-atom addition and abstraction reactions during the formation of CH$_3$OH from CO and H$_2$CO. These reactions also make the link between glycolaldehyde and ethylene glycol stronger -- possibly accounting for the higher abundances of these species relative to methyl formate between IRAS16293B on one side and SgrB2(N) on the other.

Concerning acetic acid and acetaldehyde the agreement between the IRAS16293B and SgrB2(N) abundances is noteworthy. Although acetic acid is an isomer of methyl formate and glycolaldehyde, its formation is not expected to follow the same paths due to its different structure \citep{garrod13}. As those models show, the resulting abundances are strongly dependent on the warm-up time-scale with the slower models allowing for more build-up of acetic acid through addition of OH to CH$_3$CO that has been derived through hydrogen abstraction of acetaldehyde (CH$_3$CHO). There is, however, a priori no reason why the warm-up rates should be similar in IRAS16293B and SgrB2(N), so some other parameters must come into play and serve to regulate the abundances.

\subsection{Isotopologues of glycolaldehyde}\label{isotopologues}
A potentially very intriguing possibility offered by ALMA is to characterise the isotopic composition of different complex organics and use possible differences to show the relationships between different complex organics. \cite{charnley04}, for example, proposed a number of tests of the origins of typical hot core organics by measuring their $^{12}$C/$^{13}$C fractionation patterns in different functional groups. Another interesting culprit is the D/H ratio for different complex organics: Chemical models predict that the D/H ratios for different species reflect the conditions at the time of their formation \citep[e.g.][]{cazaux11,taquet13} and, consequently, the formation pathways for different complex organics may thereby lead to different levels of deuteration \citep[e.g.,][]{taquet14}. These effects may, for example, explain the differences in the D/H ratio of water and methanol with the former formed earlier in the evolution of prestellar cores where the D/H ratio is lower \citep[see][for a discussion]{ceccarellippvi}.

To start the fractionation study in the PILS data, we searched for the $^{13}$C- and deuterated isotopologues for glycolaldehyde in the Band~7 data. Similar to the approach above we generate synthetic spectra and match those with the lines predicted to be the brightest in the spectra. The spectroscopy for the isotopologues is available through the CDMS database: for the two $^{13}$C-isotopologues of glycolaldehyde, $^{13}$CH$_2$OHCHO and CH$_2$OH$^{13}$CHO, the data are based on measurements by \cite{haykal13}, while those for the three deuterated isotopologues, CHDOHCHO, CH$_2$ODCHO and CH$_2$OHCDO, are based on \cite{bouchez12}. The partition function for the isotopologues listed by the CDMS do not include the corrections for the vibrational contributions. Information on vibrational states of isotopic species have not been published experimentally, but quantum-chemical calculations on the main and the OD species are available as well as an investigation of the anharmonicity of the lowest two vibrational states $\varv _{18} = 1$ and $\varv _{17} = 1$ \citep{senent04}. Scaling the lowest six vibrational modes by the calculated D-to-H ratio yields a vibrational factor of 3.29 for the OD species. The increase is mainly due to the isotopic shift of the $\nu _{17}$ mode. Vibrational factors of all other isotopologues are probably closer to the value of the main isotopologue because the lowest three vibrations involve mainly the heavy atoms or the H on the OH group ($\nu _{17}$ mainly). In summary, the column density of the OD isotopologue needs to be increased by 3.29 and those of all other isotopologues by 2.80 to account for the population of vibrational states at 300~K.

Figures~\ref{chdohcho_1}--\ref{ch2odcho_1} show the spectra for the lines predicted to be the brightest for each of the deuterated isotopologues: for these species lines are clearly detected throughout the band. The lines of the $^{13}$C species are intrinsically weaker and line blending becomes more significant. Still, a number of reasonable assignments can be made that are compiled in Fig.~\ref{13glycolaldehyde}. In these plots only bright and/or relatively isolated lines are shown, corresponding to about half of the transitions predicted to be the brightest from synthetic spectra at 300~K. The derived column densities are given in Table~\ref{isotopologuetable}. For the deuterated species the immediately derived column density of the isotopologue with the deuteration in the CH$_2$ group is a factor two higher than that of the two other isotopologues. This should be expected statistically if the D/H ratio is similar for each of the hydrogen atoms across the functional groups (the CHDOHCHO and CDOHCHO versions being indistinguishable). Taking this into account, the D/H ratio for glycolaldehyde is 4.1--5.2\% where the differences between the deuteration of the different functional groups appear marginal given the uncertainties in the line identification and modeling. For the $^{13}$C species, the uncertainties are somewhat larger due to the smaller number of lines, but the modeling of both species appears consistent with a column density of 1.9--2.8$\times 10^{15}$~cm$^{-2}$, or a $^{12}$C/$^{13}$C ratio of 24--35, with a best fit column density fitting the two species simultaneously of 2.5$\times 10^{15}$~cm$^{-2}$. It is worth emphasising that these detections illustrate the great sensitivity of the ALMA observations, allowing us to measure emission from species with abundances below 0.01\% with respect to methanol (CH$_3$OH).

The D/H ratio for glycolaldehyde of $\approx$~5\% is clearly higher than the typical water deuteration levels of $\approx 0.1$\% in the warm gas on the same scales probed by interferometric observations toward IRAS~16293$-$2422 \citep[e.g.,][]{persson13} and other protostars \citep[e.g.,][]{iras4b_hdo,taquet13obs,persson14}. It appears to be lower than what has been inferred for formaldehyde, methanol and other complex organics from single-dish observations of IRAS~16293$-$2422 and other protostars \citep[e.g.,][]{parise06deuterium,demyk10,richard13}. Before concluding that this lower ratio reflects differences in the processes involved in the formation of glycolaldehyde, caution must be taken as the single-dish and interferometric data probe very different environments due to the very different beam sizes. The significant beam dilution and excitation/optical depth effects may also be an issue for the single-dish observations where many components are enclosed in one beam. Clearly, a systematic comparison of the D/H ratios for complex organics on the scales of the ALMA observations will be critical.

It is particularly noteworthy that no differences are seen in the deuteration of the three different functional groups in glycolaldehyde. For deuterated methanol the CH$_3$-group has been found to be significantly enhanced compared to the OH group with the D/H ratios of 12\% for CH$_2$DOH (ratio of the column densities of CH$_2$DOH/CH$_3$OH of 37\%) and 1.8\% for CH$_3$OD \citep{parise04}. This led \cite{faure15} to propose that the functional groups able to establish hydrogen bonds (-OH, -NH) are expected to equilibrate with the D/H ratio in water ice during heating events. In this scenario, the OH group of glycolaldehyde should show a significantly lower D/H ratio than the CH$_2$ and CHO groups, in contrast to what is observed. A separate analysis of formamide, NH$_2$CHO, based on the PILS data \citep{coutens16}, shows a similar effect with the three deuterated isotopologues NH$_2$CDO and cis-/trans-NHDCHO all having similar degrees of deuteration.

The inferred $^{12}$C/$^{13}$C ratio of 24--35 is lower than the ratio of 68$\pm$15 of the local interstellar medium \citep[][]{milam05}. This appears to be a real difference and not reflect, e.g., high optical depth of the lines of the main isotopologue, given that a relatively high number of significant, low optical depth lines are identified. It is in contrast to recent measurements of the $^{12}$C/$^{13}$C ratio for methyl formate toward a sample of high-mass star forming regions (including Orion-KL) by \cite{favre14mf}, who found similar isotope ratios for complex organic molecules as for CO. The $^{13}$CO enhancement could result from low temperature ion-molecule reactions in the gas increasing the gaseous $^{13}$CO abundance at 10~K just prior to freeze-out \citep{langer84}. Once on the grains, the enhanced $^{13}$CO can be further incorporated into complex organic molecules \citep{charnley04}. This could be a simple explanation for the measured ratio for glycolaldehyde: if formed efficiently from CO ices at low temperatures ($T \approx 10-15$~K) on the grains such as proposed by \cite{fedoseev15} and \cite{chuang16} based on laboratory experiments, glycolaldehyde ice would naturally inherit any $^{13}$C-enhancement (i.e., low $^{12}$C:$^{13}$C isotopologue ratio) from the CO ices, which would be preserved upon sublimation into the gas-phase. In that case many of the other complex organics formed at low temperatures in the ices from CO, should show comparable (low) ratios between the $^{12}$C and $^{13}$C isotopologues as glycolaldehyde, whereas molecules with significant contributions from other C-containing radicals or from gas-phase routes may not \citep{charnley04}. The more ``standard'' $^{12}$C:$^{13}$C ratio in the high-mass star forming regions would then reflect the shorter time-scales and higher temperatures there, with 13CO less enhanced, similar to what has been invoked to explain the differences in the deuteration.

An alternative is that the $^{12}$CO/$^{13}$CO ratio in the ices is affected by slightly different binding energies of the two isotopologues. \cite{smith15} recently measured the $^{12}$CO/$^{13}$CO ratio in the gas-phase for a sample of young stellar objects (including four in Ophiuchus) through high spectral resolution infrared observations and generally found very high $^{12}$CO/$^{13}$CO values ranging from 85 to 165. \citeauthor{smith15} discuss several options for these high gas-phase values,
 including possible differential sublimation of $^{12}$CO versus $^{13}$CO due to
 slightly different binding energies and isotope selective
 photodissociation coupled with low temperature chemistry. It should be noted that no strong evidence for such differences in binding energies has yet been found in the laboratory experiments \citep[e.g.,][]{bisschop06,acharyya07}. However, if the binding energies were in fact different, it would lead $^{12}$CO to sublimate at slightly lower temperatures than $^{13}$CO, thus leading to an enhancement of the latter $^{13}$C-isotopologue in the ice and, vice versa, an enhancement of the $^{12}$C-isotopologue in the gas-phase. Further observations of the carbon-isotopic composition of complex organics could shed more light on the underlying mechanism and further extend its use in determining the formation routes for the different complex organics.
\begin{table}
\caption{Derived column densities for the isotopologues of glycolaldehyde and abundances relative to the main isotopologue.}\label{isotopologuetable}
\begin{tabular}{lll}\hline\hline
Species            & $N$$^{a}$               & $N / N({\rm CH_2OHCHO})$$^{b}$ \\
                   & [cm$^{-2}$]             & \\ \hline 
CH$_2$OHCHO        & $6.8\times 10^{16}$     & $\ldots$       \\ 
CHDOHCHO           & $7.1\times 10^{15}$     & 0.10 (9.6)     \\
CH$_2$ODCHO        & $3.2\times 10^{15}$     & 0.047 (21)     \\
CH$_2$OHCDO        & $3.5\times 10^{15}$     & 0.052 (19)     \\
$^{13}$CH$_2$OHCHO & $2.5\times 10^{15}$$^c$ & 0.037 (27)$^c$ \\
CH$_2$OH$^{13}$CHO & $2.5\times 10^{15}$$^c$ & 0.037 (27)$^c$ \\ \hline
\end{tabular}

$^{a}$Column density of the isotopologue corrected for the vibrational contributions at 300~K. $^{b}$Column density of isotopologue relative to that of the main species. Number in parenthesis gives the inverse ratio: traditionally the former is quoted for measurements of the deuterated species, while the latter is typically quoted for $^{12}$C/$^{13}$C measurements. $^c$The $^{13}$CH$_2$OHCHO and CH$_2$OH$^{13}$CHO fitted with a common column density for both due to the relatively low number of lines seen for the two species. 
\end{table}

\section{Summary}\label{summary}
We have presented an overview and some of the first results from a large unbiased spectral survey of the protostellar binary IRAS~16293$-$2422 using ALMA. The full frequency window from 329 to 363~GHz is covered with a spectral resolution of 0.2~km~s$^{-1}$ and beam size of 0.5$''$ (60~AU diameter at the distance of IRAS~16293$-$2422) in addition to three selected settings at 3.0 and 1.3~mm. The main findings of this paper are:

\begin{itemize}
\item The continuum is well detected in each band with different signatures toward the two protostars: clear elongated emission seen toward IRAS16293A in the direction of a previously reported velocity gradient, making it appear as a flattened edge-on structure. The binarity of this source at submillimeter wavelengths previously reported is not confirmed in these data. In contrast, the emission for IRAS16293B is clearly optical thick out to wavelengths of approximately 1~mm (and possibly beyond). The optical thickness of radiation toward IRAS16293B confirms that the emission has its origin in a different component than the larger scale envelope with a density larger than $3\times 10^{10}$~cm$^{-3}$.
\item More than 10,000 lines are seen in spectra toward IRAS16293B. The high gas density implied by the continuum radiation means that lines of typical complex organic molecules are thermalised and LTE is thus a valid approximation for their analysis.
\item The spectra at 3.0, 1.3 and 0.8~mm provide strong confirmations of the previous detection of glycolaldehyde as well as the derived excitation temperature of 300~K based on ALMA Science Verification data \citep{jorgensen12}. In addition the spectra show detections of acetic acid (isomer of glycolaldehyde) and two conformers of ethylene glycol (the reduced alcohol of glycolaldehyde) with the detection of one of these, the $gGg'$ conformer, the first reported detection in the ISM. The excitation temperatures of these species are consistent, $\approx$~300~K. The abundance of glycolaldehyde is comparable to that of the main conformer of ethylene glycol with the second conformer not much rarer, as one would expect given the high temperatures in the gas. Small differences between the relative glycolaldehyde and ethylene glycol abundances in data of different beam sizes possibly reflect glycolaldehyde being slightly more centrally concentrated than ethylene glycol.
\item Relative to methanol (determined from observations of optically thin lines of the CH$_3^{18}$OH isotopologue), the abundances of glycolaldehyde and related species are between 0.03\% and a~few~\%. Glycolaldehyde and ethylene glycol are clearly more abundant relative to methanol toward IRAS16293B compared to the Galactic Center source SgrB2(N) \citep{belloche13}, whereas the abundances of the glycolaldehyde isomers (methyl formate and acetic acid) and acetaldehyde are comparable for the two sources. A possible explanation for this is the formation of glycolaldehyde from CO at low temperatures in ices toward IRAS~16293$-$2422, in agreement with recent laboratory experiments -- a route that is unlikely to apply for the warmer Galactic Center.
\item The data also show detections of two $^{13}$C-isotopologues of glycolaldehyde as well as three deuterated isotopologues. These are the first detections of these five isotopologues reported for the ISM, enabled by the narrow line-widths toward IRAS16293B. The D/H ratio for glycolaldehyde is $\approx$~5\% for all three deuterated isotopologues with no measurable differences for the deuteration of the different functional groups. These ratios are higher than in water, but lower than reported D/H ratios for methanol, formaldehyde and other complex organics, although those should be revisited at the same scales. The derived $^{12}$C/$^{13}$C ratio of glycolaldehyde $\approx$~30 with our data is lower than the canonical ISM. This may reflect a low $^{12}$CO/$^{13}$CO ratio in ice from which it is formed, either due to ion-molecule reactions in the gas or differences in binding energies for the different CO isotopologues.
\end{itemize}

This first presentation of data has only scratched the surface of all the information available in the survey, but already raised a number of new questions concerning, in particular, the formation of complex organic molecules around protostars. Moving forward it is clear that the possibility to systematically derive accurate (relative) abundances of different organic molecules (and their isotopologues) will be an important tool. In this respect, many of the answers to the questions concerning the origin of complex, prebiotic, molecules may be hidden in this rich ALMA dataset. 

\begin{acknowledgements}

We are grateful to the staff at the Nordic ALMA Regional Center (ARC) node at Onsala Space Observatory, i.p. Ivan Mart\'i-Vidal and S\'ebastien Muller, for their support in setting-up the program and assistance with the data reduction. We also thank the anonymous referee for many comments that helped improving the presentation. Thanks also go to G. Fedoseev, K.-J. Chuang and H. Linnartz for experimental information and discussions on the formation of glycolaldehyde and ethylene glycol in ices at low temperatures. This paper makes use of the following ALMA data: ADS/JAO.ALMA\#2012.1.00712.S and ADS/JAO.ALMA\#2013.1.00278.S. ALMA is a partnership of ESO (representing its member states), NSF (USA) and NINS (Japan), together with NRC (Canada), NSC and ASIAA (Taiwan), and KASI (Republic of Korea), in cooperation with the Republic of Chile. The Joint ALMA Observatory is operated by ESO, AUI/NRAO and NAOJ. The group of JKJ acknowledges support from a Lundbeck Foundation Group Leader Fellowship as well as the European Research Council (ERC) under the
European Union's Horizon 2020 research and innovation programme (grant agreement No~646908) through ERC Consolidator Grant ``S4F''. Research at Centre for Star and Planet Formation is funded by the Danish National Research Foundation. The work of AC was funded by a STFC grant. AC thanks the COST action CM1401 ``Our Astrochemical History'' for additional financial support. The group of EvD acknowledges A-ERC grant 291141 CHEMPLAN.

\end{acknowledgements}

\bibliographystyle{aa}

\begin{thebibliography}{196}
\expandafter\ifx\csname natexlab\endcsname\relax\def\natexlab#1{#1}\fi

\bibitem[{{Acharyya} {et~al.}(2007){Acharyya}, {Fuchs}, {Fraser}, {van
  Dishoeck}, \& {Linnartz}}]{acharyya07}
{Acharyya}, K., {Fuchs}, G.~W., {Fraser}, H.~J., {van Dishoeck}, E.~F., \&
  {Linnartz}, H. 2007, \aap, 466, 1005

\bibitem[{{Adams} \& {Shu}(1986)}]{adams86}
{Adams}, F.~C. \& {Shu}, F.~H. 1986, \apj, 308, 836

\bibitem[{{ALMA Partnership} {et~al.}(2015){ALMA Partnership}, {Brogan},
  {P{\'e}rez}, {Hunter}, {Dent}, {Hales}, {Hills}, {Corder}, {Fomalont},
  {Vlahakis}, {Asaki}, {Barkats}, {Hirota}, {Hodge}, {Impellizzeri}, {Kneissl},
  {Liuzzo}, {Lucas}, {Marcelino}, {Matsushita}, {Nakanishi}, {Phillips},
  {Richards}, {Toledo}, {Aladro}, {Broguiere}, {Cortes}, {Cortes}, {Espada},
  {Galarza}, {Garcia-Appadoo}, {Guzman-Ramirez}, {Humphreys}, {Jung}, {Kameno},
  {Laing}, {Leon}, {Marconi}, {Mignano}, {Nikolic}, {Nyman}, {Radiszcz},
  {Remijan}, {Rod{\'o}n}, {Sawada}, {Takahashi}, {Tilanus}, {Vila Vilaro},
  {Watson}, {Wiklind}, {Akiyama}, {Chapillon}, {de Gregorio-Monsalvo}, {Di
  Francesco}, {Gueth}, {Kawamura}, {Lee}, {Nguyen Luong}, {Mangum}, {Pietu},
  {Sanhueza}, {Saigo}, {Takakuwa}, {Ubach}, {van Kempen}, {Wootten},
  {Castro-Carrizo}, {Francke}, {Gallardo}, {Garcia}, {Gonzalez}, {Hill},
  {Kaminski}, {Kurono}, {Liu}, {Lopez}, {Morales}, {Plarre}, {Schieven},
  {Testi}, {Videla}, {Villard}, {Andreani}, {Hibbard}, \& {Tatematsu}}]{hltau}
{ALMA Partnership}, {Brogan}, C.~L., {P{\'e}rez}, L.~M., {et~al.} 2015, \apjl,
  808, L3

\bibitem[{{Andr\'e} {et~al.}(1993){Andr\'e}, {Ward-Thompson}, \&
  {Barsony}}]{andre93}
{Andr\'e}, P., {Ward-Thompson}, D., \& {Barsony}, M. 1993, \apj, 406, 122

\bibitem[{{Arce} {et~al.}(2008){Arce}, {Santiago-Garc{\'{\i}}a},
  {J{\o}rgensen}, {Tafalla}, \& {Bachiller}}]{arce08}
{Arce}, H.~G., {Santiago-Garc{\'{\i}}a}, J., {J{\o}rgensen}, J.~K., {Tafalla},
  M., \& {Bachiller}, R. 2008, \apjl, 681, L21

\bibitem[{{Bachiller} \& {P\'erez Guti\'errez}(1997)}]{bachiller97}
{Bachiller}, R. \& {P\'erez Guti\'errez}, M. 1997, \apjl, 487, L93

\bibitem[{{Bacmann} {et~al.}(2010){Bacmann}, {Caux}, {Hily-Blant}, {Parise},
  {Pagani}, {Bottinelli}, {Maret}, {Vastel}, {Ceccarelli}, {Cernicharo},
  {Henning}, {Castets}, {Coutens}, {Bergin}, {Blake}, {Crimier}, {Demyk},
  {Dominik}, {Gerin}, {Hennebelle}, {Kahane}, {Klotz}, {Melnick}, {Schilke},
  {Wakelam}, {Walters}, {Baudry}, {Bell}, {Benedettini}, {Boogert}, {Cabrit},
  {Caselli}, {Codella}, {Comito}, {Encrenaz}, {Falgarone}, {Fuente},
  {Goldsmith}, {Helmich}, {Herbst}, {Jacq}, {Kama}, {Langer}, {Lefloch}, {Lis},
  {Lord}, {Lorenzani}, {Neufeld}, {Nisini}, {Pacheco}, {Pearson}, {Phillips},
  {Salez}, {Saraceno}, {Schuster}, {Tielens}, {van der Tak}, {van der Wiel},
  {Viti}, {Wyrowski}, {Yorke}, {Faure}, {Benz}, {Coeur-Joly}, {Cros},
  {G{\"u}sten}, \& {Ravera}}]{bacmann10}
{Bacmann}, A., {Caux}, E., {Hily-Blant}, P., {et~al.} 2010, \aap, 521, L42

\bibitem[{{Bacmann} {et~al.}(2012){Bacmann}, {Taquet}, {Faure}, {Kahane}, \&
  {Ceccarelli}}]{bacmann12}
{Bacmann}, A., {Taquet}, V., {Faure}, A., {Kahane}, C., \& {Ceccarelli}, C.
  2012, \aap, 541, L12

\bibitem[{{Baryshev} {et~al.}(2015){Baryshev}, {Hesper}, {Mena}, {Klapwijk},
  {van Kempen}, {Hogerheijde}, {Jackson}, {Adema}, {Gerlofsma}, {Bekema},
  {Barkhof}, {de Haan-Stijkel}, {van den Bemt}, {Koops}, {Keizer}, {Pieters},
  {Koops van het Jagt}, {Schaeffer}, {Zijlstra}, {Kroug}, {Lodewijk},
  {Wielinga}, {Boland}, {de Graauw}, {van Dishoeck}, {Jager}, \&
  {Wild}}]{baryshev15}
{Baryshev}, A.~M., {Hesper}, R., {Mena}, F.~P., {et~al.} 2015, \aap, 577, A129

\bibitem[{{Belloche} {et~al.}(2016){Belloche}, {M{\"u}ller}, {Garrod}, \&
  {Menten}}]{belloche16}
{Belloche}, A., {M{\"u}ller}, H.~S.~P., {Garrod}, R.~T., \& {Menten}, K.~M.
  2016, \aap, 587, A91

\bibitem[{{Belloche} {et~al.}(2013){Belloche}, {M{\"u}ller}, {Menten},
  {Schilke}, \& {Comito}}]{belloche13}
{Belloche}, A., {M{\"u}ller}, H.~S.~P., {Menten}, K.~M., {Schilke}, P., \&
  {Comito}, C. 2013, \aap, 559, A47

\bibitem[{{Beltr{\'a}n} {et~al.}(2009){Beltr{\'a}n}, {Codella}, {Viti}, {Neri},
  \& {Cesaroni}}]{beltran09}
{Beltr{\'a}n}, M.~T., {Codella}, C., {Viti}, S., {Neri}, R., \& {Cesaroni}, R.
  2009, \apjl, 690, L93

\bibitem[{{Bisschop} {et~al.}(2006){Bisschop}, {Fraser}, {{\"O}berg}, {van
  Dishoeck}, \& {Schlemmer}}]{bisschop06}
{Bisschop}, S.~E., {Fraser}, H.~J., {{\"O}berg}, K.~I., {van Dishoeck}, E.~F.,
  \& {Schlemmer}, S. 2006, \aap, 449, 1297

\bibitem[{{Bisschop} {et~al.}(2008){Bisschop}, {J{\o}rgensen}, {Bourke},
  {Bottinelli}, \& {van Dishoeck}}]{bisschop08}
{Bisschop}, S.~E., {J{\o}rgensen}, J.~K., {Bourke}, T.~L., {Bottinelli}, S., \&
  {van Dishoeck}, E.~F. 2008, \aap, 488, 959

\bibitem[{{Bisschop} {et~al.}(2007){Bisschop}, {J{\o}rgensen}, {van Dishoeck},
  \& {de Wachter}}]{bisschop07}
{Bisschop}, S.~E., {J{\o}rgensen}, J.~K., {van Dishoeck}, E.~F., \& {de
  Wachter}, E.~B.~M. 2007, \aap, 465, 913

\bibitem[{{Biver} {et~al.}(2014){Biver}, {Bockel{\'e}e-Morvan}, {Debout},
  {Crovisier}, {Boissier}, {Lis}, {Dello Russo}, {Moreno}, {Colom}, {Paubert},
  {Vervack}, \& {Weaver}}]{biver14}
{Biver}, N., {Bockel{\'e}e-Morvan}, D., {Debout}, V., {et~al.} 2014, \aap, 566,
  L5

\bibitem[{{Biver} {et~al.}(2015){Biver}, {Bockelee-Morvan}, {Moreno},
  {Crovisier}, {Colom}, {Lis}, {Sandqvist}, {Boissier}, {Despois}, \&
  {Milam}}]{biver15}
{Biver}, N., {Bockelee-Morvan}, D., {Moreno}, R., {et~al.} 2015, Science
  Advances, 1, e1500863

\bibitem[{{Blake} {et~al.}(1987){Blake}, {Sutton}, {Masson}, \&
  {Phillips}}]{blake87}
{Blake}, G.~A., {Sutton}, E.~C., {Masson}, C.~R., \& {Phillips}, T.~G. 1987,
  \apj, 315, 621

\bibitem[{{Blake} {et~al.}(1994){Blake}, {van Dishoek}, {Jansen}, {Groesbeck},
  \& {Mundy}}]{blake94}
{Blake}, G.~A., {van Dishoek}, E.~F., {Jansen}, D.~J., {Groesbeck}, T.~D., \&
  {Mundy}, L.~G. 1994, \apj, 428, 680

\bibitem[{{Bottinelli} {et~al.}(2004{\natexlab{a}}){Bottinelli}, {Ceccarelli},
  {Lefloch}, {Williams}, {Castets}, {Caux}, {Cazaux}, {Maret}, {Parise}, \&
  {Tielens}}]{bottinelli04n1333i4a}
{Bottinelli}, S., {Ceccarelli}, C., {Lefloch}, B., {et~al.} 2004{\natexlab{a}},
  \apj, 615, 354

\bibitem[{{Bottinelli} {et~al.}(2004{\natexlab{b}}){Bottinelli}, {Ceccarelli},
  {Neri}, {Williams}, {Caux}, {Cazaux}, {Lefloch}, {Maret}, \&
  {Tielens}}]{bottinelli04iras16293}
{Bottinelli}, S., {Ceccarelli}, C., {Neri}, R., {et~al.} 2004{\natexlab{b}},
  \apjl, 617, L69

\bibitem[{{Bottinelli} {et~al.}(2014){Bottinelli}, {Wakelam}, {Caux}, {Vastel},
  {Aikawa}, \& {Ceccarelli}}]{bottinelli14}
{Bottinelli}, S., {Wakelam}, V., {Caux}, E., {et~al.} 2014, \mnras, 441, 1964

\bibitem[{{Bouchez} {et~al.}(2012){Bouchez}, {Margul{\`e}s}, {Motiyenko},
  {Guillemin}, {Walters}, {Bottinelli}, {Ceccarelli}, \& {Kahane}}]{bouchez12}
{Bouchez}, A., {Margul{\`e}s}, L., {Motiyenko}, R.~A., {et~al.} 2012, \aap,
  540, A51

\bibitem[{{Brouillet} {et~al.}(2015){Brouillet}, {Despois}, {Lu}, {Baudry},
  {Cernicharo}, {Bockel{\'e}e-Morvan}, {Crovisier}, \& {Biver}}]{brouillet15}
{Brouillet}, N., {Despois}, D., {Lu}, X.-H., {et~al.} 2015, \aap, 576, A129

\bibitem[{{Brown} {et~al.}(1975){Brown}, {Crofts}, {Godfrey}, {Gardner},
  {Robinson}, \& {Whiteoak}}]{brown75}
{Brown}, R.~D., {Crofts}, J.~G., {Godfrey}, P.~D., {et~al.} 1975, \apjl, 197,
  L29

\bibitem[{{Buckley} \& {Gigu{\`e}re}(1967)}]{buckley67}
{Buckley}, P. \& {Gigu{\`e}re}, P.~A. 1967, Can. J. Chem., 45, 397

\bibitem[{{Butler} {et~al.}(2001){Butler}, {De Lucia}, {Petkie},
  {M{\o}llendal}, {Horn}, \& {Herbst}}]{butler01}
{Butler}, R.~A.~H., {De Lucia}, F.~C., {Petkie}, D.~T., {et~al.} 2001, \apjs,
  134, 319

\bibitem[{{Butner} {et~al.}(2007){Butner}, {Charnley}, {Ceccarelli}, {Rodgers},
  {Pardo}, {Parise}, {Cernicharo}, \& {Davis}}]{butner07}
{Butner}, H.~M., {Charnley}, S.~B., {Ceccarelli}, C., {et~al.} 2007, \apjl,
  659, L137

\bibitem[{{Calcutt} {et~al.}(2014){Calcutt}, {Viti}, {Codella}, {Beltr{\'a}n},
  {Fontani}, \& {Woods}}]{calcutt14}
{Calcutt}, H., {Viti}, S., {Codella}, C., {et~al.} 2014, \mnras, 443, 3157

\bibitem[{{Carroll} {et~al.}(2010){Carroll}, {Drouin}, \& {Widicus
  Weaver}}]{carroll10}
{Carroll}, P.~B., {Drouin}, B.~J., \& {Widicus Weaver}, S.~L. 2010, \apj, 723,
  845

\bibitem[{{Caselli} {et~al.}(2003){Caselli}, {van der Tak}, {Ceccarelli}, \&
  {Bacmann}}]{caselli03}
{Caselli}, P., {van der Tak}, F.~F.~S., {Ceccarelli}, C., \& {Bacmann}, A.
  2003, \aap, 403, L37

\bibitem[{{Castets} {et~al.}(2001){Castets}, {Ceccarelli}, {Loinard}, {Caux},
  \& {Lefloch}}]{castets01}
{Castets}, A., {Ceccarelli}, C., {Loinard}, L., {Caux}, E., \& {Lefloch}, B.
  2001, \aap, 375, 40

\bibitem[{{Caux} {et~al.}(2011){Caux}, {Kahane}, {Castets}, {Coutens},
  {Ceccarelli}, {Bacmann}, {Bisschop}, {Bottinelli}, {Comito}, {Helmich},
  {Lefloch}, {Parise}, {Schilke}, {Tielens}, {van Dishoeck}, {Vastel},
  {Wakelam}, \& {Walters}}]{caux11}
{Caux}, E., {Kahane}, C., {Castets}, A., {et~al.} 2011, \aap, 532, A23

\bibitem[{{Cazaux} {et~al.}(2011){Cazaux}, {Caselli}, \& {Spaans}}]{cazaux11}
{Cazaux}, S., {Caselli}, P., \& {Spaans}, M. 2011, \apjl, 741, L34

\bibitem[{{Cazaux} {et~al.}(2003){Cazaux}, {Tielens}, {Ceccarelli}, {Castets},
  {Wakelam}, {Caux}, {Parise}, \& {Teyssier}}]{cazaux03}
{Cazaux}, S., {Tielens}, A.~G.~G.~M., {Ceccarelli}, C., {et~al.} 2003, \apjl,
  593, L51

\bibitem[{{Ceccarelli} {et~al.}(2010){Ceccarelli}, {Bacmann}, {Boogert},
  {Caux}, {Dominik}, {Lefloch}, {Lis}, {Schilke}, {van der Tak}, {Caselli},
  {Cernicharo}, {Codella}, {Comito}, {Fuente}, {Baudry}, {Bell}, {Benedettini},
  {Bergin}, {Blake}, {Bottinelli}, {Cabrit}, {Castets}, {Coutens}, {Crimier},
  {Demyk}, {Encrenaz}, {Falgarone}, {Gerin}, {Goldsmith}, {Helmich},
  {Hennebelle}, {Henning}, {Herbst}, {Hily-Blant}, {Jacq}, {Kahane}, {Kama},
  {Klotz}, {Langer}, {Lord}, {Lorenzani}, {Maret}, {Melnick}, {Neufeld},
  {Nisini}, {Pacheco}, {Pagani}, {Parise}, {Pearson}, {Phillips}, {Salez},
  {Saraceno}, {Schuster}, {Tielens}, {van der Wiel}, {Vastel}, {Viti},
  {Wakelam}, {Walters}, {Wyrowski}, {Yorke}, {Liseau}, {Olberg}, {Szczerba},
  {Benz}, \& {Melchior}}]{ceccarelli10}
{Ceccarelli}, C., {Bacmann}, A., {Boogert}, A., {et~al.} 2010, \aap, 521, L22

\bibitem[{{Ceccarelli} {et~al.}(2014){Ceccarelli}, {Caselli},
  {Bockel{\'e}e-Morvan}, {Mousis}, {Pizzarello}, {Robert}, \&
  {Semenov}}]{ceccarellippvi}
{Ceccarelli}, C., {Caselli}, P., {Bockel{\'e}e-Morvan}, D., {et~al.} 2014,
  Protostars and Planets VI, 859

\bibitem[{{Ceccarelli} {et~al.}(2000{\natexlab{a}}){Ceccarelli}, {Castets},
  {Caux}, {Hollenbach}, {Loinard}, {Molinari}, \&
  {Tielens}}]{ceccarelli00model}
{Ceccarelli}, C., {Castets}, A., {Caux}, E., {et~al.} 2000{\natexlab{a}}, \aap,
  355, 1129

\bibitem[{{Ceccarelli} {et~al.}(1998{\natexlab{a}}){Ceccarelli}, {Castets},
  {Loinard}, {Caux}, \& {Tielens}}]{ceccarelli98d2co}
{Ceccarelli}, C., {Castets}, A., {Loinard}, L., {Caux}, E., \& {Tielens},
  A.~G.~G.~M. 1998{\natexlab{a}}, \aap, 338, L43

\bibitem[{{Ceccarelli} {et~al.}(1999){Ceccarelli}, {Caux}, {Loinard},
  {Castets}, {Tielens}, {Molinari}, {Liseau}, {Saraceno}, {Smith}, \&
  {White}}]{ceccarelli99}
{Ceccarelli}, C., {Caux}, E., {Loinard}, L., {et~al.} 1999, \aap, 342, L21

\bibitem[{{Ceccarelli} {et~al.}(1998{\natexlab{b}}){Ceccarelli}, {Caux},
  {White}, {Molinari}, {Furniss}, {Liseau}, {Nisini}, {Saraceno}, {Spinoglio},
  \& {Wolfire}}]{ceccarelli98h2o}
{Ceccarelli}, C., {Caux}, E., {White}, G.~J., {et~al.} 1998{\natexlab{b}},
  \aap, 331, 372

\bibitem[{{Ceccarelli} {et~al.}(2000{\natexlab{b}}){Ceccarelli}, {Loinard},
  {Castets}, {Faure}, \& {Lefloch}}]{ceccarelli00glycine}
{Ceccarelli}, C., {Loinard}, L., {Castets}, A., {Faure}, A., \& {Lefloch}, B.
  2000{\natexlab{b}}, \aap, 362, 1122

\bibitem[{{Ceccarelli} {et~al.}(2000{\natexlab{c}}){Ceccarelli}, {Loinard},
  {Castets}, {Tielens}, \& {Caux}}]{ceccarelli00h2co}
{Ceccarelli}, C., {Loinard}, L., {Castets}, A., {Tielens}, A.~G.~G.~M., \&
  {Caux}, E. 2000{\natexlab{c}}, \aap, 357, L9

\bibitem[{{Ceccarelli} {et~al.}(2001){Ceccarelli}, {Loinard}, {Castets},
  {Tielens}, {Caux}, {Lefloch}, \& {Vastel}}]{ceccarelli01d2co}
{Ceccarelli}, C., {Loinard}, L., {Castets}, A., {et~al.} 2001, \aap, 372, 998

\bibitem[{{Ceccarelli} {et~al.}(2003){Ceccarelli}, {Maret}, {Tielens},
  {Castets}, \& {Caux}}]{ceccarelli03}
{Ceccarelli}, C., {Maret}, S., {Tielens}, A.~G.~G.~M., {Castets}, A., \&
  {Caux}, E. 2003, \aap, 410, 587

\bibitem[{{Chandler} {et~al.}(2005){Chandler}, {Brogan}, {Shirley}, \&
  {Loinard}}]{chandler05}
{Chandler}, C.~J., {Brogan}, C.~L., {Shirley}, Y.~L., \& {Loinard}, L. 2005,
  \apj, 632, 371

\bibitem[{{Charnley} {et~al.}(2004){Charnley}, {Ehrenfreund}, {Millar},
  {Boogert}, {Markwick}, {Butner}, {Ruiterkamp}, \& {Rodgers}}]{charnley04}
{Charnley}, S.~B., {Ehrenfreund}, P., {Millar}, T.~J., {et~al.} 2004, \mnras,
  347, 157

\bibitem[{{Christen} \& {M{\"u}ller}(2003)}]{christen03}
{Christen}, D. \& {M{\"u}ller}, H.~S.~P. 2003, Physical Chemistry Chemical
  Physics (Incorporating Faraday Transactions), 5

\bibitem[{{Chuang} {et~al.}(2016){Chuang}, {Fedoseev}, {Ioppolo}, {van
  Dishoeck}, \& {Linnartz}}]{chuang16}
{Chuang}, K.-J., {Fedoseev}, G., {Ioppolo}, S., {van Dishoeck}, E.~F., \&
  {Linnartz}, H. 2016, \mnras, 455, 1702

\bibitem[{{Codella} {et~al.}(2014){Codella}, {Cabrit}, {Gueth}, {Podio},
  {Leurini}, {Bachiller}, {Gusdorf}, {Lefloch}, {Nisini}, {Tafalla}, \&
  {Yvart}}]{codella14}
{Codella}, C., {Cabrit}, S., {Gueth}, F., {et~al.} 2014, \aap, 568, L5

\bibitem[{{Correia} {et~al.}(2004){Correia}, {Griffin}, \&
  {Saraceno}}]{correia04}
{Correia}, J.~C., {Griffin}, M., \& {Saraceno}, P. 2004, \aap, 418, 607

\bibitem[{{Coutens} {et~al.}({2016}){Coutens}, {J{\o}rgensen}, {van der Wiel},
  {Author}, {Author}, {Author}, {Author}, {Author}, {Author}, {Author}, \&
  {Author}}]{coutens16}
{Coutens}, A., {J{\o}rgensen}, J.~K., {van der Wiel}, M.~H.~D., {et~al.}
  {2016}, \aap, {submitted}

\bibitem[{{Coutens} {et~al.}(2015){Coutens}, {Persson}, {J{\o}rgensen},
  {Wampfler}, \& {Lykke}}]{coutens15}
{Coutens}, A., {Persson}, M.~V., {J{\o}rgensen}, J.~K., {Wampfler}, S.~F., \&
  {Lykke}, J.~M. 2015, \aap, 576, A5

\bibitem[{{Coutens} {et~al.}(2012){Coutens}, {Vastel}, {Caux}, {Ceccarelli},
  {Bottinelli}, {Wiesenfeld}, {Faure}, {Scribano}, \& {Kahane}}]{coutens12}
{Coutens}, A., {Vastel}, C., {Caux}, E., {et~al.} 2012, \aap, 539, A132

\bibitem[{{Coutens} {et~al.}(2013){Coutens}, {Vastel}, {Cazaux}, {Bottinelli},
  {Caux}, {Ceccarelli}, {Demyk}, {Taquet}, \& {Wakelam}}]{coutens13}
{Coutens}, A., {Vastel}, C., {Cazaux}, S., {et~al.} 2013, \aap, 553, A75

\bibitem[{{Crovisier} {et~al.}(2004){Crovisier}, {Bockel{\'e}e-Morvan},
  {Biver}, {Colom}, {Despois}, \& {Lis}}]{crovisier04}
{Crovisier}, J., {Bockel{\'e}e-Morvan}, D., {Biver}, N., {et~al.} 2004, \aap,
  418, L35

\bibitem[{{Demyk} {et~al.}(2010){Demyk}, {Bottinelli}, {Caux}, {Vastel},
  {Ceccarelli}, {Kahane}, \& {Castets}}]{demyk10}
{Demyk}, K., {Bottinelli}, S., {Caux}, E., {et~al.} 2010, \aap, 517, A17

\bibitem[{{Doty} {et~al.}(2004){Doty}, {Sch\"{o}ier}, \& {van
  Dishoeck}}]{doty04}
{Doty}, S.~D., {Sch\"{o}ier}, F.~L., \& {van Dishoeck}, E.~F. 2004, \aap, 418,
  1021

\bibitem[{{Ellder} {et~al.}(1980){Ellder}, {Friberg}, {Hjalmarson}, {Hoglund},
  {Johansson}, {Olofsson}, {Rydbeck}, {Rydbeck}, {Guelin}, \&
  {Irvine}}]{ellder80}
{Ellder}, J., {Friberg}, P., {Hjalmarson}, A., {et~al.} 1980, \apjl, 242, L93

\bibitem[{{Faure} {et~al.}(2015){Faure}, {Faure}, {Theul{\'e}}, {Quirico}, \&
  {Schmitt}}]{faure15}
{Faure}, A., {Faure}, M., {Theul{\'e}}, P., {Quirico}, E., \& {Schmitt}, B.
  2015, \aap, 584, A98

\bibitem[{{Favre} {et~al.}(2014{\natexlab{a}}){Favre}, {Carvajal}, {Field},
  {J{\o}rgensen}, {Bisschop}, {Brouillet}, {Despois}, {Baudry}, {Kleiner},
  {Bergin}, {Crockett}, {Neill}, {Margul{\`e}s}, {Huet}, \&
  {Demaison}}]{favre14mf}
{Favre}, C., {Carvajal}, M., {Field}, D., {et~al.} 2014{\natexlab{a}}, \apjs,
  215, 25

\bibitem[{{Favre} {et~al.}(2014{\natexlab{b}}){Favre}, {J{\o}rgensen}, {Field},
  {Brinch}, {Bisschop}, {Bourke}, {Hogerheijde}, \& {Frieswijk}}]{favre14eSMA}
{Favre}, C., {J{\o}rgensen}, J.~K., {Field}, D., {et~al.} 2014{\natexlab{b}},
  \apj, 790, 55

\bibitem[{{Fedoseev} {et~al.}(2015){Fedoseev}, {Cuppen}, {Ioppolo}, {Lamberts},
  \& {Linnartz}}]{fedoseev15}
{Fedoseev}, G., {Cuppen}, H.~M., {Ioppolo}, S., {Lamberts}, T., \& {Linnartz},
  H. 2015, \mnras, 448, 1288

\bibitem[{{Fraser} {et~al.}(2001){Fraser}, {Collings}, {McCoustra}, \&
  {Williams}}]{fraser01}
{Fraser}, H.~J., {Collings}, M.~P., {McCoustra}, M.~R.~S., \& {Williams}, D.~A.
  2001, \mnras, 327, 1165

\bibitem[{{Friesen} {et~al.}(2014){Friesen}, {Di Francesco}, {Bourke},
  {Caselli}, {J{\o}rgensen}, {Pineda}, \& {Wong}}]{friesen14}
{Friesen}, R.~K., {Di Francesco}, J., {Bourke}, T.~L., {et~al.} 2014, \apj,
  797, 27

\bibitem[{{Fuente} {et~al.}(2014){Fuente}, {Cernicharo}, {Caselli}, {McCoey},
  {Johnstone}, {Fich}, {van Kempen}, {Palau}, {Y{\i}ld{\i}z}, {Tercero}, \&
  {L{\'o}pez}}]{fuente14}
{Fuente}, A., {Cernicharo}, J., {Caselli}, P., {et~al.} 2014, \aap, 568, A65

\bibitem[{{Fukui} {et~al.}(1986){Fukui}, {Sugitani}, {Takaba}, {Iwata},
  {Mizuno}, {Ogawa}, \& {Kawabata}}]{fukui86}
{Fukui}, Y., {Sugitani}, K., {Takaba}, H., {et~al.} 1986, \apjl, 311, L85

\bibitem[{{Furuya} {et~al.}(2001){Furuya}, {Kitamura}, {Wootten}, {Claussen},
  \& {Kawabe}}]{furuya01}
{Furuya}, R.~S., {Kitamura}, Y., {Wootten}, H.~A., {Claussen}, M.~J., \&
  {Kawabe}, R. 2001, \apjl, 559, L143

\bibitem[{{Garay} {et~al.}(2002){Garay}, {Mardones}, {Rodr{\' i}guez},
  {Caselli}, \& {Bourke}}]{garay02}
{Garay}, G., {Mardones}, D., {Rodr{\' i}guez}, L.~F., {Caselli}, P., \&
  {Bourke}, T.~L. 2002, \apj, 567, 980

\bibitem[{{Garrod}(2013)}]{garrod13}
{Garrod}, R.~T. 2013, \apj, 765, 60

\bibitem[{{Gerin} {et~al.}(2006){Gerin}, {Lis}, {Philipp}, {G{\"u}sten},
  {Roueff}, \& {Reveret}}]{gerin06}
{Gerin}, M., {Lis}, D.~C., {Philipp}, S., {et~al.} 2006, \aap, 454, L63

\bibitem[{{Gibb} {et~al.}(2000){Gibb}, {Nummelin}, {Irvine}, {Whittet}, \&
  {Bergman}}]{gibb00}
{Gibb}, E., {Nummelin}, A., {Irvine}, W.~M., {Whittet}, D.~C.~B., \& {Bergman},
  P. 2000, \apj, 545, 309

\bibitem[{Girart {et~al.}(2014)Girart, Estalella, Palau, Torrelles, \&
  Rao}]{girart14}
Girart, J.~M., Estalella, R., Palau, A., Torrelles, J.~M., \& Rao, R. 2014,
  \apjl, 780, L11

\bibitem[{{Goldsmith} {et~al.}(1999){Goldsmith}, {Langer}, \&
  {Velusamy}}]{goldsmith99}
{Goldsmith}, P.~F., {Langer}, W.~D., \& {Velusamy}, T. 1999, \apjl, 519, L173

\bibitem[{{Halfen} {et~al.}(2006){Halfen}, {Apponi}, {Woolf}, {Polt}, \&
  {Ziurys}}]{halfen06}
{Halfen}, D.~T., {Apponi}, A.~J., {Woolf}, N., {Polt}, R., \& {Ziurys}, L.~M.
  2006, \apj, 639, 237

\bibitem[{{Haykal} {et~al.}(2013){Haykal}, {Motiyenko}, {Margul{\`e}s}, \&
  {Huet}}]{haykal13}
{Haykal}, I., {Motiyenko}, R.~A., {Margul{\`e}s}, L., \& {Huet}, T.~R. 2013,
  \aap, 549, A96

\bibitem[{{Herbst} \& {van Dishoeck}(2009)}]{herbst09}
{Herbst}, E. \& {van Dishoeck}, E.~F. 2009, \araa, 47, 427

\bibitem[{{Hily-Blant} {et~al.}(2010){Hily-Blant}, {Maret}, {Bacmann},
  {Bottinelli}, {Parise}, {Caux}, {Faure}, {Bergin}, {Blake}, {Castets},
  {Ceccarelli}, {Cernicharo}, {Coutens}, {Crimier}, {Demyk}, {Dominik},
  {Gerin}, {Hennebelle}, {Henning}, {Kahane}, {Klotz}, {Melnick}, {Pagani},
  {Schilke}, {Vastel}, {Wakelam}, {Walters}, {Baudry}, {Bell}, {Benedettini},
  {Boogert}, {Cabrit}, {Caselli}, {Codella}, {Comito}, {Encrenaz}, {Falgarone},
  {Fuente}, {Goldsmith}, {Helmich}, {Herbst}, {Jacq}, {Kama}, {Langer},
  {Lefloch}, {Lis}, {Lord}, {Lorenzani}, {Neufeld}, {Nisini}, {Pacheco},
  {Phillips}, {Salez}, {Saraceno}, {Schuster}, {Tielens}, {van der Tak}, {van
  der Wiel}, {Viti}, {Wyrowski}, \& {Yorke}}]{hilyblant10}
{Hily-Blant}, P., {Maret}, S., {Bacmann}, A., {et~al.} 2010, \aap, 521, L52

\bibitem[{{Hirano} {et~al.}(2001){Hirano}, {Mikami}, {Umemoto}, {Yamamoto}, \&
  {Taniguchi}}]{hirano01}
{Hirano}, N., {Mikami}, H., {Umemoto}, T., {Yamamoto}, S., \& {Taniguchi}, Y.
  2001, \apj, 547, 899

\bibitem[{Hollis {et~al.}(2004)Hollis, Jewell, Lovas, \& Remijan}]{hollis04}
Hollis, J.~M., Jewell, P.~R., Lovas, F.~J., \& Remijan, A. 2004, ApJ, 613, L45

\bibitem[{{Hollis} {et~al.}(2000){Hollis}, {Lovas}, \& {Jewell}}]{hollis00}
{Hollis}, J.~M., {Lovas}, F.~J., \& {Jewell}, P.~R. 2000, \apjl, 540, L107

\bibitem[{{Hollis} {et~al.}(2002){Hollis}, {Lovas}, {Jewell}, \&
  {Coudert}}]{hollis02}
{Hollis}, J.~M., {Lovas}, F.~J., {Jewell}, P.~R., \& {Coudert}, L.~H. 2002,
  \apjl, 571, L59

\bibitem[{{Hollis} {et~al.}(2001){Hollis}, {Vogel}, {Snyder}, {Jewell}, \&
  {Lovas}}]{hollis01}
{Hollis}, J.~M., {Vogel}, S.~N., {Snyder}, L.~E., {Jewell}, P.~R., \& {Lovas},
  F.~J. 2001, \apjl, 554, L81

\bibitem[{{Howard} {et~al.}(2005){Howard}, {J{\o}rgensen}, \&
  {Kjaergaard}}]{howard05}
{Howard}, D.~L., {J{\o}rgensen}, P., \& {Kjaergaard}, H.~G. 2005,
  J.~A.~Chem.~Soc, 127, 17096

\bibitem[{{Ilyushin} {et~al.}(2009){Ilyushin}, {Kryvda}, \&
  {Alekseev}}]{ilyushin09}
{Ilyushin}, V., {Kryvda}, A., \& {Alekseev}, E. 2009, Journal of Molecular
  Spectroscopy, 255, 32

\bibitem[{{Jaber} {et~al.}(2014){Jaber}, {Ceccarelli}, {Kahane}, \&
  {Caux}}]{jaber14}
{Jaber}, A.~A., {Ceccarelli}, C., {Kahane}, C., \& {Caux}, E. 2014, \apj, 791,
  29

\bibitem[{{Johnson} {et~al.}(2013){Johnson}, {Sams}, {Profeta}, {Akagi},
  {Burling}, {Yokelson}, \& {Williams}}]{johnson13}
{Johnson}, T.~J., {Sams}, R.~L., {Profeta}, L.~T.~M., {et~al.} 2013, J. Phys.
  Chem. A., 117, 4096

\bibitem[{{J{\o}rgensen} {et~al.}(2005{\natexlab{a}}){J{\o}rgensen}, {Bourke},
  {Myers}, {Sch\"{o}ier}, {van Dishoeck}, \& {Wilner}}]{iras2sma}
{J{\o}rgensen}, J.~K., {Bourke}, T.~L., {Myers}, P.~C., {et~al.}
  2005{\natexlab{a}}, \apj, {632}, 973

\bibitem[{{J{\o}rgensen} {et~al.}(2011){J{\o}rgensen}, {Bourke}, {Nguyen
  Luong}, \& {Takakuwa}}]{iras16293sma}
{J{\o}rgensen}, J.~K., {Bourke}, T.~L., {Nguyen Luong}, Q., \& {Takakuwa}, S.
  2011, \aap, {534}, {A100}

\bibitem[{{J{\o}rgensen} {et~al.}(2012){J{\o}rgensen}, {Favre}, {Bisschop},
  {Bourke}, {van Dishoeck}, \& {Schmalzl}}]{jorgensen12}
{J{\o}rgensen}, J.~K., {Favre}, C., {Bisschop}, S.~E., {et~al.} 2012, \apjl,
  757, L4

\bibitem[{{J{\o}rgensen} {et~al.}(2004){J{\o}rgensen}, {Hogerheijde}, {van
  Dishoeck}, {Blake}, \& {Sch\"{o}ier}}]{n1333i2art}
{J{\o}rgensen}, J.~K., {Hogerheijde}, M.~R., {van Dishoeck}, E.~F., {Blake},
  G.~A., \& {Sch\"{o}ier}, F.~L. 2004, \aap, 413, 993

\bibitem[{{J{\o}rgensen} {et~al.}(2008){J{\o}rgensen}, {Johnstone}, {Kirk},
  {Myers}, {Allen}, \& {Shirley}}]{jorgensen08}
{J{\o}rgensen}, J.~K., {Johnstone}, D., {Kirk}, H., {et~al.} 2008, \apj, 683,
  822

\bibitem[{{J{\o}rgensen} {et~al.}(2005{\natexlab{b}}){J{\o}rgensen}, {Lahuis},
  {Sch{\"o}ier}, {van Dishoeck}, {Blake}, {Boogert}, {Dullemond}, {Evans},
  {Kessler-Silacci}, \& {Pontoppidan}}]{iras16293letter}
{J{\o}rgensen}, J.~K., {Lahuis}, F., {Sch{\"o}ier}, F.~L., {et~al.}
  2005{\natexlab{b}}, \apjl, 631, L77

\bibitem[{{J{\o}rgensen} {et~al.}(2002){J{\o}rgensen}, {Sch{\" o}ier}, \& {van
  Dishoeck}}]{jorgensen02}
{J{\o}rgensen}, J.~K., {Sch{\" o}ier}, F.~L., \& {van Dishoeck}, E.~F. 2002,
  \aap, 389, 908

\bibitem[{{J{\o}rgensen} {et~al.}(2005{\natexlab{c}}){J{\o}rgensen},
  {Sch\"{o}ier}, \& {van Dishoeck}}]{hotcoresample}
{J{\o}rgensen}, J.~K., {Sch\"{o}ier}, F.~L., \& {van Dishoeck}, E.~F.
  2005{\natexlab{c}}, \aap, 437, 501

\bibitem[{{J{\o}rgensen} \& {van Dishoeck}(2010)}]{iras4b_hdo}
{J{\o}rgensen}, J.~K. \& {van Dishoeck}, E.~F. 2010, \apjl, 725, L172

\bibitem[{{J{\o}rgensen} {et~al.}({2013}){J{\o}rgensen}, {Visser}, {Sakai},
  {Author}, {Author}, {Author}, {Author}, {Author}, {Author}, {Author}, \&
  {Author}}]{jorgensen13}
{J{\o}rgensen}, J.~K., {Visser}, R., {Sakai}, N., {et~al.} {2013}, \apjl,
  {779}, {L22}

\bibitem[{{Kahane} {et~al.}(2013){Kahane}, {Ceccarelli}, {Faure}, \&
  {Caux}}]{kahane13}
{Kahane}, C., {Ceccarelli}, C., {Faure}, A., \& {Caux}, E. 2013, \apjl, 763,
  L38

\bibitem[{{Knude} \& {H{\o}g}(1998)}]{knude98}
{Knude}, J. \& {H{\o}g}, E. 1998, \aap, 338, 897

\bibitem[{{Kristensen} {et~al.}(2013){Kristensen}, {Klaassen}, {Mottram},
  {Schmalzl}, \& {Hogerheijde}}]{kristensen13}
{Kristensen}, L.~E., {Klaassen}, P.~D., {Mottram}, J.~C., {Schmalzl}, M., \&
  {Hogerheijde}, M.~R. 2013, \aap, 549, L6

\bibitem[{{Kuan} {et~al.}(2004){Kuan}, {Huang}, {Charnley}, {Hirano},
  {Takakuwa}, {Wilner}, {Liu}, {Ohashi}, {Bourke}, {Qi}, \& {Zhang}}]{kuan04}
{Kuan}, Y., {Huang}, H., {Charnley}, S.~B., {et~al.} 2004, \apjl, 616, L27

\bibitem[{{Lada} \& {Wilking}(1984)}]{lada84}
{Lada}, C.~J. \& {Wilking}, B.~A. 1984, \apj, 287, 610

\bibitem[{{Langer} {et~al.}(1984){Langer}, {Graedel}, {Frerking}, \&
  {Armentrout}}]{langer84}
{Langer}, W.~D., {Graedel}, T.~E., {Frerking}, M.~A., \& {Armentrout}, P.~B.
  1984, \apj, 277, 581

\bibitem[{{Lindberg} {et~al.}(2014){Lindberg}, {J{\o}rgensen}, {Brinch},
  {Haugb{\o}lle}, {Bergin}, {Harsono}, {Persson}, {Visser}, \&
  {Yamamoto}}]{lindberg14alma}
{Lindberg}, J.~E., {J{\o}rgensen}, J.~K., {Brinch}, C., {et~al.} 2014, \aap,
  566, A74

\bibitem[{{Lis} {et~al.}(2006){Lis}, {Gerin}, {Roueff}, {Vastel}, \&
  {Phillips}}]{lis06}
{Lis}, D.~C., {Gerin}, M., {Roueff}, E., {Vastel}, C., \& {Phillips}, T.~G.
  2006, \apj, 636, 916

\bibitem[{{Loinard} {et~al.}(2001){Loinard}, {Castets}, {Ceccarelli}, {Caux},
  \& {Tielens}}]{loinard01}
{Loinard}, L., {Castets}, A., {Ceccarelli}, C., {Caux}, E., \& {Tielens},
  A.~G.~G.~M. 2001, \apjl, 552, L163

\bibitem[{{Loinard} {et~al.}(2000){Loinard}, {Castets}, {Ceccarelli},
  {Tielens}, {Faure}, {Caux}, \& {Duvert}}]{loinard00}
{Loinard}, L., {Castets}, A., {Ceccarelli}, C., {et~al.} 2000, \aap, 359, 1169

\bibitem[{Loinard {et~al.}(2007)Loinard, Chandler, Rodríguez, D’Alessio,
  Brogan, Wilner, \& Ho}]{loinard07}
Loinard, L., Chandler, C.~J., Rodríguez, L.~F., {et~al.} 2007, The
  Astrophysical Journal, 670, 1353

\bibitem[{{Loinard} {et~al.}(2008){Loinard}, {Torres}, {Mioduszewski}, \&
  {Rodr{\'{\i}}guez}}]{loinard08}
{Loinard}, L., {Torres}, R.~M., {Mioduszewski}, A.~J., \& {Rodr{\'{\i}}guez},
  L.~F. 2008, \apjl, 675, L29

\bibitem[{{Loinard} {et~al.}(2013){Loinard}, {Zapata}, {Rodr{\'{\i}}guez},
  {Pech}, {Chandler}, {Brogan}, {Wilner}, {Ho}, {Parise}, {Hartmann}, {Zhu},
  {Takahashi}, \& {Trejo}}]{loinard13}
{Loinard}, L., {Zapata}, L.~A., {Rodr{\'{\i}}guez}, L.~F., {et~al.} 2013,
  \mnras, 430, L10

\bibitem[{{Lombardi} {et~al.}(2008){Lombardi}, {Lada}, \& {Alves}}]{lombardi08}
{Lombardi}, M., {Lada}, C.~J., \& {Alves}, J. 2008, \aap, 480, 785

\bibitem[{{Looney} {et~al.}(2000){Looney}, {Mundy}, \& {Welch}}]{looney00}
{Looney}, L.~W., {Mundy}, L.~G., \& {Welch}, W.~J. 2000, \apj, 529, 477

\bibitem[{{Lykke} {et~al.}({2016}){Lykke}, {Coutens}, {J{\o}rgensen}, {Author},
  {Author}, {Author}, {Author}, {Author}, {Author}, {Author}, \&
  {Author}}]{lykke16}
{Lykke}, J.~M., {Coutens}, A., {J{\o}rgensen}, J.~K., {et~al.} {2016}, \aap,
  {in prep.}

\bibitem[{{Lykke} {et~al.}(2015){Lykke}, {Favre}, {Bergin}, \&
  {J{\o}rgensen}}]{lykke15}
{Lykke}, J.~M., {Favre}, C., {Bergin}, E.~A., \& {J{\o}rgensen}, J.~K. 2015,
  \aap, 582, A64

\bibitem[{{MacDonald} {et~al.}(1996){MacDonald}, {Gibb}, {Habing}, \&
  {Millar}}]{macdonald96}
{MacDonald}, G.~H., {Gibb}, A.~G., {Habing}, R.~J., \& {Millar}, T.~J. 1996,
  \aaps, 119, 333

\bibitem[{{Makiwa}(2014)}]{makiwa14}
{Makiwa}, G. 2014, PhD thesis, University of Lethbridge, Canada

\bibitem[{{Maret} {et~al.}(2004){Maret}, {Ceccarelli}, {Caux}, {Tielens},
  {J{\o}rgensen}, {van Dishoeck}, {Bacmann}, {Castets}, {Lefloch}, {Loinard},
  {Parise}, \& {Sch\"oier}}]{maret04}
{Maret}, S., {Ceccarelli}, C., {Caux}, E., {et~al.} 2004, \aap, 416, 577

\bibitem[{{Maret} {et~al.}(2005){Maret}, {Ceccarelli}, {Tielens}, {Caux},
  {Lefloch}, {Faure}, {Castets}, \& {Flower}}]{maret05}
{Maret}, S., {Ceccarelli}, C., {Tielens}, A.~G.~G.~M., {et~al.} 2005, \aap,
  442, 527

\bibitem[{{Mehringer} {et~al.}(1997){Mehringer}, {Snyder}, {Miao}, \&
  {Lovas}}]{mehringer97}
{Mehringer}, D.~M., {Snyder}, L.~E., {Miao}, Y., \& {Lovas}, F.~J. 1997, \apjl,
  480, L71

\bibitem[{{Mendoza} {et~al.}(2014){Mendoza}, {Lefloch}, {L{\'o}pez-Sepulcre},
  {Ceccarelli}, {Codella}, {Boechat-Roberty}, \& {Bachiller}}]{mendoza14}
{Mendoza}, E., {Lefloch}, B., {L{\'o}pez-Sepulcre}, A., {et~al.} 2014, \mnras,
  445, 151

\bibitem[{{Menten} {et~al.}(1987){Menten}, {Serabyn}, {Guesten}, \&
  {Wilson}}]{menten87}
{Menten}, K.~M., {Serabyn}, E., {Guesten}, R., \& {Wilson}, T.~L. 1987, \aap,
  177, L57

\bibitem[{{Milam} {et~al.}(2005){Milam}, {Savage}, {Brewster}, {Ziurys}, \&
  {Wyckoff}}]{milam05}
{Milam}, S.~N., {Savage}, C., {Brewster}, M.~A., {Ziurys}, L.~M., \& {Wyckoff},
  S. 2005, \apj, 634, 1126

\bibitem[{{Mizuno} {et~al.}(1990){Mizuno}, {Fukui}, {Iwata}, {Nozawa}, \&
  {Takano}}]{mizuno90}
{Mizuno}, A., {Fukui}, Y., {Iwata}, T., {Nozawa}, S., \& {Takano}, T. 1990,
  \apj, 356, 184

\bibitem[{{M{\"u}ller} {et~al.}(2016){M{\"u}ller}, {Belloche}, {Xu}, {Lees},
  {Garrod}, {Walters}, {van Wijngaarden}, {Lewen}, {Schlemmer}, \&
  {Menten}}]{muller16}
{M{\"u}ller}, H.~S.~P., {Belloche}, A., {Xu}, L.-H., {et~al.} 2016, \aap, 587,
  A92

\bibitem[{{M{\"u}ller} \& {Christen}(2004)}]{muller04}
{M{\"u}ller}, H.~S.~P. \& {Christen}, D. 2004, Journal of Molecular
  Spectroscopy, 228, 298

\bibitem[{{M{\"u}ller} {et~al.}(2005){M{\"u}ller}, {Schl{\"o}der}, {Stutzki},
  \& {Winnewisser}}]{cdms2}
{M{\"u}ller}, H.~S.~P., {Schl{\"o}der}, F., {Stutzki}, J., \& {Winnewisser}, G.
  2005, Journal of Molecular Structure, 742, 215

\bibitem[{{M{\"u}ller} {et~al.}(2001){M{\"u}ller}, {Thorwirth}, {Roth}, \&
  {Winnewisser}}]{cdms1}
{M{\"u}ller}, H.~S.~P., {Thorwirth}, S., {Roth}, D.~A., \& {Winnewisser}, G.
  2001, \aap, 370, L49

\bibitem[{{Mundy} {et~al.}(1986){Mundy}, {Myers}, \& {Wilking}}]{mundy86}
{Mundy}, L.~G., {Myers}, S.~T., \& {Wilking}, B.~A. 1986, \apjl, 311, L75

\bibitem[{{Mundy} {et~al.}(1992){Mundy}, {Wootten}, {Wilking}, {Blake}, \&
  {Sargent}}]{mundy92}
{Mundy}, L.~G., {Wootten}, A., {Wilking}, B.~A., {Blake}, G.~A., \& {Sargent},
  A.~I. 1992, \apj, 385, 306

\bibitem[{{Mundy} {et~al.}(1990){Mundy}, {Wootten}, \& {Wilking}}]{mundy90}
{Mundy}, L.~G., {Wootten}, H.~A., \& {Wilking}, B.~A. 1990, \apj, 352, 159

\bibitem[{{Murillo} {et~al.}(2015){Murillo}, {Bruderer}, {van Dishoeck},
  {Walsh}, {Harsono}, {Lai}, \& {Fuchs}}]{murillo15}
{Murillo}, N.~M., {Bruderer}, S., {van Dishoeck}, E.~F., {et~al.} 2015, \aap,
  579, A114

\bibitem[{{Narayanan} {et~al.}(1998){Narayanan}, {Walker}, \&
  {Buckley}}]{narayanan98}
{Narayanan}, G., {Walker}, C.~K., \& {Buckley}, H.~D. 1998, \apj, 496, 292

\bibitem[{{Nisini} {et~al.}(2002){Nisini}, {Giannini}, \&
  {Lorenzetti}}]{nisini02}
{Nisini}, B., {Giannini}, T., \& {Lorenzetti}, D. 2002, \apj, 574, 246

\bibitem[{{Nutter} {et~al.}(2006){Nutter}, {Ward-Thompson}, \&
  {Andr{\'e}}}]{nutter06}
{Nutter}, D., {Ward-Thompson}, D., \& {Andr{\'e}}, P. 2006, \mnras, 368, 1833

\bibitem[{{{\"O}berg} {et~al.}(2010){{\"O}berg}, {Bottinelli}, {J{\o}rgensen},
  \& {van Dishoeck}}]{oberg10}
{{\"O}berg}, K.~I., {Bottinelli}, S., {J{\o}rgensen}, J.~K., \& {van Dishoeck},
  E.~F. 2010, \apj, 716, 825

\bibitem[{{Ossenkopf} \& {Henning}(1994)}]{ossenkopf94}
{Ossenkopf}, V. \& {Henning}, T. 1994, \aap, 291, 943

\bibitem[{{Oya} {et~al.}(2016){Oya}, {Sakai}, {L{\'o}pez-Sepulcre}, {Watanabe},
  {Ceccarelli}, {Lefloch}, {Favre}, \& {Yamamoto}}]{oya16}
{Oya}, Y., {Sakai}, N., {L{\'o}pez-Sepulcre}, A., {et~al.} 2016, \apj, 824, 88

\bibitem[{{Oya} {et~al.}(2014){Oya}, {Sakai}, {Sakai}, {Watanabe}, {Hirota},
  {Lindberg}, {Bisschop}, {J{\o}rgensen}, {van Dishoeck}, \&
  {Yamamoto}}]{oya14}
{Oya}, Y., {Sakai}, N., {Sakai}, T., {et~al.} 2014, \apj, 795, 152

\bibitem[{{Padgett} {et~al.}(2008){Padgett}, {Rebull}, {Stapelfeldt},
  {Chapman}, {Lai}, {Mundy}, {Evans}, {Brooke}, {Cieza}, {Spiesman},
  {Noriega-Crespo}, {McCabe}, {Allen}, {Blake}, {Harvey}, {Huard},
  {J{\o}rgensen}, {Koerner}, {Myers}, {Sargent}, {Teuben}, {van Dishoeck},
  {Wahhaj}, \& {Young}}]{padgett08}
{Padgett}, D.~L., {Rebull}, L.~M., {Stapelfeldt}, K.~R., {et~al.} 2008, \apj,
  672, 1013

\bibitem[{{Parise} {et~al.}(2004){Parise}, {Castets}, {Herbst}, {Caux},
  {Ceccarelli}, {Mukhopadhyay}, \& {Tielens}}]{parise04}
{Parise}, B., {Castets}, A., {Herbst}, E., {et~al.} 2004, \aap, 416, 159

\bibitem[{{Parise} {et~al.}(2006){Parise}, {Ceccarelli}, {Tielens}, {Castets},
  {Caux}, {Lefloch}, \& {Maret}}]{parise06deuterium}
{Parise}, B., {Ceccarelli}, C., {Tielens}, A.~G.~G.~M., {et~al.} 2006, \aap,
  453, 949

\bibitem[{{Parise} {et~al.}(2002){Parise}, {Ceccarelli}, {Tielens}, {Herbst},
  {Lefloch}, {Caux}, {Castets}, {Mukhopadhyay}, {Pagani}, \&
  {Loinard}}]{parise02}
{Parise}, B., {Ceccarelli}, C., {Tielens}, A.~G.~G.~M., {et~al.} 2002, \aap,
  393, L49

\bibitem[{{Parise} {et~al.}(2012){Parise}, {Du}, {Liu}, {Belloche},
  {Wiesemeyer}, {G{\"u}sten}, {Menten}, {H{\"u}bers}, \& {Klein}}]{parise12}
{Parise}, B., {Du}, F., {Liu}, F.-C., {et~al.} 2012, \aap, 542, L5

\bibitem[{{Pech} {et~al.}(2010){Pech}, {Loinard}, {Chandler},
  {Rodr{\'{\i}}guez}, {D'Alessio}, {Brogan}, {Wilner}, \& {Ho}}]{pech10}
{Pech}, G., {Loinard}, L., {Chandler}, C.~J., {et~al.} 2010, \apj, 712, 1403

\bibitem[{{Persson} {et~al.}(2013){Persson}, {J{\o}rgensen}, \& {van
  Dishoeck}}]{persson13}
{Persson}, M.~V., {J{\o}rgensen}, J.~K., \& {van Dishoeck}, E.~F. 2013, \aap,
  549, L3

\bibitem[{{Persson} {et~al.}(2014){Persson}, {J{\o}rgensen}, {van Dishoeck}, \&
  {Harsono}}]{persson14}
{Persson}, M.~V., {J{\o}rgensen}, J.~K., {van Dishoeck}, E.~F., \& {Harsono},
  D. 2014, \aap, 563, A74

\bibitem[{{Pickett} {et~al.}(1998){Pickett}, {Poynter}, {Cohen}, {Delitsky},
  {Pearson}, \& {Muller}}]{jpl}
{Pickett}, H.~M., {Poynter}, I.~R.~L., {Cohen}, E.~A., {et~al.} 1998, \jqsrt,
  60, 883

\bibitem[{{Pineda} {et~al.}(2012){Pineda}, {Maury}, {Fuller}, {Testi},
  {Garc{\'{\i}}a-Appadoo}, {Peck}, {Villard}, {Corder}, {van Kempen}, {Turner},
  {Tachihara}, \& {Dent}}]{pineda12}
{Pineda}, J.~E., {Maury}, A.~J., {Fuller}, G.~A., {et~al.} 2012, \aap, 544, L7

\bibitem[{{Podio} {et~al.}(2015){Podio}, {Codella}, {Gueth}, {Cabrit},
  {Bachiller}, {Gusdorf}, {Lee}, {Lefloch}, {Leurini}, {Nisini}, \&
  {Tafalla}}]{podio15}
{Podio}, L., {Codella}, C., {Gueth}, F., {et~al.} 2015, \aap, 581, A85

\bibitem[{{Rabli} \& {Flower}(2010)}]{rabli10}
{Rabli}, D. \& {Flower}, D.~R. 2010, \mnras, 406, 95

\bibitem[{{Remijan} {et~al.}(2002){Remijan}, {Snyder}, {Liu}, {Mehringer}, \&
  {Kuan}}]{remijan02}
{Remijan}, A., {Snyder}, L.~E., {Liu}, S.-Y., {Mehringer}, D., \& {Kuan}, Y.-J.
  2002, \apj, 576, 264

\bibitem[{{Remijan} {et~al.}(2007){Remijan}, {Markwick-Kemper}, \& {ALMA
  Working Group on Spectral Line Frequencies}}]{remijan07}
{Remijan}, A.~J., {Markwick-Kemper}, A., \& {ALMA Working Group on Spectral
  Line Frequencies}. 2007, in Bulletin of the American Astronomical Society,
  Vol.~39, American Astronomical Society Meeting Abstracts, 963

\bibitem[{{Requena-Torres} {et~al.}(2008){Requena-Torres},
  {Mart{\'{\i}}n-Pintado}, {Mart{\'{\i}}n}, \& {Morris}}]{requenatorres08}
{Requena-Torres}, M.~A., {Mart{\'{\i}}n-Pintado}, J., {Mart{\'{\i}}n}, S., \&
  {Morris}, M.~R. 2008, \apj, 672, 352

\bibitem[{{Richard} {et~al.}(2013){Richard}, {Margul{\`e}s}, {Caux}, {Kahane},
  {Ceccarelli}, {Guillemin}, {Motiyenko}, {Vastel}, \& {Groner}}]{richard13}
{Richard}, C., {Margul{\`e}s}, L., {Caux}, E., {et~al.} 2013, \aap, 552, A117

\bibitem[{{Rivera} {et~al.}(2015){Rivera}, {Loinard}, {Dzib}, {Ortiz-Le{\'o}n},
  {Rodr{\'{\i}}guez}, \& {Torres}}]{rivera15}
{Rivera}, J.~L., {Loinard}, L., {Dzib}, S.~A., {et~al.} 2015, \apj, 807, 119

\bibitem[{{Roberts} {et~al.}(2003){Roberts}, {Herbst}, \& {Millar}}]{roberts03}
{Roberts}, H., {Herbst}, E., \& {Millar}, T.~J. 2003, \apjl, 591, L41

\bibitem[{{Roberts} \& {Millar}(2000)}]{roberts00}
{Roberts}, H. \& {Millar}, T.~J. 2000, \aap, 361, 388

\bibitem[{{Roueff} {et~al.}(2005){Roueff}, {Lis}, {van der Tak}, {Gerin}, \&
  {Goldsmith}}]{roueff05}
{Roueff}, E., {Lis}, D.~C., {van der Tak}, F.~F.~S., {Gerin}, M., \&
  {Goldsmith}, P.~F. 2005, \aap, 438, 585

\bibitem[{{Sakai} {et~al.}(2014){Sakai}, {Sakai}, {Hirota}, {Watanabe},
  {Ceccarelli}, {Kahane}, {Bottinelli}, {Caux}, {Demyk}, {Vastel}, {Coutens},
  {Taquet}, {Ohashi}, {Takakuwa}, {Yen}, {Aikawa}, \& {Yamamoto}}]{sakai14}
{Sakai}, N., {Sakai}, T., {Hirota}, T., {et~al.} 2014, \nat, 507, 78

\bibitem[{{Sakai} {et~al.}(2006){Sakai}, {Sakai}, \& {Yamamoto}}]{sakai06}
{Sakai}, N., {Sakai}, T., \& {Yamamoto}, S. 2006, \pasj, 58, L15

\bibitem[{{Sch\"{o}ier} {et~al.}(2002){Sch\"{o}ier}, {J{\o}rgensen}, {van
  Dishoeck}, \& {Blake}}]{schoeier02}
{Sch\"{o}ier}, F.~L., {J{\o}rgensen}, J.~K., {van Dishoeck}, E.~F., \& {Blake},
  G.~A. 2002, \aap, 390, 1001

\bibitem[{{Sch\"{o}ier} {et~al.}(2004){Sch\"{o}ier}, {J{\o}rgensen}, {van
  Dishoeck}, \& {Blake}}]{hotcorepaper}
{Sch\"{o}ier}, F.~L., {J{\o}rgensen}, J.~K., {van Dishoeck}, E.~F., \& {Blake},
  G.~A. 2004, \aap, 418, 185

\bibitem[{{Sch{\"o}ier} {et~al.}(2004){Sch{\"o}ier}, {J{\o}rgensen}, {van
  Dishoeck}, \& {Blake}}]{schoeier04}
{Sch{\"o}ier}, F.~L., {J{\o}rgensen}, J.~K., {van Dishoeck}, E.~F., \& {Blake},
  G.~A. 2004, \aap, 418, 185

\bibitem[{{Sch\"{o}ier} {et~al.}(2005){Sch\"{o}ier}, {van der Tak}, {van
  Dishoeck}, \& {Black}}]{lamda}
{Sch\"{o}ier}, F.~L., {van der Tak}, F.~F.~S., {van Dishoeck}, E.~F., \&
  {Black}, J.~H. 2005, \aap, 432, 369

\bibitem[{{Senent}(2004)}]{senent04}
{Senent}, M.~L. 2004, Journal of Physical Chemistry A, 108, 6286

\bibitem[{{Shiao} {et~al.}(2010){Shiao}, {Looney}, {Remijan}, {Snyder}, \&
  {Friedel}}]{shiao10}
{Shiao}, Y.-S.~J., {Looney}, L.~W., {Remijan}, A.~J., {Snyder}, L.~E., \&
  {Friedel}, D.~N. 2010, \apj, 716, 286

\bibitem[{{Shu} {et~al.}(1987){Shu}, {Adams}, \& {Lizano}}]{shu87}
{Shu}, F.~H., {Adams}, F.~C., \& {Lizano}, S. 1987, \araa, 25, 23

\bibitem[{{Smith} {et~al.}(2015){Smith}, {Pontoppidan}, {Young}, \&
  {Morris}}]{smith15}
{Smith}, R.~L., {Pontoppidan}, K.~M., {Young}, E.~D., \& {Morris}, M.~R. 2015,
  \apj, 813, 120

\bibitem[{{Stark} {et~al.}(2004){Stark}, {Sandell}, {Beck}, {Hogerheijde}, {van
  Dishoeck}, {van der Wal}, {van der Tak}, {Schaefer}, {Melnick}, {Ashby}, \&
  {de Lange}}]{stark04}
{Stark}, R., {Sandell}, G., {Beck}, S.~C., {et~al.} 2004, \apj, 608, 341

\bibitem[{{Stark} {et~al.}(1999){Stark}, {van der Tak}, \& {van
  Dishoeck}}]{stark99}
{Stark}, R., {van der Tak}, F.~F.~S., \& {van Dishoeck}, E.~F. 1999, \apjl,
  521, L67

\bibitem[{{Sugimura} {et~al.}(2011){Sugimura}, {Yamaguchi}, {Sakai}, {Umemoto},
  {Sakai}, {Takano}, {Aikawa}, {Hirano}, {Liu}, {Millar}, {Nomura}, {Su},
  {Takakuwa}, \& {Yamamoto}}]{sugimura11}
{Sugimura}, M., {Yamaguchi}, T., {Sakai}, T., {et~al.} 2011, \pasj, 63, 459

\bibitem[{{Tachihara} {et~al.}(2000){Tachihara}, {Mizuno}, \&
  {Fukui}}]{tachihara00}
{Tachihara}, K., {Mizuno}, A., \& {Fukui}, Y. 2000, \apj, 528, 817

\bibitem[{{Taquet} {et~al.}(2014){Taquet}, {Charnley}, \&
  {Sipil{\"a}}}]{taquet14}
{Taquet}, V., {Charnley}, S.~B., \& {Sipil{\"a}}, O. 2014, \apj, 791, 1

\bibitem[{{Taquet} {et~al.}(2015){Taquet}, {L{\'o}pez-Sepulcre}, {Ceccarelli},
  {Neri}, {Kahane}, \& {Charnley}}]{taquet15}
{Taquet}, V., {L{\'o}pez-Sepulcre}, A., {Ceccarelli}, C., {et~al.} 2015, \apj,
  804, 81

\bibitem[{{Taquet} {et~al.}(2013{\natexlab{a}}){Taquet}, {L{\'o}pez-Sepulcre},
  {Ceccarelli}, {Neri}, {Kahane}, {Coutens}, \& {Vastel}}]{taquet13}
{Taquet}, V., {L{\'o}pez-Sepulcre}, A., {Ceccarelli}, C., {et~al.}
  2013{\natexlab{a}}, \apjl, 768, L29

\bibitem[{{Taquet} {et~al.}(2013{\natexlab{b}}){Taquet}, {L{\'o}pez-Sepulcre},
  {Ceccarelli}, {Neri}, {Kahane}, {Coutens}, \& {Vastel}}]{taquet13obs}
{Taquet}, V., {L{\'o}pez-Sepulcre}, A., {Ceccarelli}, C., {et~al.}
  2013{\natexlab{b}}, \apjl, 768, L29

\bibitem[{{Terebey} {et~al.}(1984){Terebey}, {Shu}, \& {Cassen}}]{terebey84}
{Terebey}, S., {Shu}, F.~H., \& {Cassen}, P. 1984, \apj, 286, 529

\bibitem[{{van der Tak} {et~al.}(2007){van der Tak}, {Black}, {Sch{\"o}ier},
  {Jansen}, \& {van Dishoeck}}]{vandertak07radex}
{van der Tak}, F.~F.~S., {Black}, J.~H., {Sch{\"o}ier}, F.~L., {Jansen}, D.~J.,
  \& {van Dishoeck}, E.~F. 2007, \aap, 468, 627

\bibitem[{{van der Tak} {et~al.}(2000){van der Tak}, {van Dishoeck}, {Evans},
  \& {Blake}}]{vandertak00}
{van der Tak}, F.~F.~S., {van Dishoeck}, E.~F., {Evans}, N.~J., \& {Blake},
  G.~A. 2000, \apj, 537, 283

\bibitem[{{van Dishoeck} {et~al.}(1993){van Dishoeck}, {Blake}, {Draine}, \&
  {Lunine}}]{vandishoeckppiii}
{van Dishoeck}, E.~F., {Blake}, G.~A., {Draine}, B.~T., \& {Lunine}, J.~I.
  1993, in Protostars and Planets III, ed. E.~H. {Levy} \& J.~I. {Lunine},
  163--241

\bibitem[{{van Dishoeck} {et~al.}(1995){van Dishoeck}, {Blake}, {Jansen}, \&
  {Groesbeck}}]{vandishoeck95}
{van Dishoeck}, E.~F., {Blake}, G.~A., {Jansen}, D.~J., \& {Groesbeck}, T.~D.
  1995, \apj, 447, 760

\bibitem[{{Vastel} {et~al.}(2010){Vastel}, {Ceccarelli}, {Caux}, {Coutens},
  {Cernicharo}, {Bottinelli}, {Demyk}, {Faure}, {Wiesenfeld}, {Scribano},
  {Bacmann}, {Hily-Blant}, {Maret}, {Walters}, {Bergin}, {Blake}, {Castets},
  {Crimier}, {Dominik}, {Encrenaz}, {G{\'e}rin}, {Hennebelle}, {Kahane},
  {Klotz}, {Melnick}, {Pagani}, {Parise}, {Schilke}, {Wakelam}, {Baudry},
  {Bell}, {Benedettini}, {Boogert}, {Cabrit}, {Caselli}, {Codella}, {Comito},
  {Falgarone}, {Fuente}, {Goldsmith}, {Helmich}, {Henning}, {Herbst}, {Jacq},
  {Kama}, {Langer}, {Lefloch}, {Lis}, {Lord}, {Lorenzani}, {Neufeld}, {Nisini},
  {Pacheco}, {Pearson}, {Phillips}, {Salez}, {Saraceno}, {Schuster}, {Tielens},
  {van der Tak}, {van der Wiel}, {Viti}, {Wyrowski}, {Yorke}, {Cais}, {Krieg},
  {Olberg}, \& {Ravera}}]{vastel10}
{Vastel}, C., {Ceccarelli}, C., {Caux}, E., {et~al.} 2010, \aap, 521, L31

\bibitem[{{Vastel} {et~al.}(2014){Vastel}, {Ceccarelli}, {Lefloch}, \&
  {Bachiller}}]{vastel14}
{Vastel}, C., {Ceccarelli}, C., {Lefloch}, B., \& {Bachiller}, R. 2014, \apjl,
  795, L2

\bibitem[{{Vastel} {et~al.}(2004){Vastel}, {Phillips}, \& {Yoshida}}]{vastel04}
{Vastel}, C., {Phillips}, T.~G., \& {Yoshida}, H. 2004, \apjl, 606, L127

\bibitem[{{Walker} {et~al.}(1990){Walker}, {Adams}, \& {Lada}}]{walker90}
{Walker}, C.~K., {Adams}, F.~C., \& {Lada}, C.~J. 1990, \apj, 349, 515

\bibitem[{{Walker} {et~al.}(1986){Walker}, {Lada}, {Young}, {Maloney}, \&
  {Wilking}}]{walker86}
{Walker}, C.~K., {Lada}, C.~J., {Young}, E.~T., {Maloney}, P.~R., \& {Wilking},
  B.~A. 1986, \apjl, 309, L47

\bibitem[{{Walker} {et~al.}(1988){Walker}, {Lada}, {Young}, \&
  {Margulis}}]{walker88}
{Walker}, C.~K., {Lada}, C.~J., {Young}, E.~T., \& {Margulis}, M. 1988, \apj,
  332, 335

\bibitem[{{Whittet} \& {van Breda}(1975)}]{whittet75}
{Whittet}, D.~C.~B. \& {van Breda}, I.~G. 1975, \apss, 38, L3

\bibitem[{{Widicus Weaver} {et~al.}(2005){Widicus Weaver}, {Butler}, {Drouin},
  {Petkie}, {Dyl}, {De Lucia}, \& {Blake}}]{widicus05}
{Widicus Weaver}, S.~L., {Butler}, R.~A.~H., {Drouin}, B.~J., {et~al.} 2005,
  \apjs, 158, 188

\bibitem[{{Wilson} \& {Rood}(1994)}]{wilson94}
{Wilson}, T.~L. \& {Rood}, R. 1994, \araa, 32, 191

\bibitem[{{Wootten}(1989)}]{wootten89}
{Wootten}, A. 1989, \apj, 337, 858

\bibitem[{{Wootten} \& {Loren}(1987)}]{wootten87}
{Wootten}, A. \& {Loren}, R.~B. 1987, \apj, 317, 220

\bibitem[{{Yeh} {et~al.}(2008){Yeh}, {Hirano}, {Bourke}, {Ho}, {Lee}, {Ohashi},
  \& {Takakuwa}}]{yeh08}
{Yeh}, S.~C.~C., {Hirano}, N., {Bourke}, T.~L., {et~al.} 2008, \apj, 675, 454

\bibitem[{{Young} {et~al.}(2006){Young}, {Enoch}, {Evans}, {Glenn}, {Sargent},
  {Huard}, {Aguirre}, {Golwala}, {Haig}, {Harvey}, {Laurent}, {Mauskopf}, \&
  {Sayers}}]{young06}
{Young}, K.~E., {Enoch}, M.~L., {Evans}, II, N.~J., {et~al.} 2006, \apj, 644,
  326

\bibitem[{{Zapata} {et~al.}(2013){Zapata}, {Loinard}, {Rodr{\'{\i}}guez},
  {Hern{\'a}ndez-Hern{\'a}ndez}, {Takahashi}, {Trejo}, \& {Parise}}]{zapata13}
{Zapata}, L.~A., {Loinard}, L., {Rodr{\'{\i}}guez}, L.~F., {et~al.} 2013,
  \apjl, 764, L14

\bibitem[{{Zhou}(1995)}]{zhou95}
{Zhou}, S. 1995, \apj, 442, 685

\end{thebibliography}

\clearpage

\appendix

\noindent\begin{minipage}{\textwidth}
    \centering
\section{Fitted lines of glycolaldehyde and ethylene glycol in Band~3 data}
\captionof{table}{Parameters for identified glycolaldehyde and ethylene glycol transitions measured in the Band~3 (3~mm) dataset.}\label{band3measurements}
\begin{tabular}{llllllll} \hline\hline
Transition & Freq & $E_{\rm up}$  & $A_{ul}$ & $g_u$ & $I$ & $V_{\rm lsr}$ & $\delta v$ \\
& [GHz] & [K]  & [s$^{-1}]$ & & [Jy~beam$^{-1}$] & [km~s$^{-1}$] & [km~s$^{-1}$] \\ \hline
\multicolumn{8}{c}{\it Glycolaldehyde} \\[0.5ex]
$20_{4,16}-20_{3,17}                  $      &  89.6164 &  130.4 &  2.00$\times 10^{-5}$ &   41 &   0.062 &   2.6 &   1.5 \\
$10_{4,7}-10_{3,8}                    $      &  89.6441 &   40.3 &  1.59$\times 10^{-5}$ &   21 &   0.057 &   2.5 &   1.6 \\
$8_{4,5}-8_{3,6}                      $      &  89.7022 &  404.1 &  1.47$\times 10^{-5}$ &   17 &   0.013 &   2.8 &   1.8 \\
$9_{0,9}-8_{1,8}                      $      &  92.8515 &  304.0 &  2.67$\times 10^{-5}$ &   19 &   0.022 &   2.6 &   1.7 \\
$12_{4,9}-12_{3,10}                   $      &  92.8539 &   53.2 &  1.83$\times 10^{-5}$ &   25 &   0.061 &   2.5 &   1.0 \\
$25_{6,19}-25_{5,20}                  $      &  92.8821 &  485.9 &  2.71$\times 10^{-5}$ &   51 &   0.023 &   2.5 &   1.7 \\
$10_{1,9}-9_{2,8}                     $      & 102.5062 &  406.1 &  2.12$\times 10^{-5}$ &   21 &   0.013 &   2.8 &   1.3 \\
$7_{2,6}-6_{1,5}                      $      & 102.5498 &   17.8 &  1.91$\times 10^{-5}$ &   15 &   0.055 &   2.7 &   1.2 \\
$14_{3,12}-14_{2,13}                  $      & 102.5729 &   63.6 &  1.82$\times 10^{-5}$ &   29 &   0.064 &   2.6 &   1.3 \\
$18_{3,15}-18_{2,16}                  $      & 102.6144 &  104.1 &  2.29$\times 10^{-5}$ &   37 &   0.073 &   2.6 &   1.2 \\
$7_{2,6}-6_{1,5}                      $      & 102.6690 &  392.0 &  1.91$\times 10^{-5}$ &   15 &   0.011 &   2.8 &   1.2 \\
$10_{0,10}-9_{1,9}$                          & 103.3913 &   28.3 &  2.78$\times 10^{-5}$ &   21 &   0.086 &   2.7 &   1.2 \\ \hline
\multicolumn{8}{c}{\it $aGg'$ ethylene glycol} \\[0.5ex]
$8_{3,5}\,\varv=1-7_{3,4}\,\varv=0      $      &  89.7082 &   22.5 &  1.45$\times 10^{-5}$ &  153 &   0.037 &   2.7 &   1.0 \\
$36_{7,29}\,\varv=1-36_{6,30}\,\varv=1  $      &  92.8604 &  357.3 &  3.53$\times 10^{-6}$ &  657 &   0.010 &   2.6 &   1.7 \\
$9_{1,9}\,\varv=1-8_{1,8}\,\varv=0      $      &  92.9759 &   21.6 &  1.80$\times 10^{-5}$ &  171 &   0.049 &   2.7 &   1.1 \\
$41_{8,33}\,\varv=0-41_{7,34}\,\varv=0  $      & 102.5189 &  461.2 &  4.51$\times 10^{-6}$ &  747 &   0.013 &   2.9 &   1.1 \\
$9_{2,7}\,\varv=1-8_{2,6}\,\varv=0      $      & 102.5394 &   25.0 &  2.38$\times 10^{-5}$ &  133 &   0.048 &   2.6 &   0.90 \\
$10_{0,10}\,\varv=1-9_{0,9}\,\varv=0   $      & 102.6898 &   26.1 &  2.63$\times 10^{-5}$ &  189 &   0.068 &   2.4 &   1.5 \\ \hline
\multicolumn{8}{c}{\it $gGg'$ ethylene glycol} \\[0.5ex]
$9_{6,4}\,\varv=1-8_{6,3}\,\varv=0       $      &  92.8017 &   39.5 &  3.95$\times 10^{-6}$ &  171 &   0.018 &   2.5 &   1.3 \\ 
$9_{6,3}\,\varv=1-8_{6,2}\,\varv=0$$^{a} $      &  92.8017 & $\ldots$ & $\ldots$ &  133 & $\ldots$ & $\ldots$ & $\ldots$ \\
$9_{5,5}\,\varv=1-8_{5,4}\,\varv=0$$^{b} $      &  92.9120 &   34.2 &  4.94$\times 10^{-6}$ &  171 &  $\ldots$ & $\ldots$ & $\ldots$ \\
$9_{5,4}\,\varv=1-8_{5,3}\,\varv=0       $      &  92.9144 &   34.2 &  4.94$\times 10^{-6}$ &  133 &   0.011 &   2.7 &   0.87 \\
$22_{6,17}\,\varv=1-22_{5,17}\,\varv=0   $      &  92.9184 &  141.6 &  4.12$\times 10^{-6}$ &  315 &   0.011 &   2.7 &   0.93 \\
$10_{4,7}\,\varv=1-9_{4,6}\,\varv=0      $      & 103.3723 &   34.7 &  8.30$\times 10^{-6}$ &  147 &   0.014 &   2.4 &   1.3 \\ \hline
\end{tabular}

$^a$Blended with above transition. $^{b}$Complete overlap with prominent acetaldehyde (CH$_3$CHO) line; excluded from fit.
\end{minipage}

\clearpage

\noindent\begin{minipage}{\textwidth}
    \centering
\section{Synthetic spectra fits for glycolaldehyde, ethylene glycol and acetic acid}\label{Band36spectra}

\resizebox{!}{23cm}{\rotatebox{90}{\includegraphics{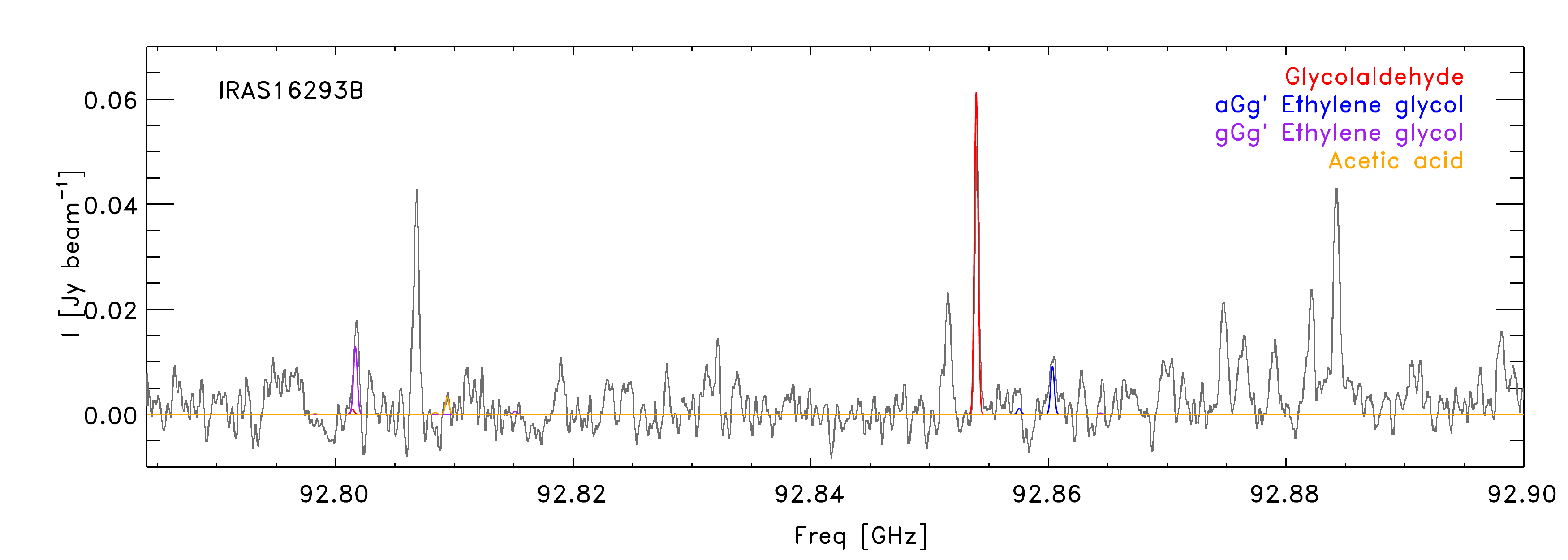}}}
\resizebox{!}{23cm}{\rotatebox{90}{\includegraphics{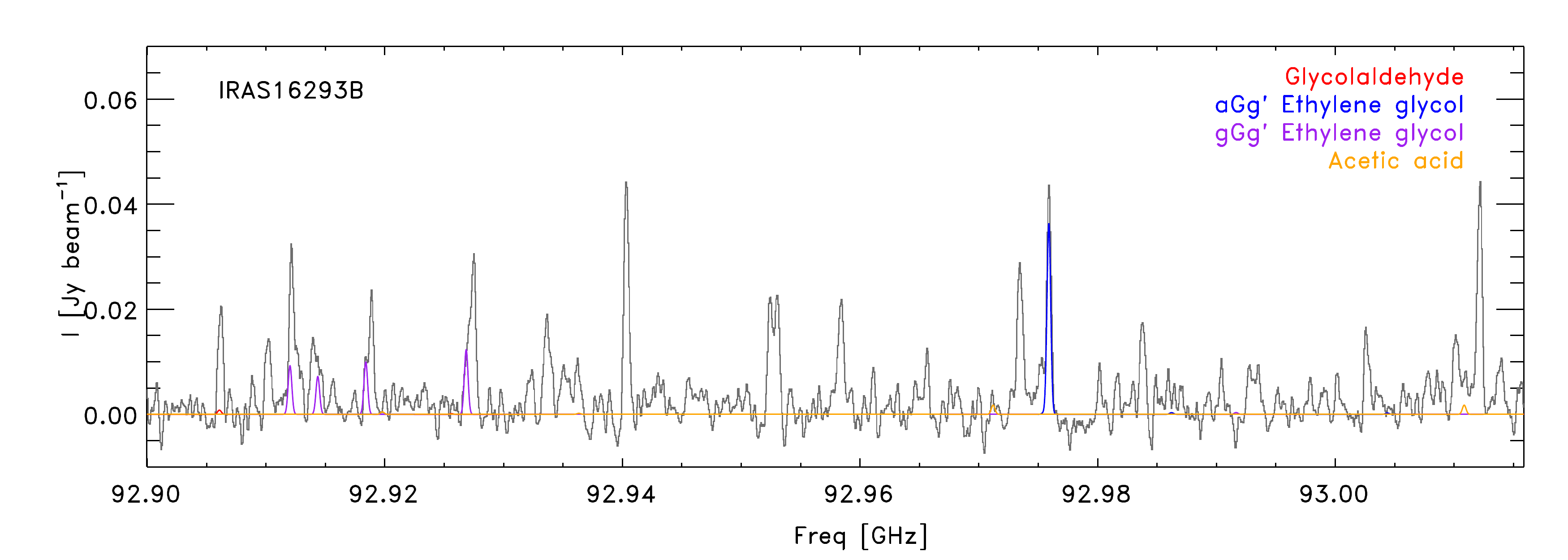}}}
\captionof{figure}{Fits to the glycolaldehyde, ethylene glycol and acetic acid lines at 3~mm (Band~3).}\label{gcaeg_band3}
\end{minipage}
\clearpage
\noindent\begin{minipage}{\textwidth}
\resizebox{!}{23cm}{\rotatebox{90}{\includegraphics{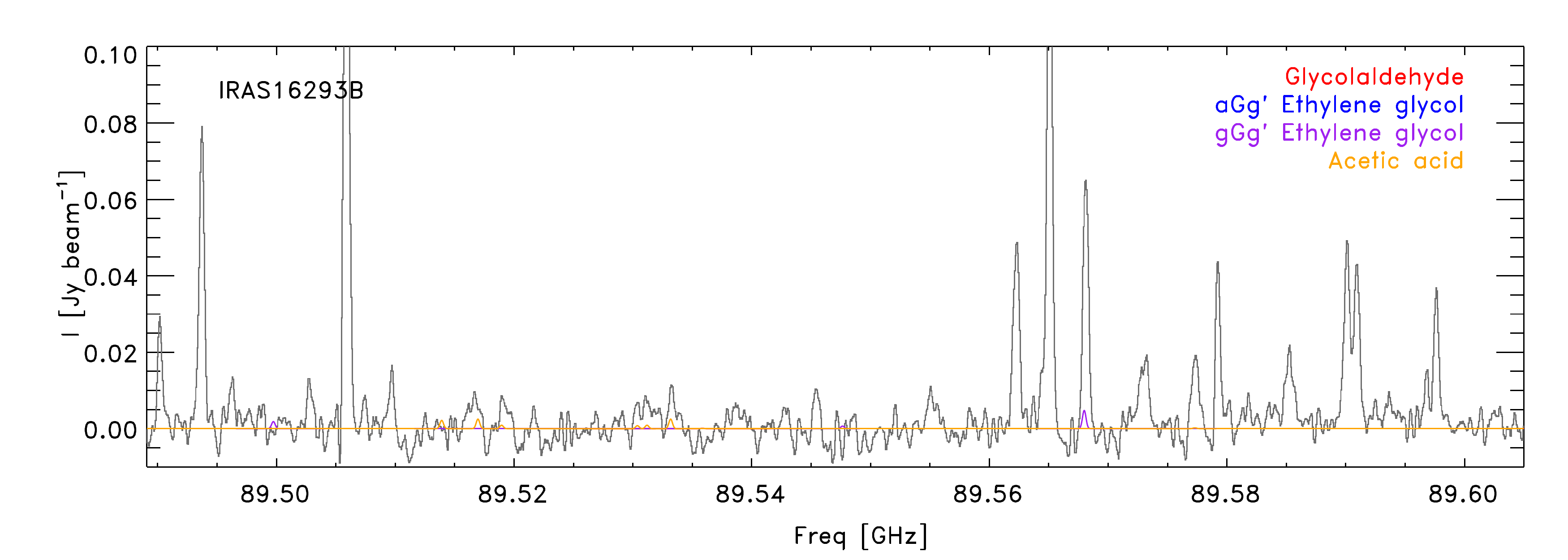}}}
\resizebox{!}{23cm}{\rotatebox{90}{\includegraphics{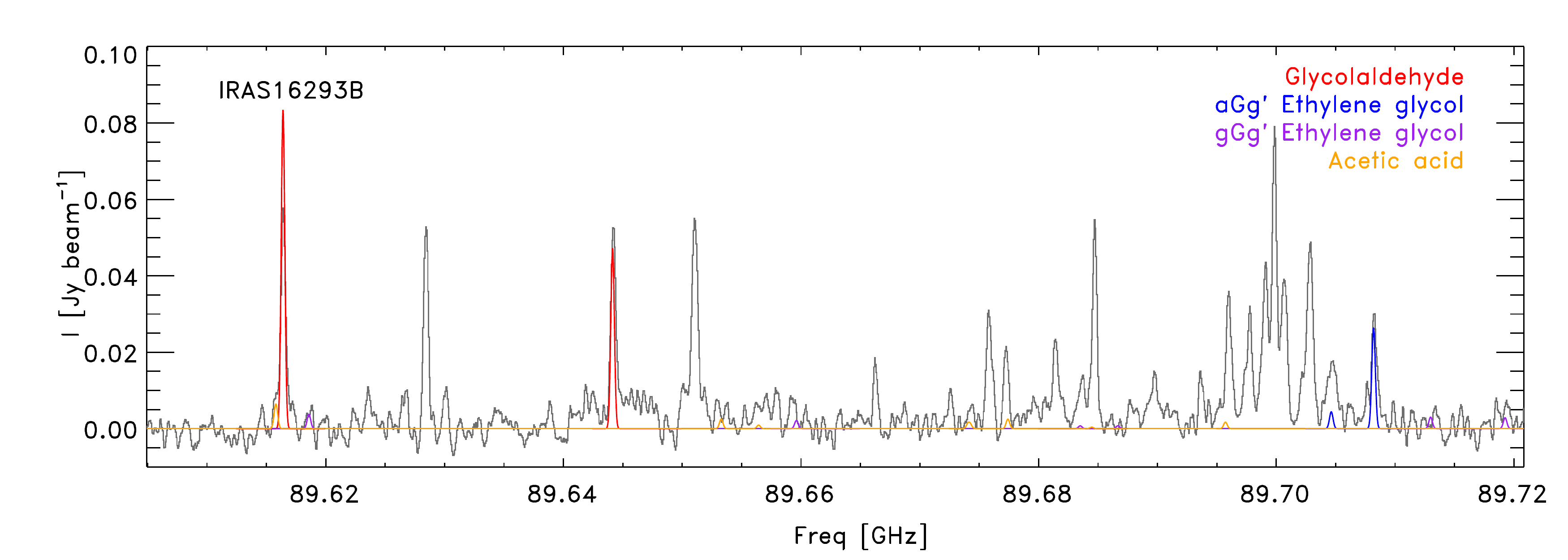}}}
\captionof{figure}{Continuation of Fig.~\ref{gcaeg_band3}.}\label{gcaeg_band3ext}
\end{minipage}
\clearpage
\noindent\begin{minipage}{\textwidth}
    \centering
\resizebox{!}{23cm}{\rotatebox{90}{\includegraphics{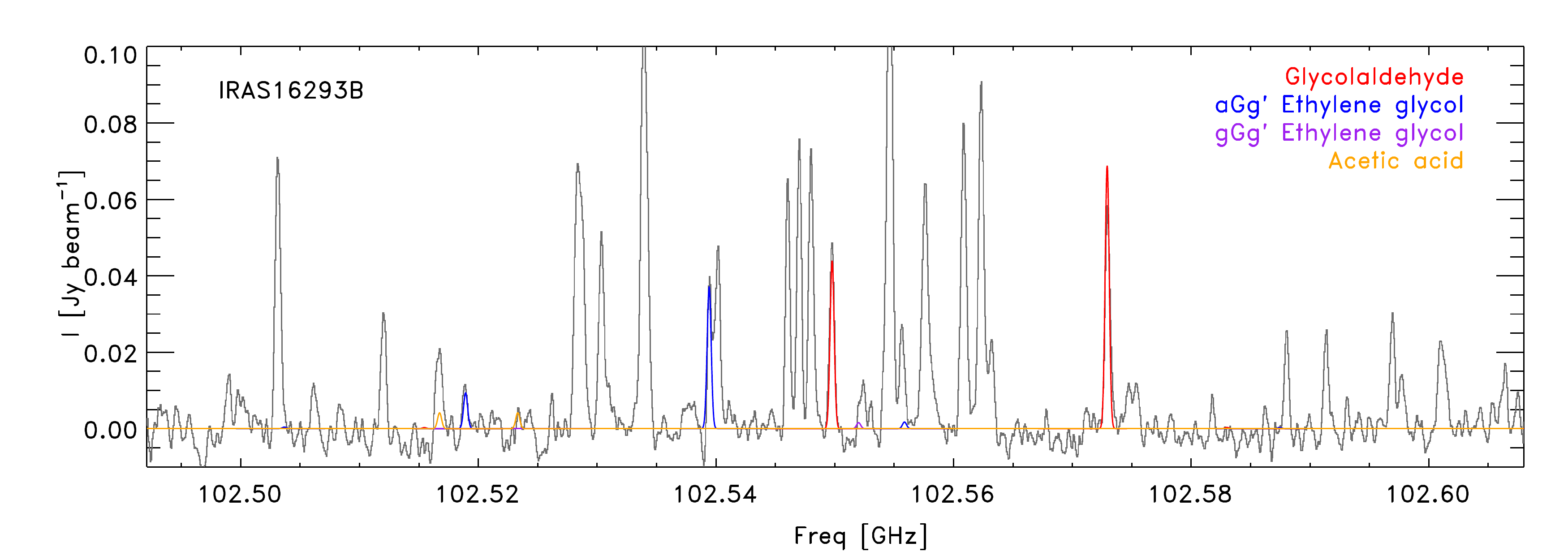}}}
\resizebox{!}{23cm}{\rotatebox{90}{\includegraphics{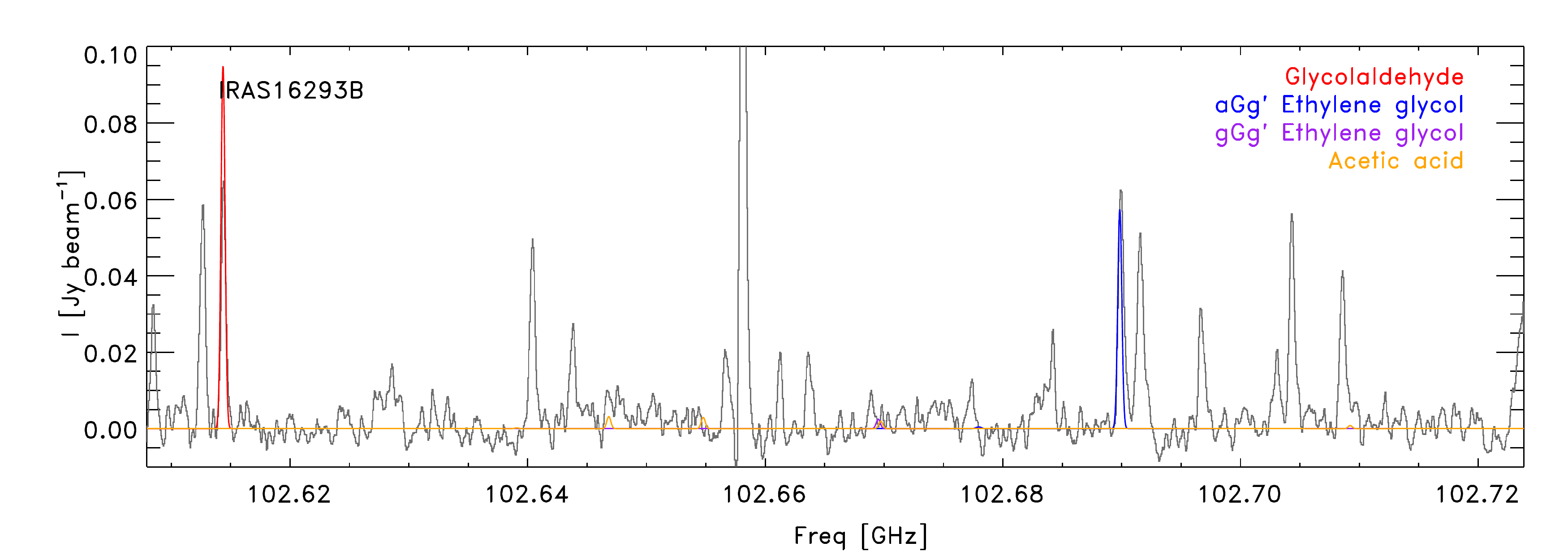}}}
\captionof{figure}{Continuation of Fig.~\ref{gcaeg_band3}.}\label{gcaeg_band3_2ext}
\end{minipage}
\clearpage
\noindent\begin{minipage}{\textwidth}
    \centering
\resizebox{!}{23cm}{\rotatebox{90}{\includegraphics{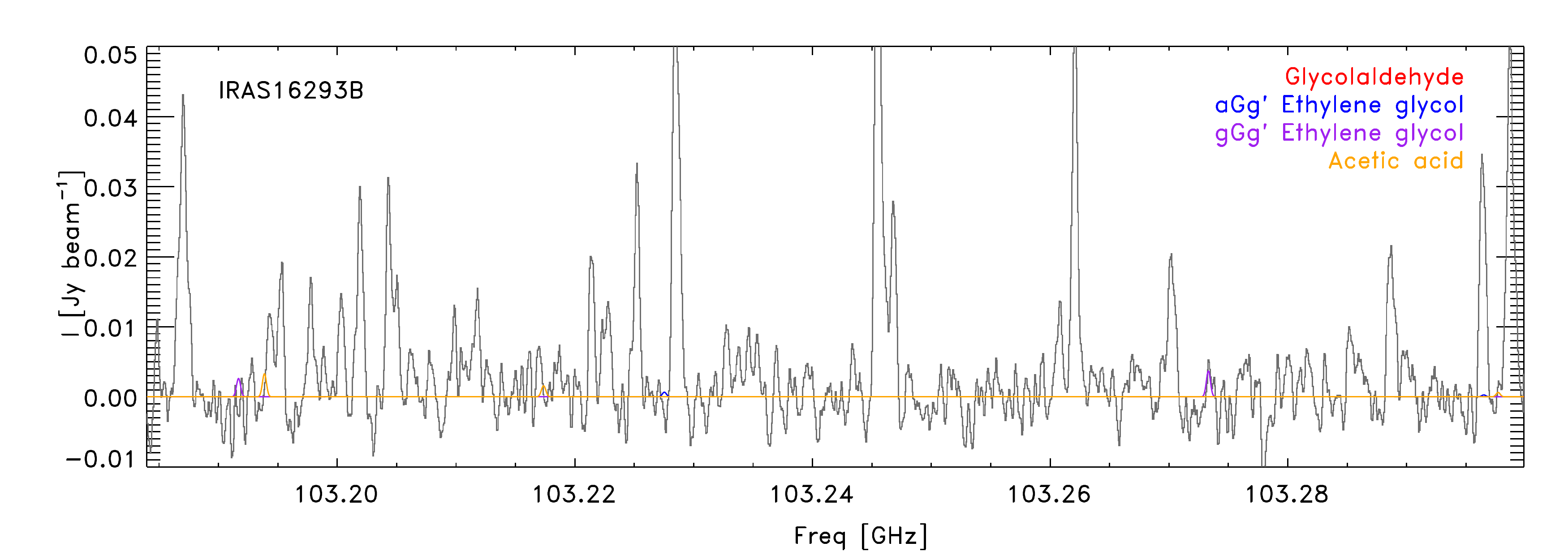}}}
\resizebox{!}{23cm}{\rotatebox{90}{\includegraphics{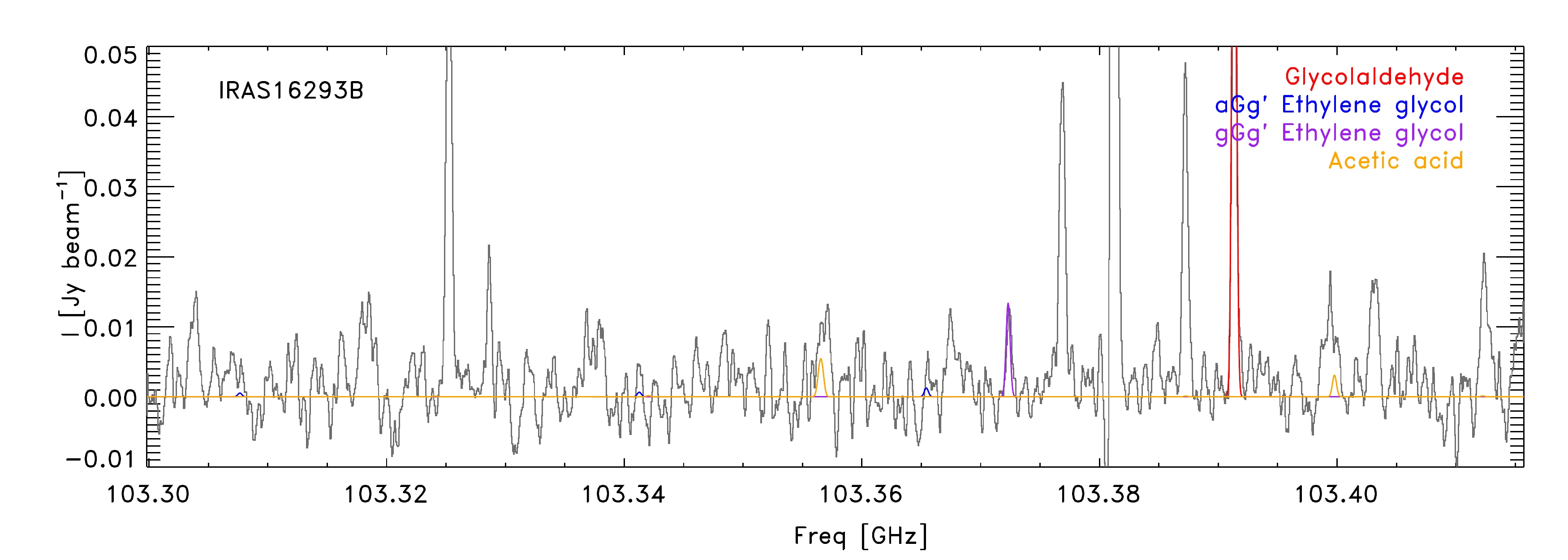}}}
\captionof{figure}{Continuation of Fig.~\ref{gcaeg_band3}.}\label{gcaeg_band3_2}
\end{minipage}
\clearpage
\noindent\begin{minipage}{\textwidth}
    \centering
\resizebox{!}{23cm}{\rotatebox{90}{\includegraphics{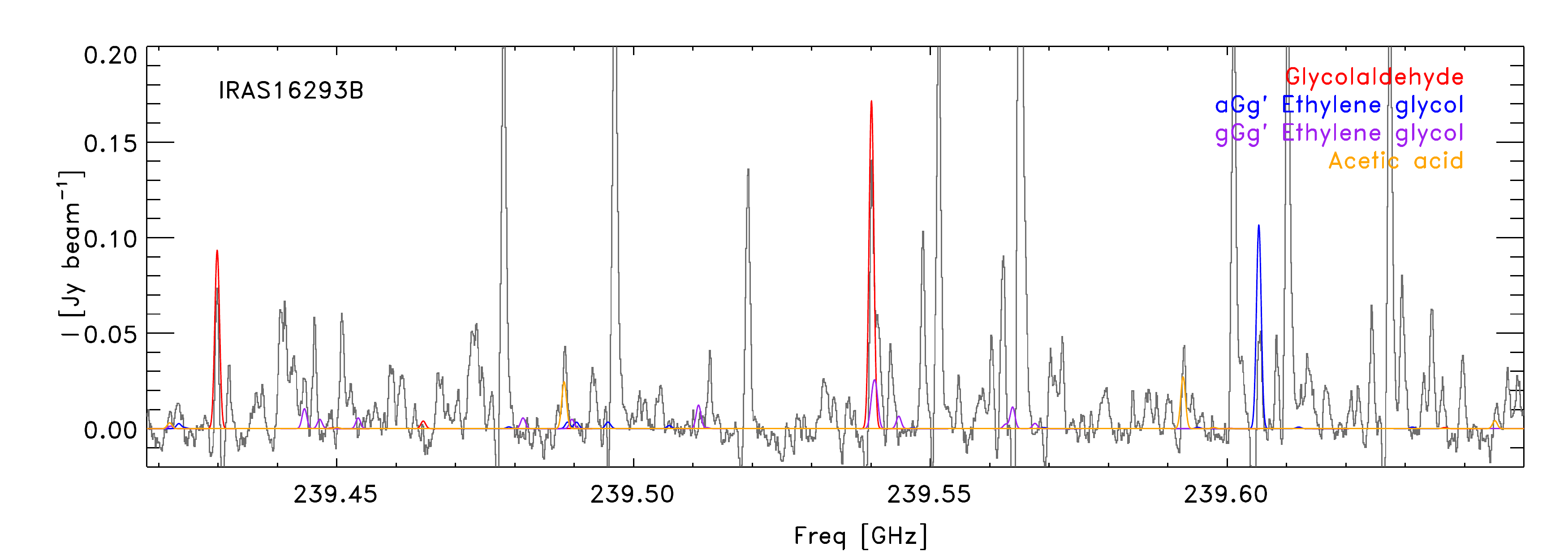}}}
\resizebox{!}{23cm}{\rotatebox{90}{\includegraphics{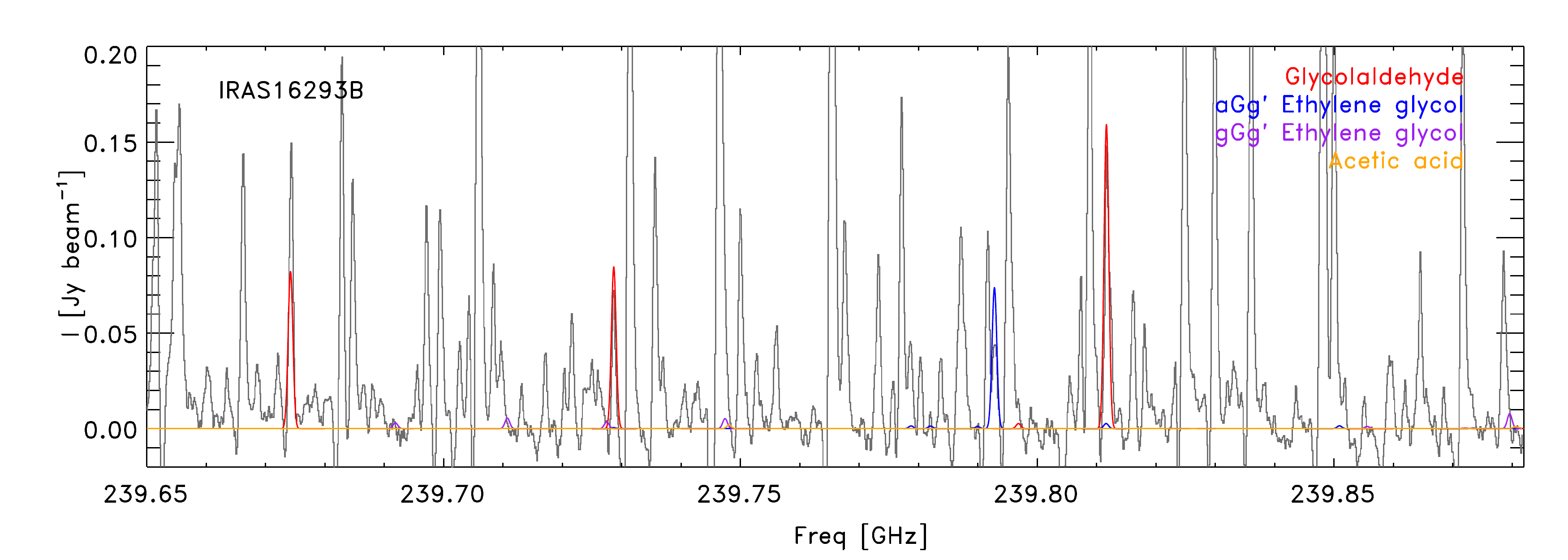}}}
\captionof{figure}{Fits to the glycolaldehyde, ethylene glycol and acetic acid lines at 1.3~mm (Band~6).}\label{gcaeg_band6-1}
\end{minipage}
\clearpage
\noindent\begin{minipage}{\textwidth}
    \centering
\resizebox{!}{23cm}{\rotatebox{90}{\includegraphics{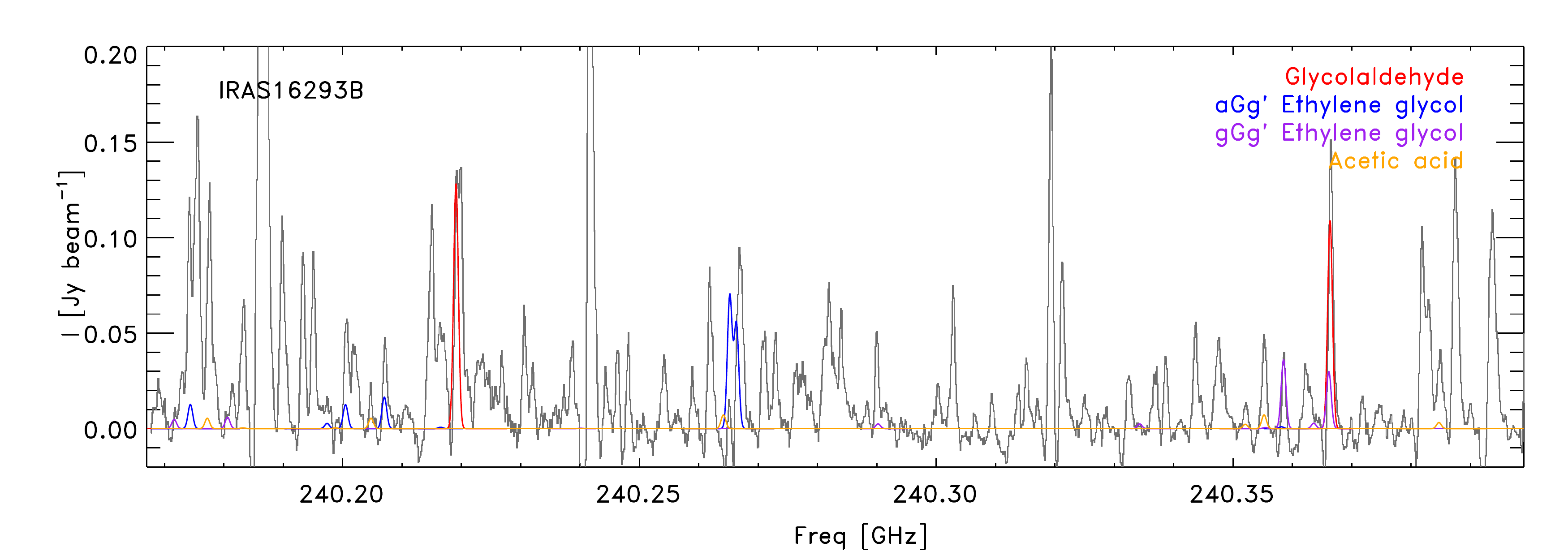}}}
\resizebox{!}{23cm}{\rotatebox{90}{\includegraphics{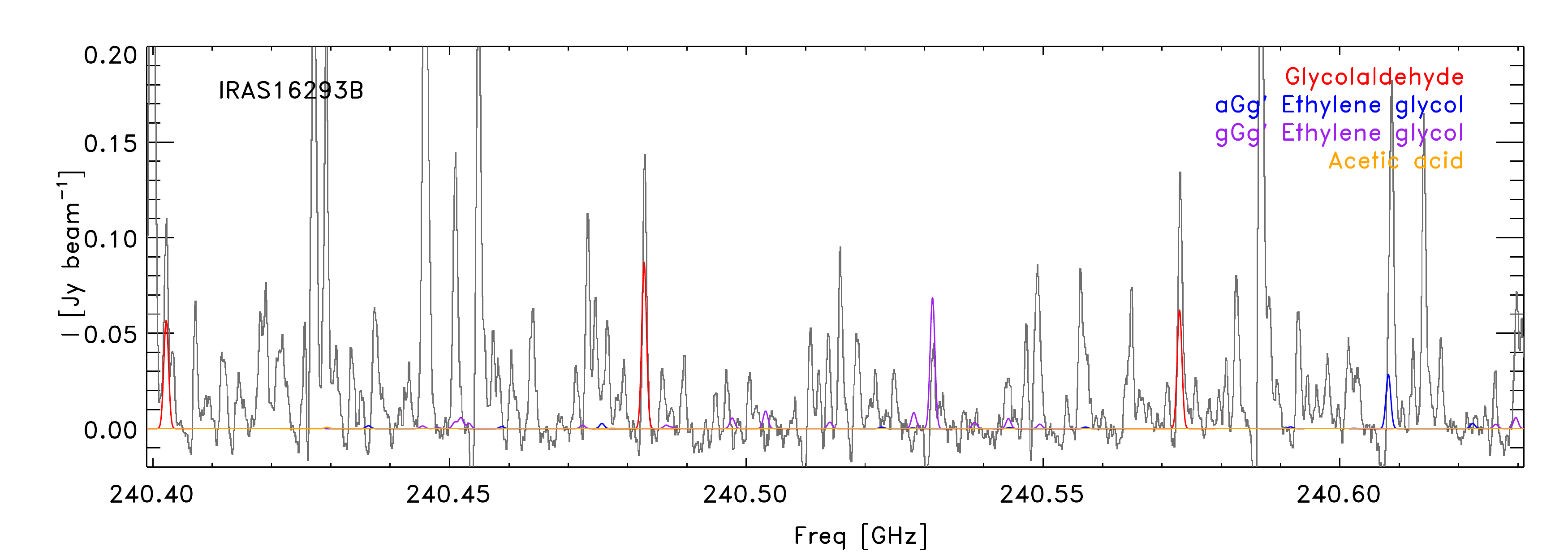}}}
\captionof{figure}{Continuation of Fig.~\ref{gcaeg_band6-1}.}
\end{minipage}
\clearpage
\noindent\begin{minipage}{\textwidth}
    \centering
\resizebox{!}{23cm}{\rotatebox{90}{\includegraphics{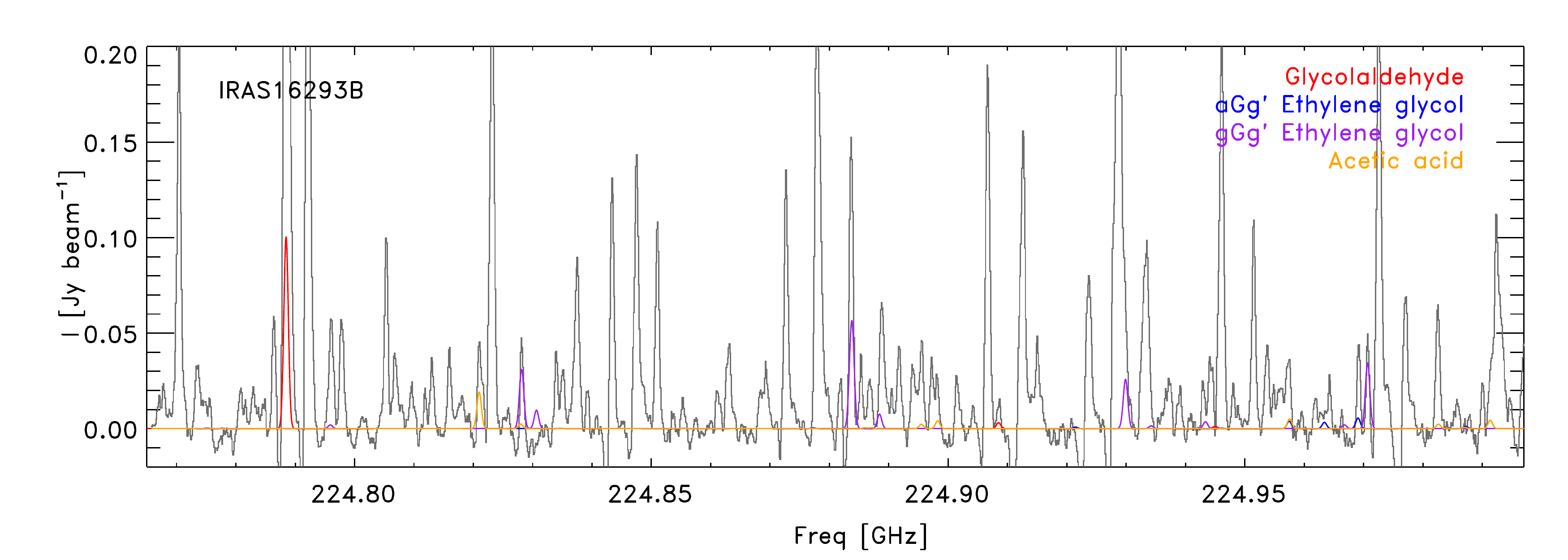}}}
\resizebox{!}{23cm}{\rotatebox{90}{\includegraphics{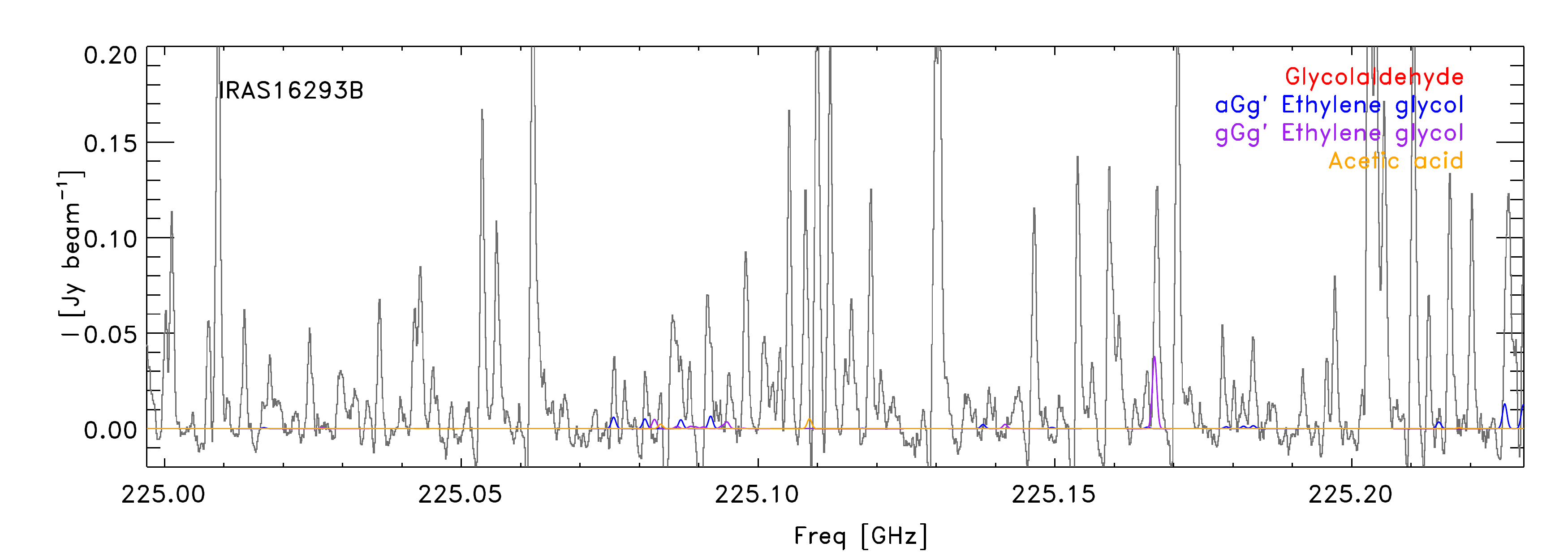}}}
\captionof{figure}{Continuation of Fig.~\ref{gcaeg_band6-1}.}
\end{minipage}
\clearpage
\noindent\begin{minipage}{\textwidth}
    \centering
\resizebox{!}{23cm}{\rotatebox{90}{\includegraphics{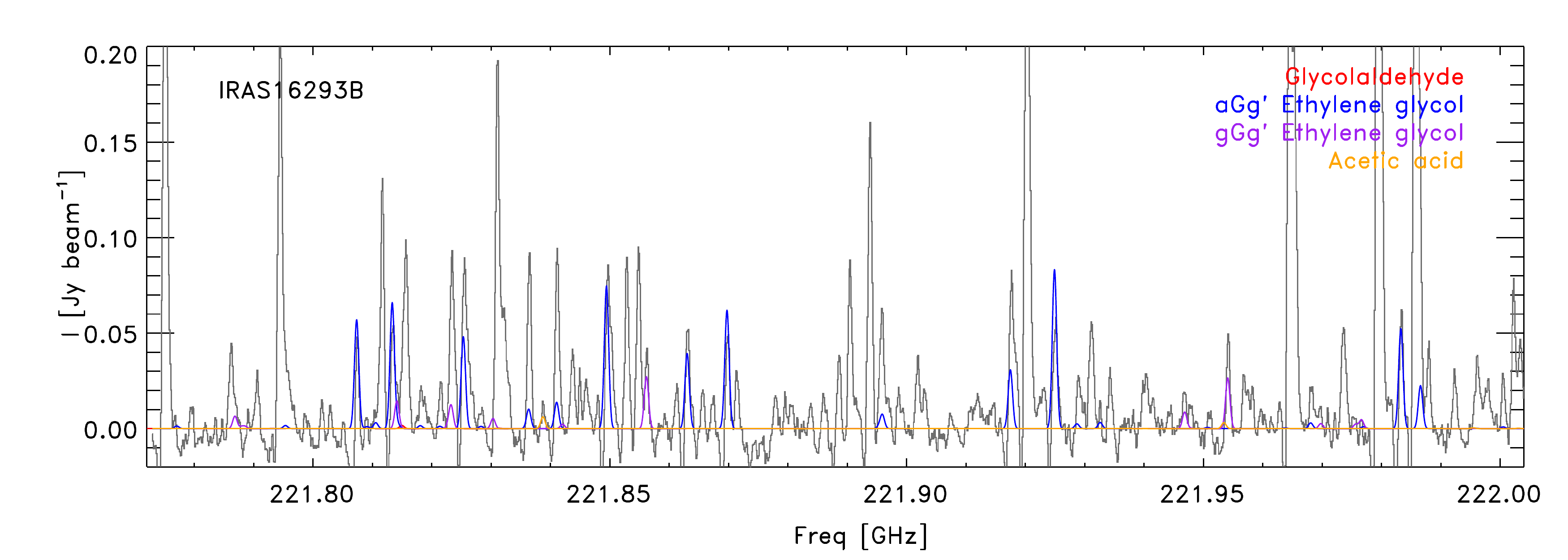}}}
\resizebox{!}{23cm}{\rotatebox{90}{\includegraphics{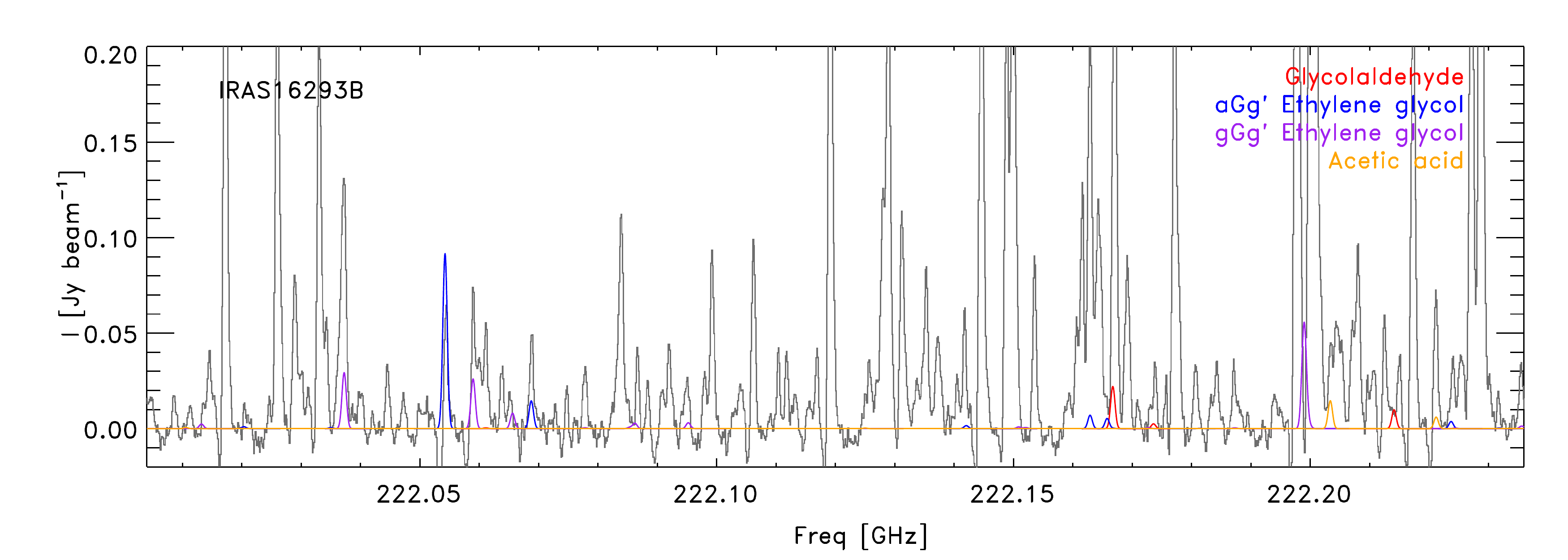}}}
\captionof{figure}{Continuation of Fig.~\ref{gcaeg_band6-1}.}\label{gcaeg_band6-2}
\end{minipage}
\clearpage
\noindent\begin{minipage}{\textwidth}
    \centering
\resizebox{!}{23cm}{\rotatebox{90}{\includegraphics{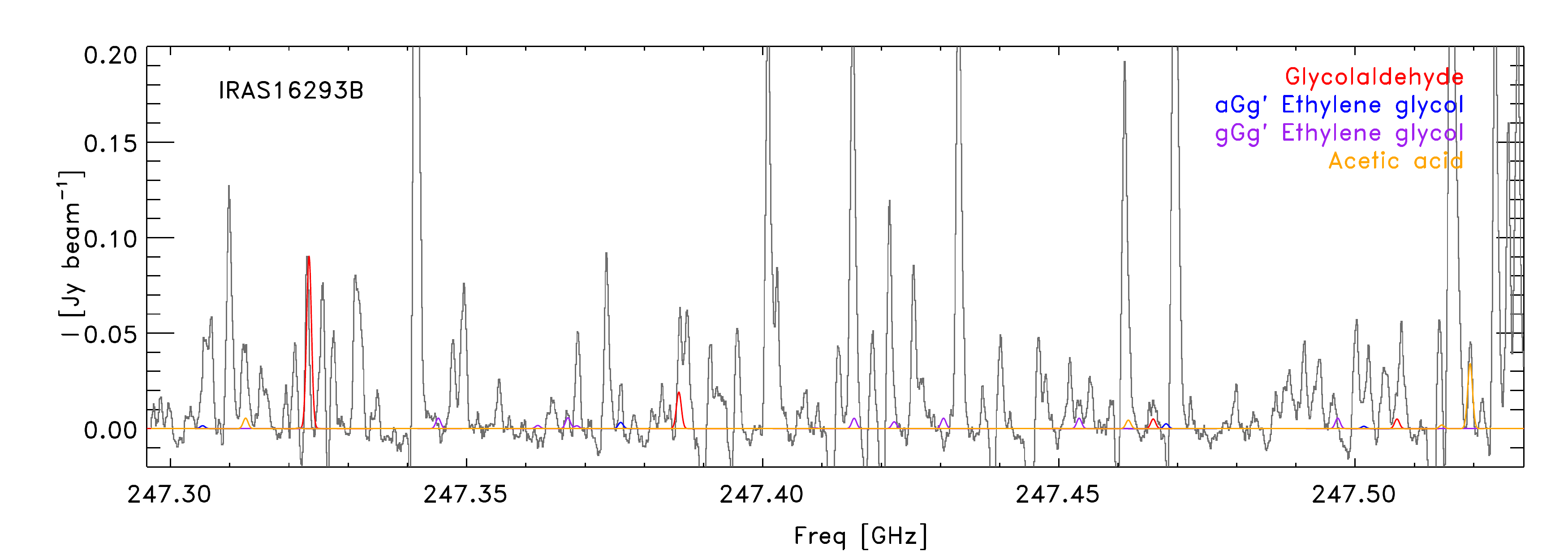}}}
\resizebox{!}{23cm}{\rotatebox{90}{\includegraphics{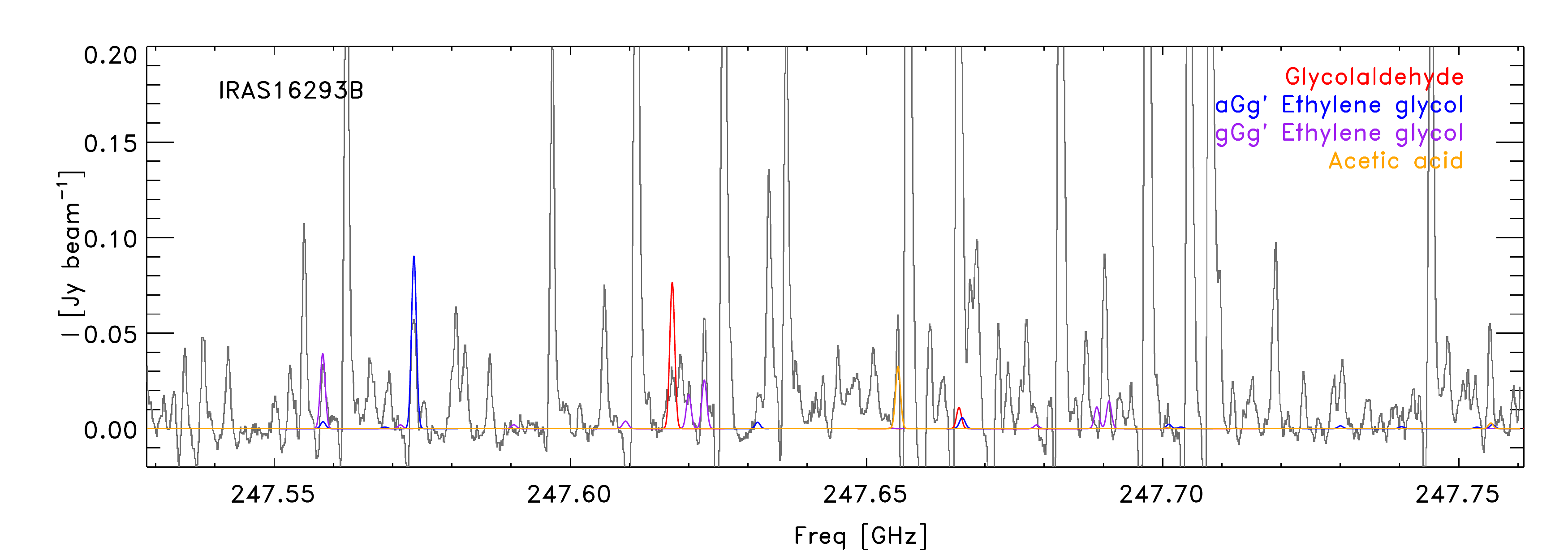}}}
\captionof{figure}{Continuation of Fig.~\ref{gcaeg_band6-2}.}\label{gcaeg_band6-3}
\end{minipage}
\clearpage
\noindent\begin{minipage}{\textwidth}
    \centering
\resizebox{!}{23cm}{\rotatebox{90}{\includegraphics{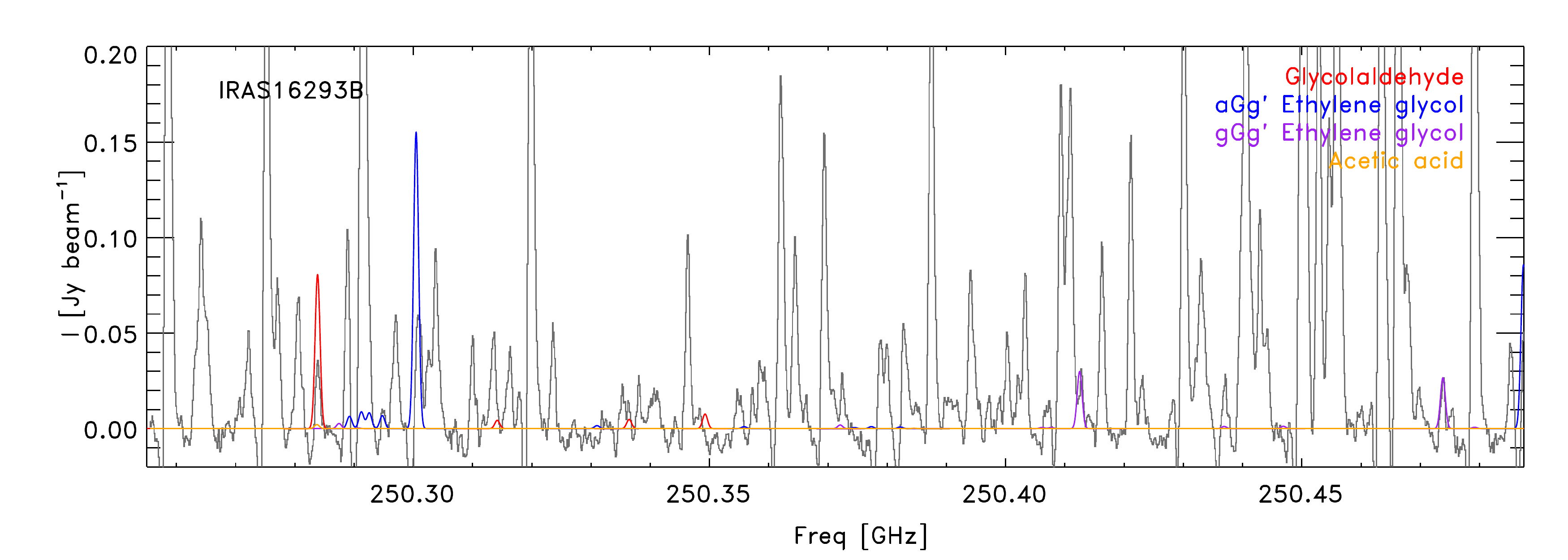}}}
\resizebox{!}{23cm}{\rotatebox{90}{\includegraphics{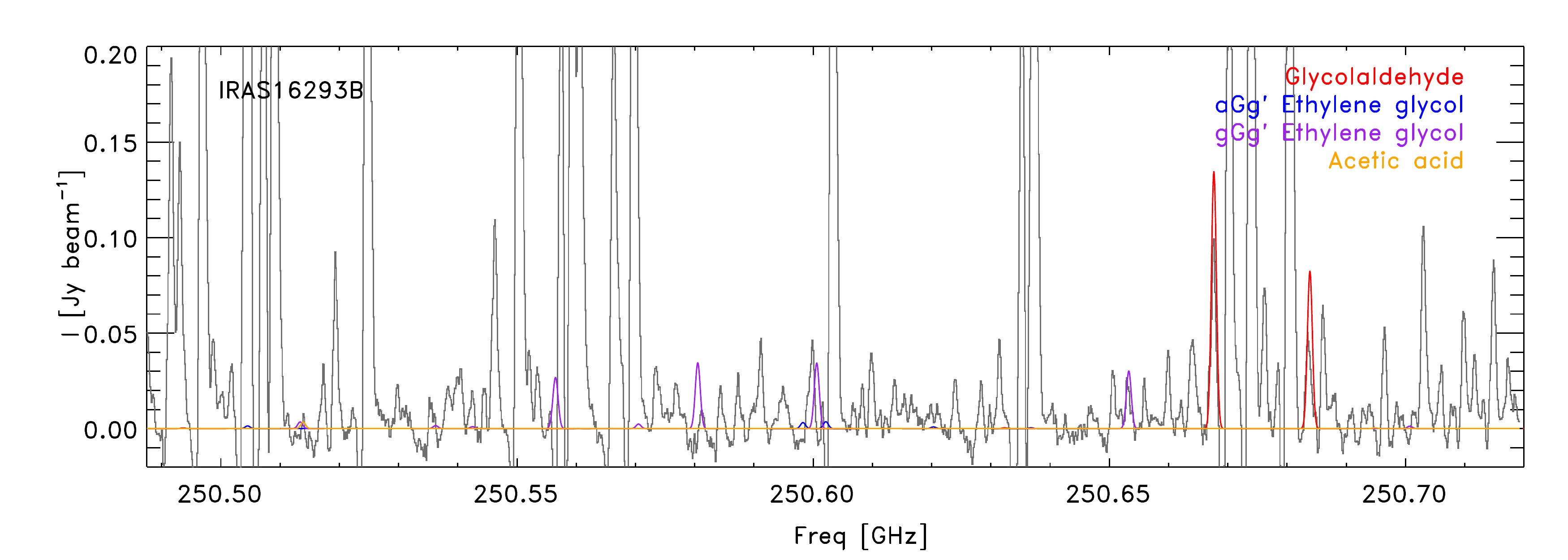}}}
\captionof{figure}{Continuation of Fig.~\ref{gcaeg_band6-3}.}
\end{minipage}
\clearpage
\noindent\begin{minipage}{\textwidth}
    \centering
\resizebox{!}{23cm}{\rotatebox{90}{\includegraphics{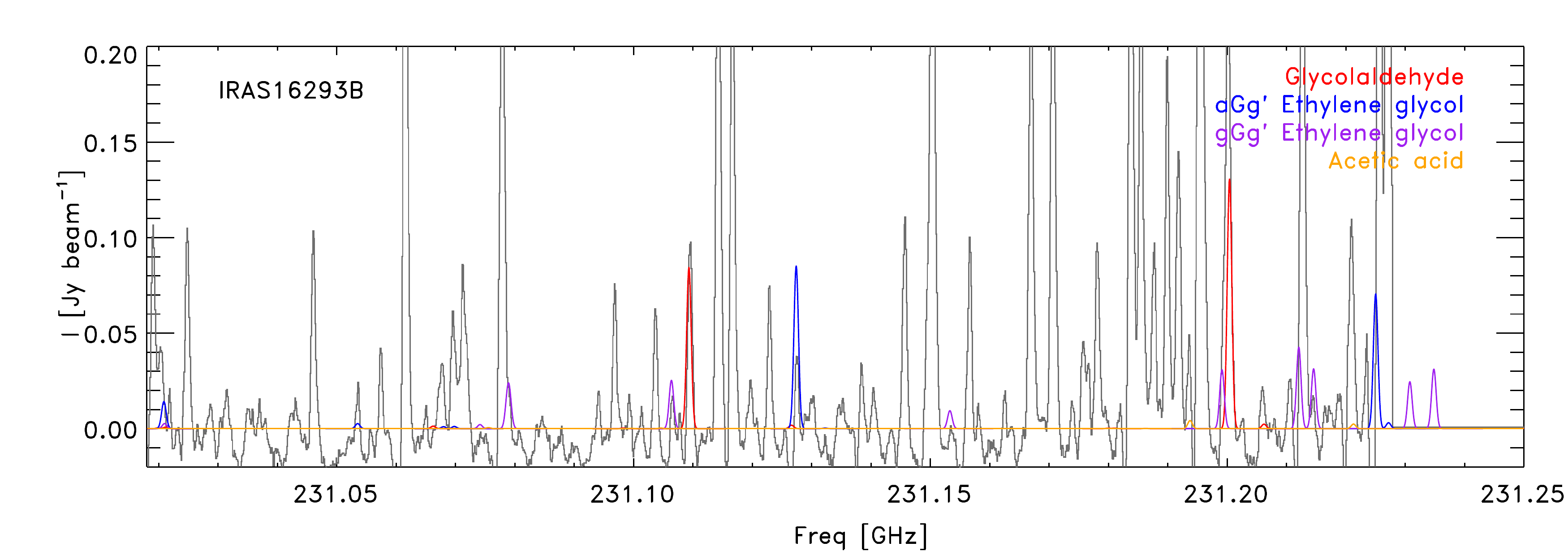}}}
\resizebox{!}{23cm}{\rotatebox{90}{\includegraphics{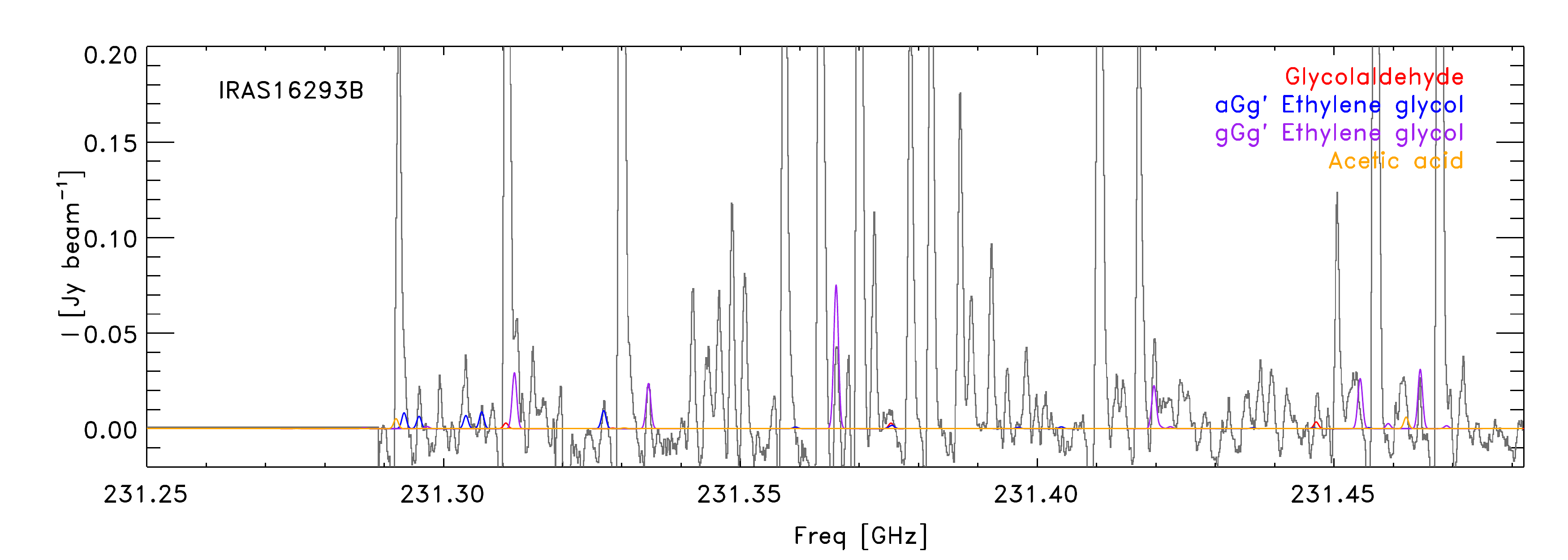}}}
\captionof{figure}{Continuation of Fig.~\ref{gcaeg_band6-3}.}
\end{minipage}
\clearpage
\noindent\begin{minipage}{\textwidth}
    \centering
\resizebox{!}{23cm}{\rotatebox{90}{\includegraphics{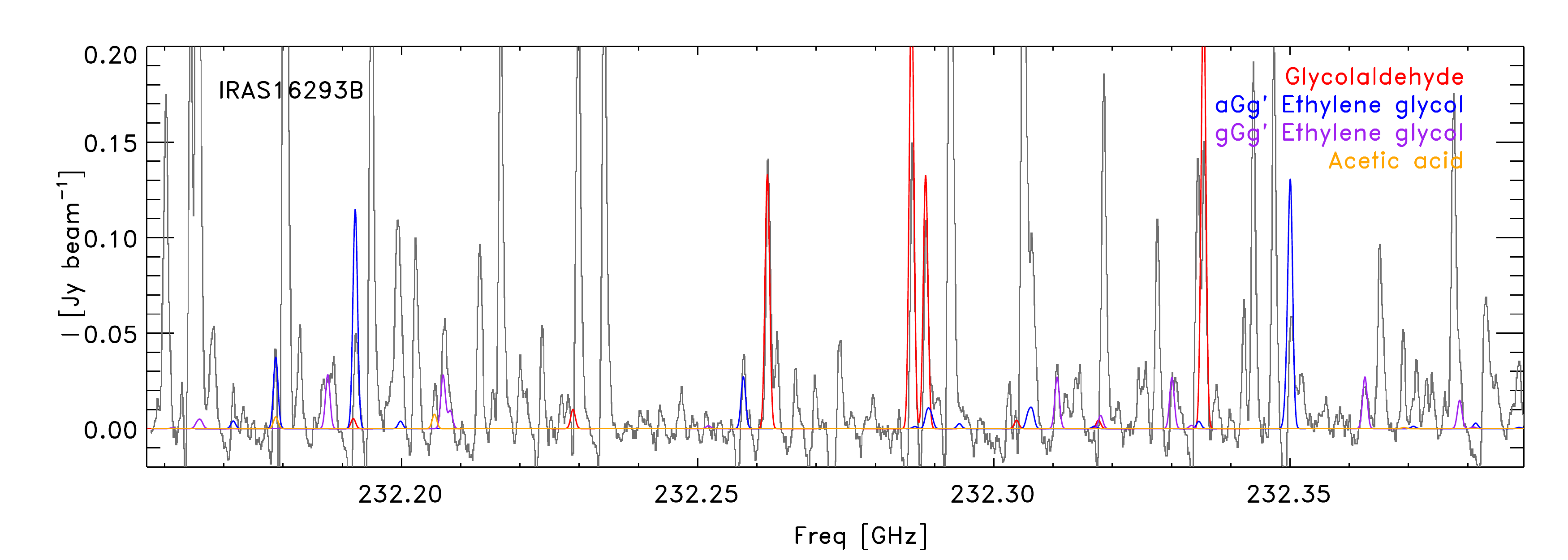}}}
\resizebox{!}{23cm}{\rotatebox{90}{\includegraphics{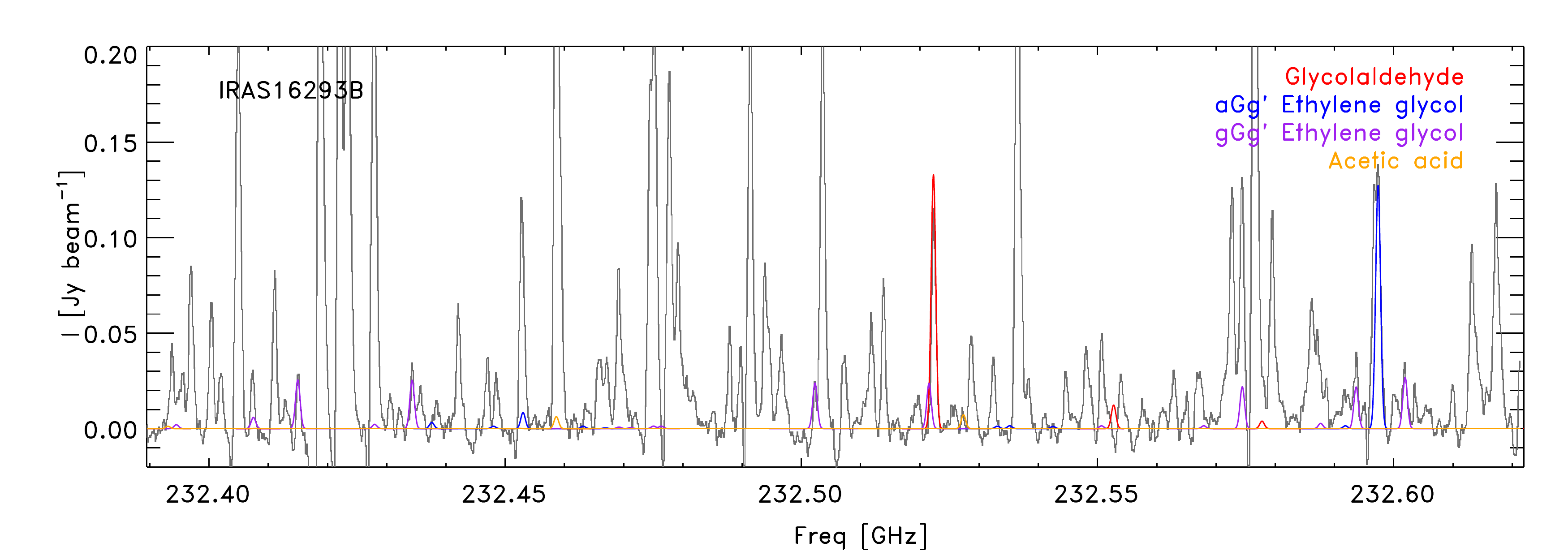}}}
\captionof{figure}{Continuation of Fig.~\ref{gcaeg_band6-3}.}\label{gcaeg_band6-4}
\end{minipage}
\clearpage

\noindent\begin{minipage}{\textwidth}
\section{Synthetic spectra fits for species in Band~7}
\subsection{Methanol (CH$_3$$^{18}$OH)}
\resizebox{0.88\textwidth}{!}{\includegraphics{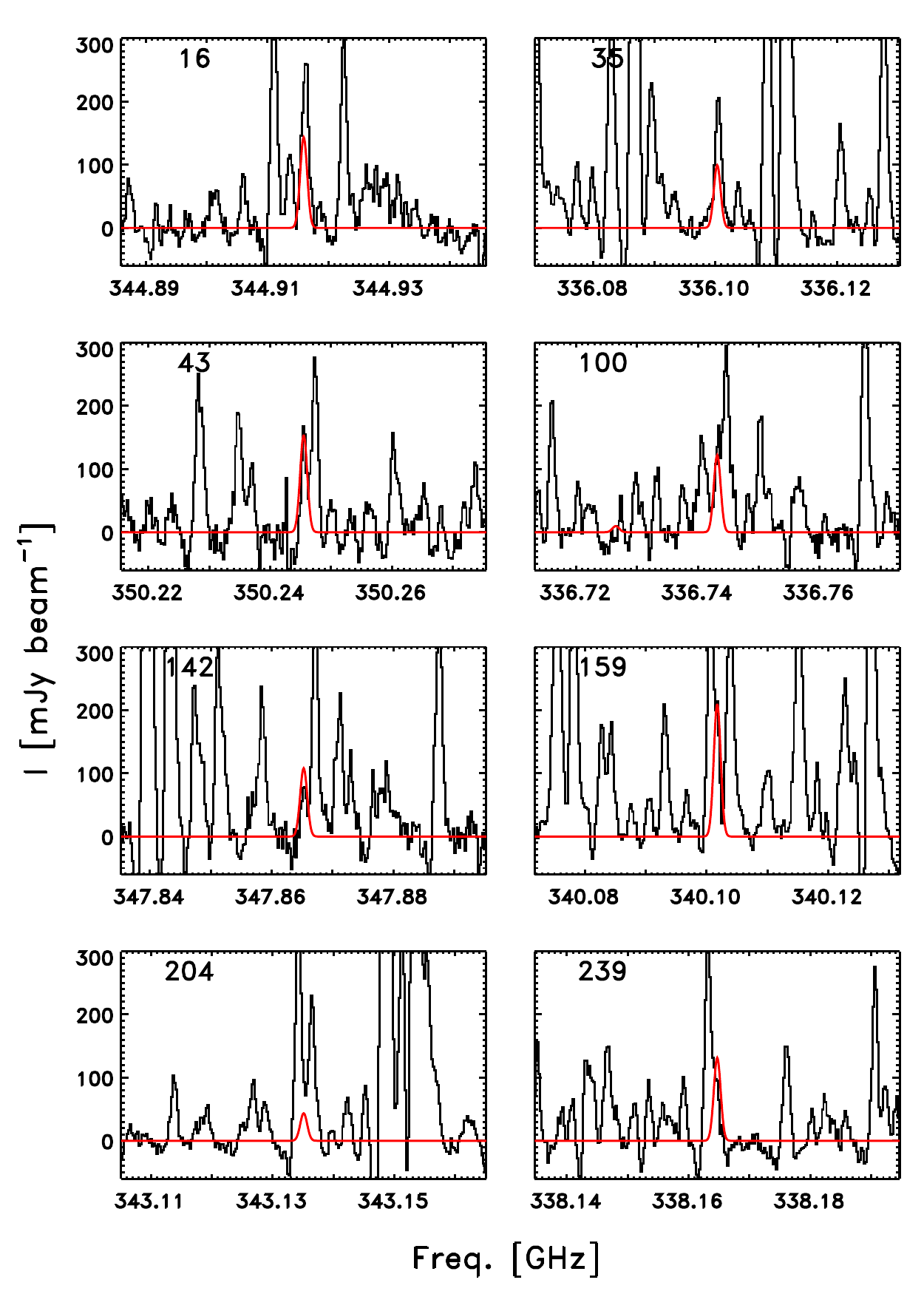}\includegraphics{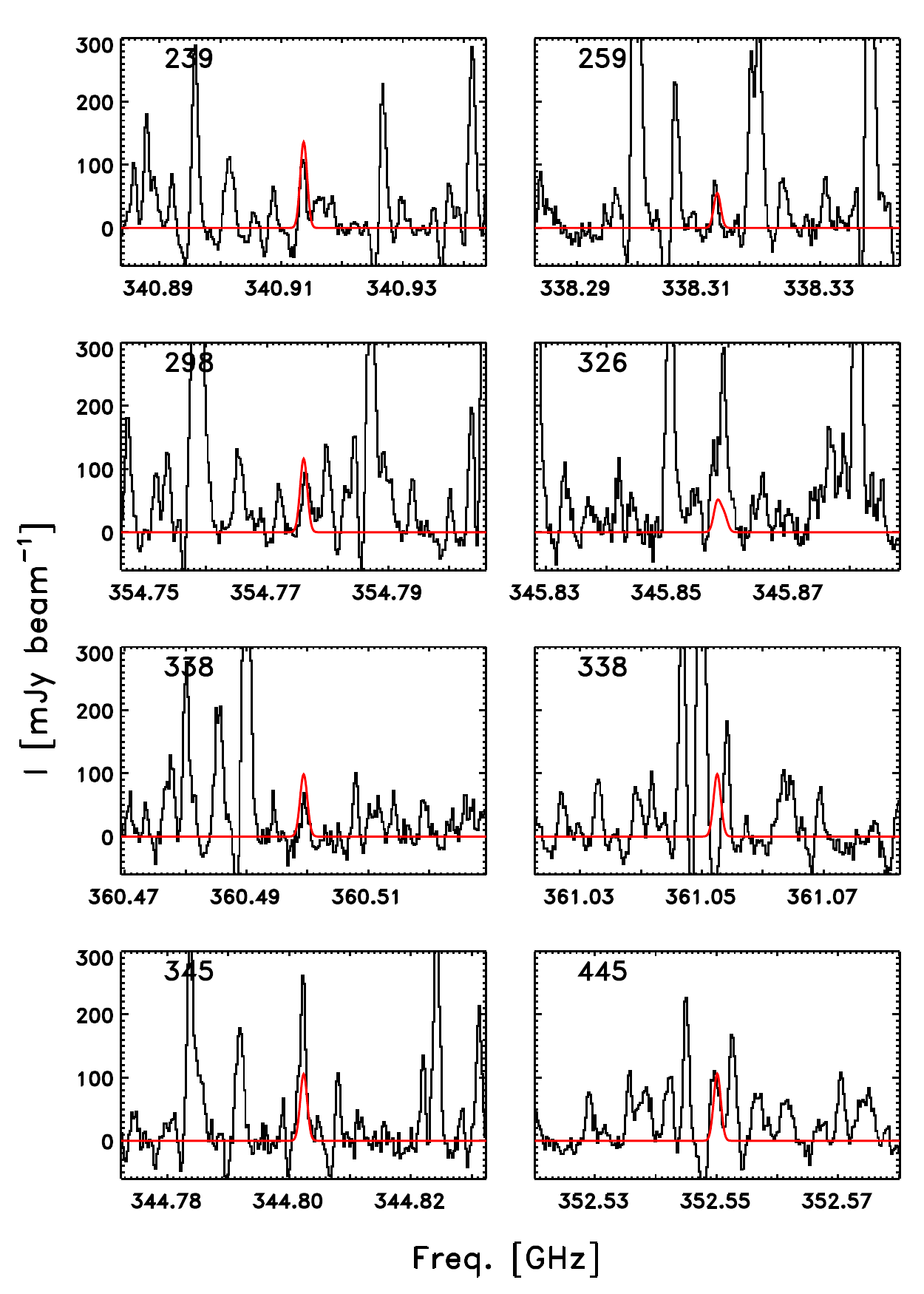}}

\resizebox{0.44\textwidth}{!}{\includegraphics{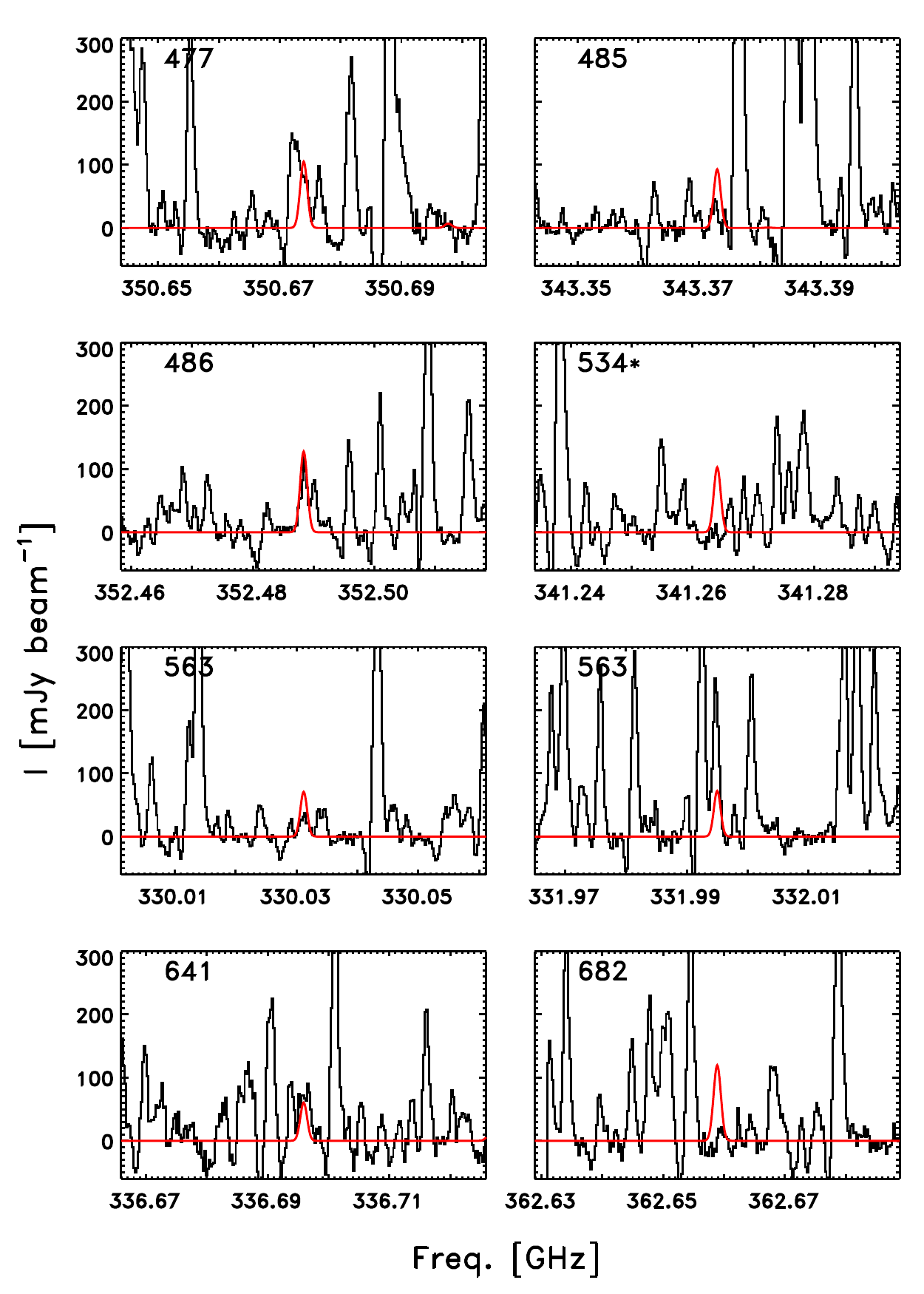}}

\captionof{figure}{The 24 brightest lines of CH$_3^{18}$OH as expected from the synthetic spectrum. These lines are sorted according to $E_{\rm up}$ given in K in the upper left corner of each panel. For frequencies where multiple transitions of CH$_3$$^{18}$OH are overlapping an "$\ast$" is added after the value for $E_{\rm up}$.}\label{18methanol_spectra1}
\end{minipage}

\clearpage

\noindent\begin{minipage}{\textwidth}

\subsection{Glycolaldehyde}
\resizebox{0.88\textwidth}{!}{\includegraphics{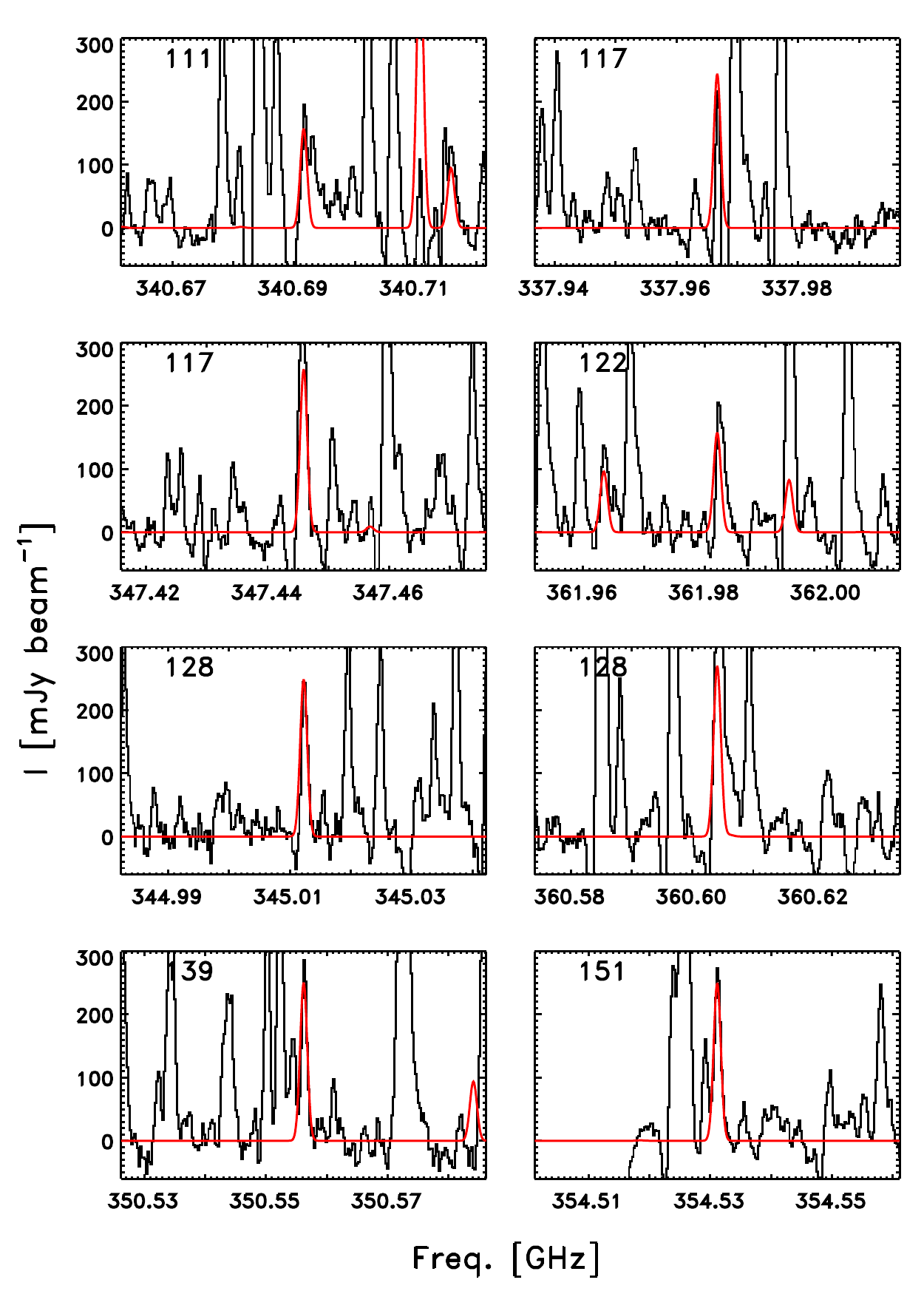}\includegraphics{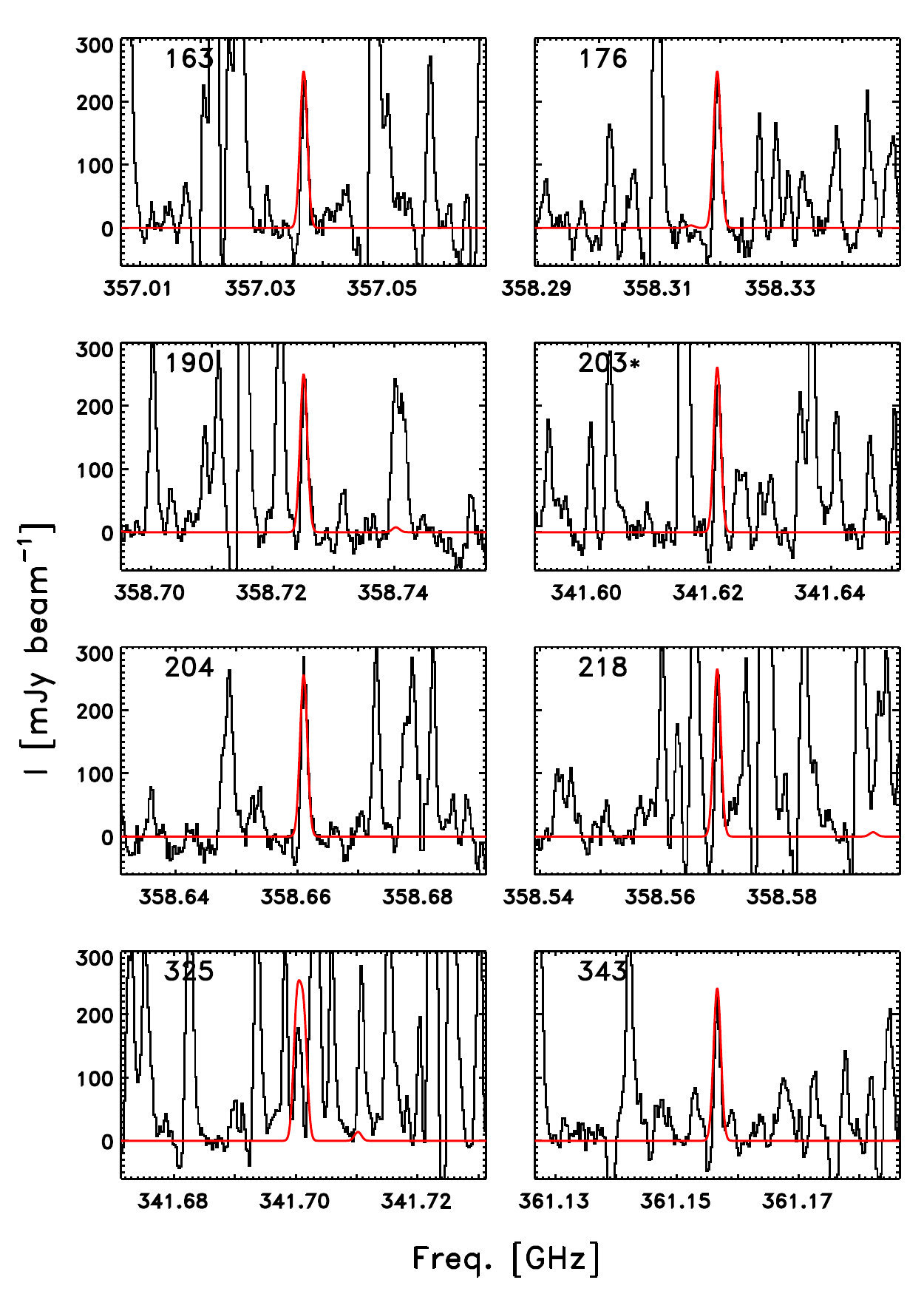}}
\resizebox{0.44\textwidth}{!}{\includegraphics{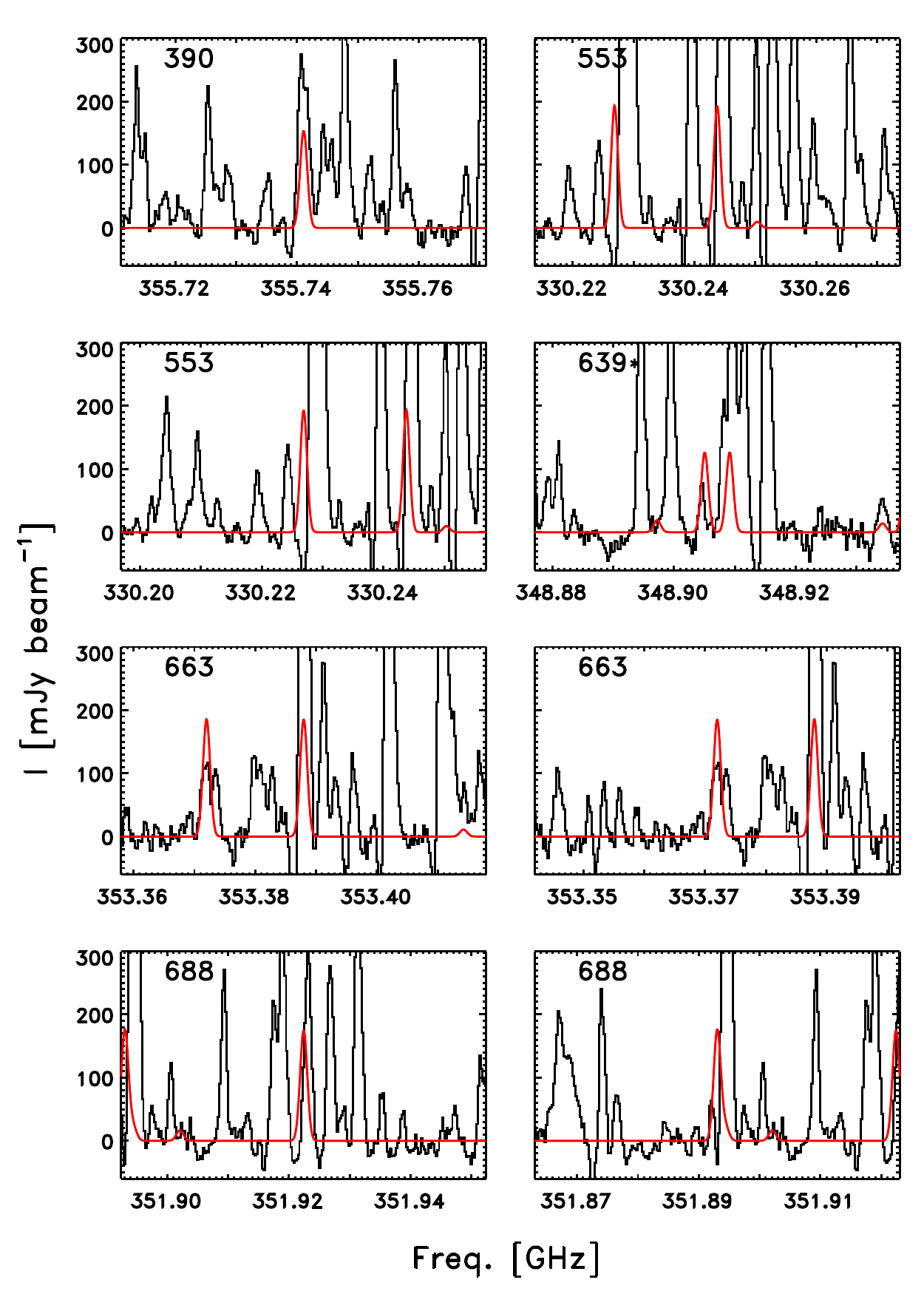}}
\captionof{figure}{As in Fig.~\ref{18methanol_spectra1} for the 24 brightest lines of glycolaldehyde with $\tau < 0.2$ as expected from the synthetic spectrum.}\label{glycolaldehyde_spectra1}
\end{minipage}

\clearpage

\noindent\begin{minipage}{\textwidth}
\subsection{$aGg'$ ethylene glycol}
\resizebox{0.88\textwidth}{!}{\includegraphics{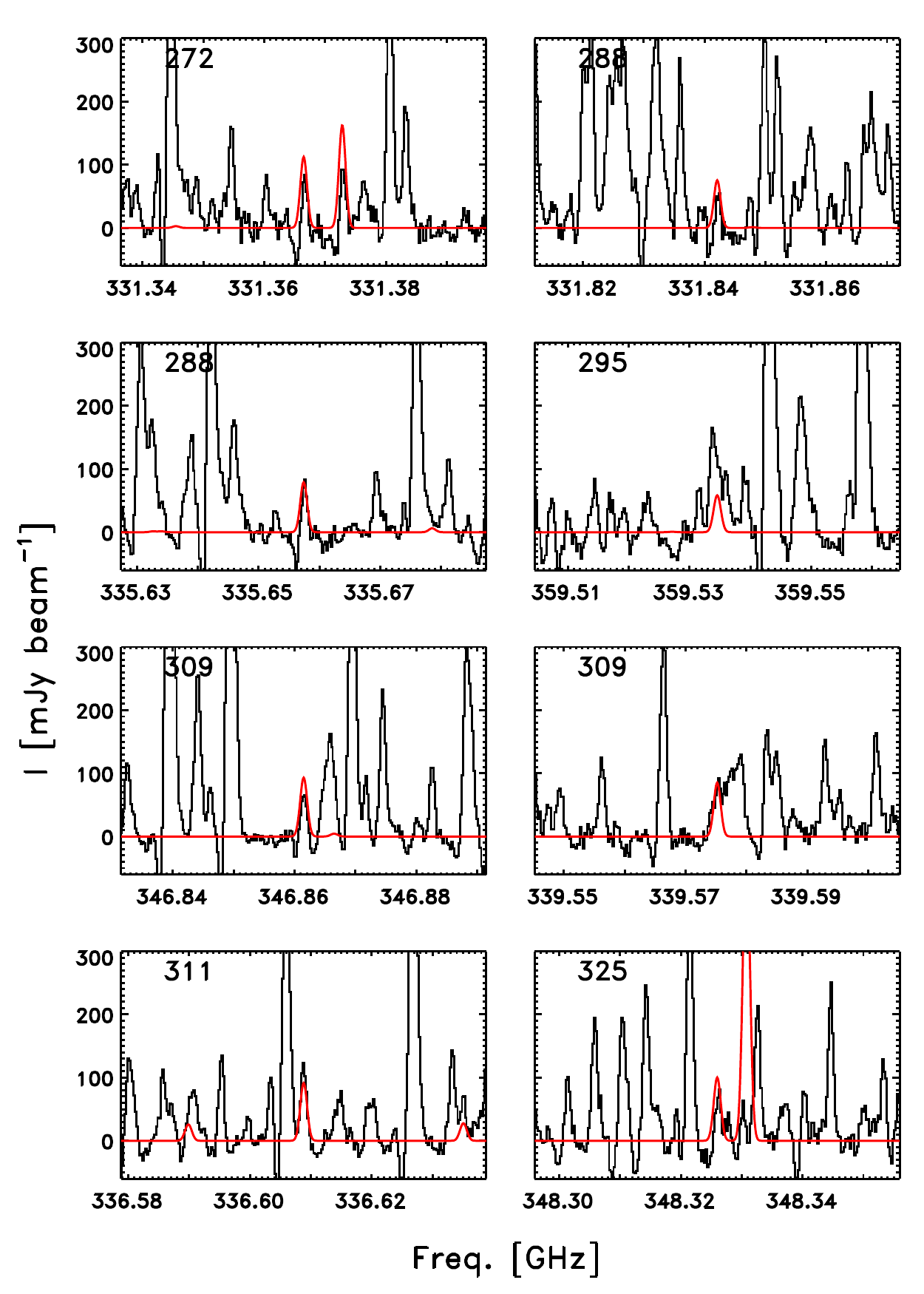}\includegraphics{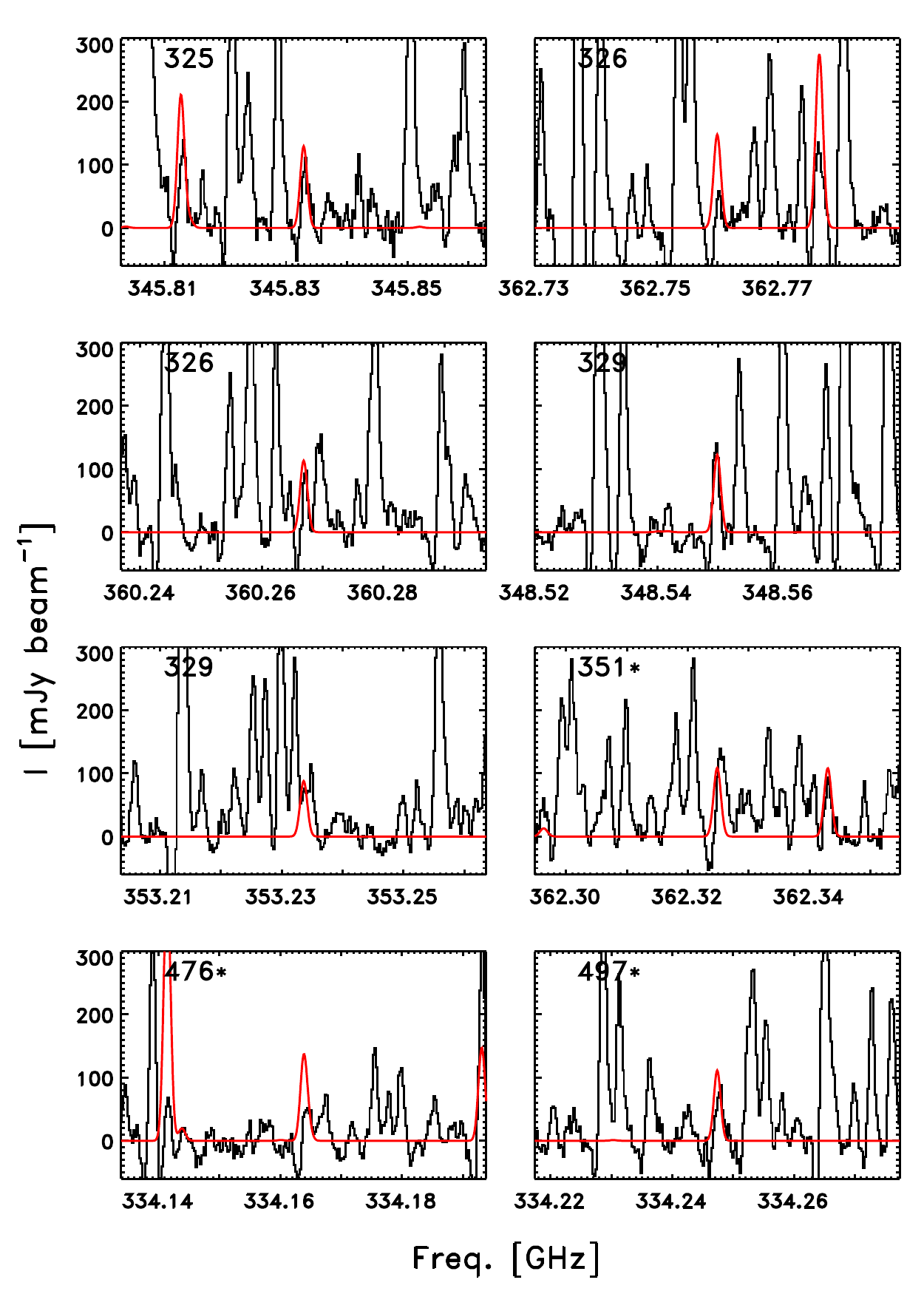}}
\resizebox{0.44\textwidth}{!}{\includegraphics{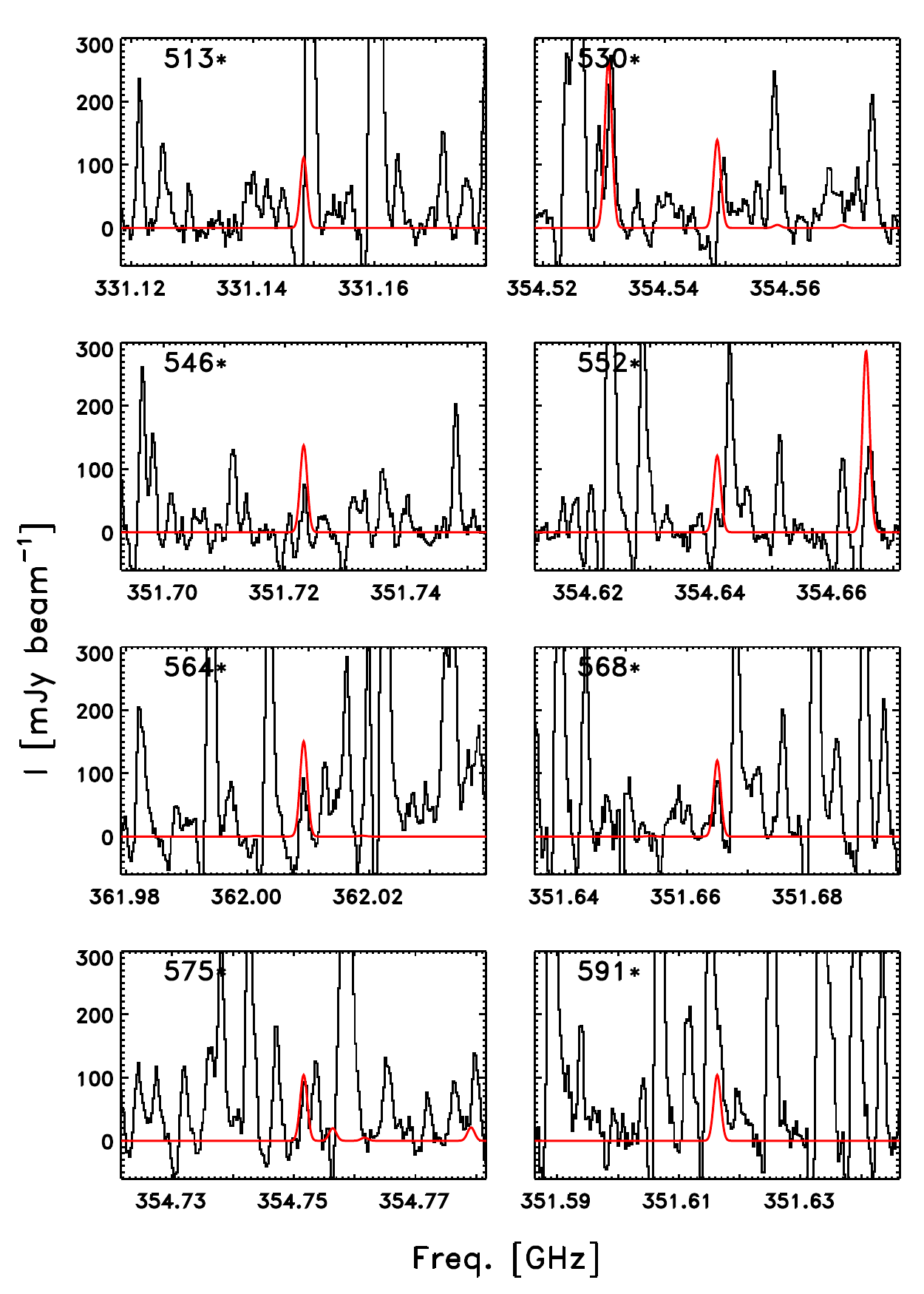}}
\captionof{figure}{As in Fig.~\ref{18methanol_spectra1} for the 24 brightest lines of $aGg'$ ethylene glycol with $\tau < 0.2$ as expected from the synthetic spectrum.}\label{Ga-ethyleneglycol_spectra1}
\end{minipage}

\clearpage

\noindent\begin{minipage}{\textwidth}
\subsection{$gGg'$ ethylene glycol}
\resizebox{0.88\textwidth}{!}{\includegraphics{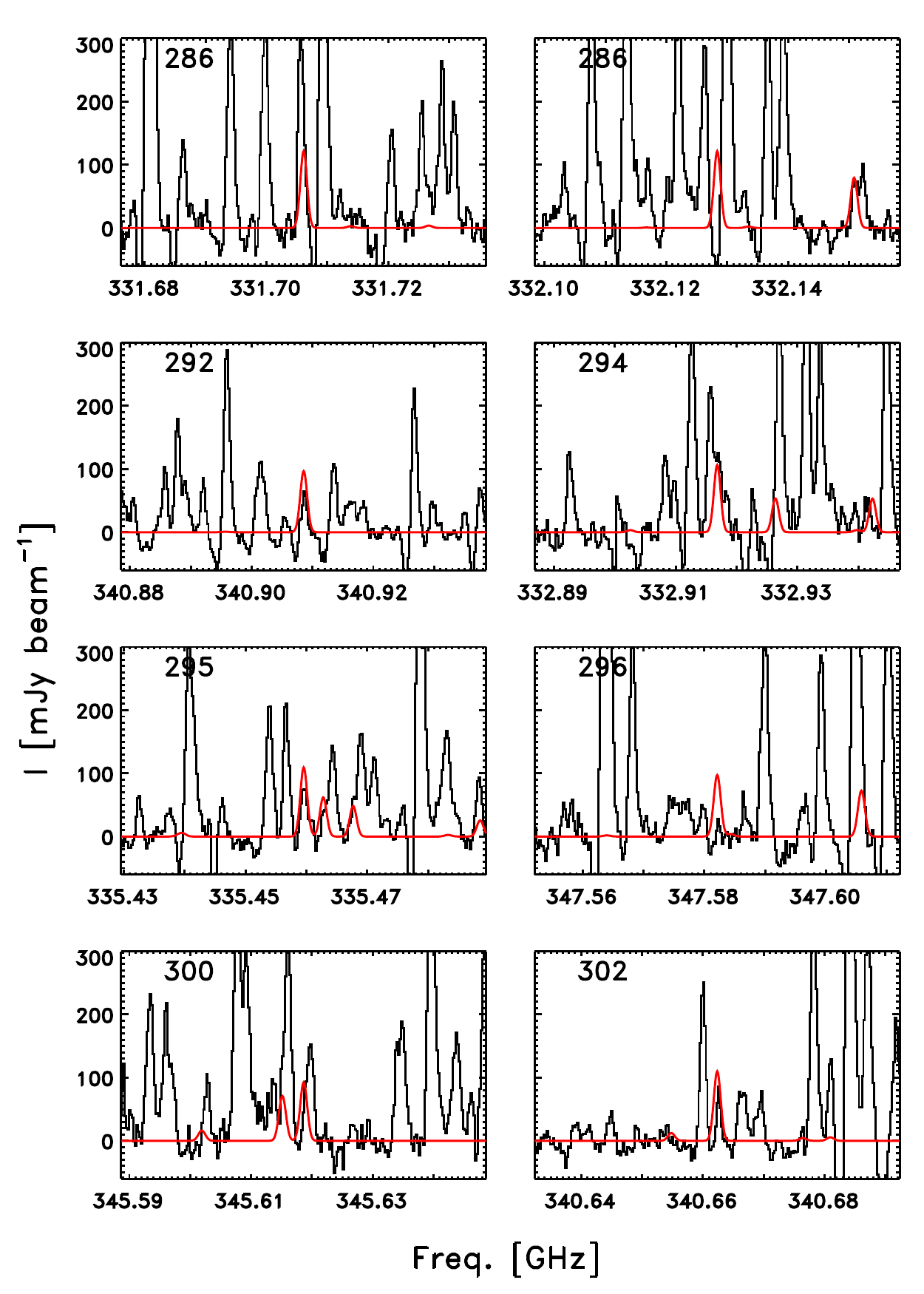}\includegraphics{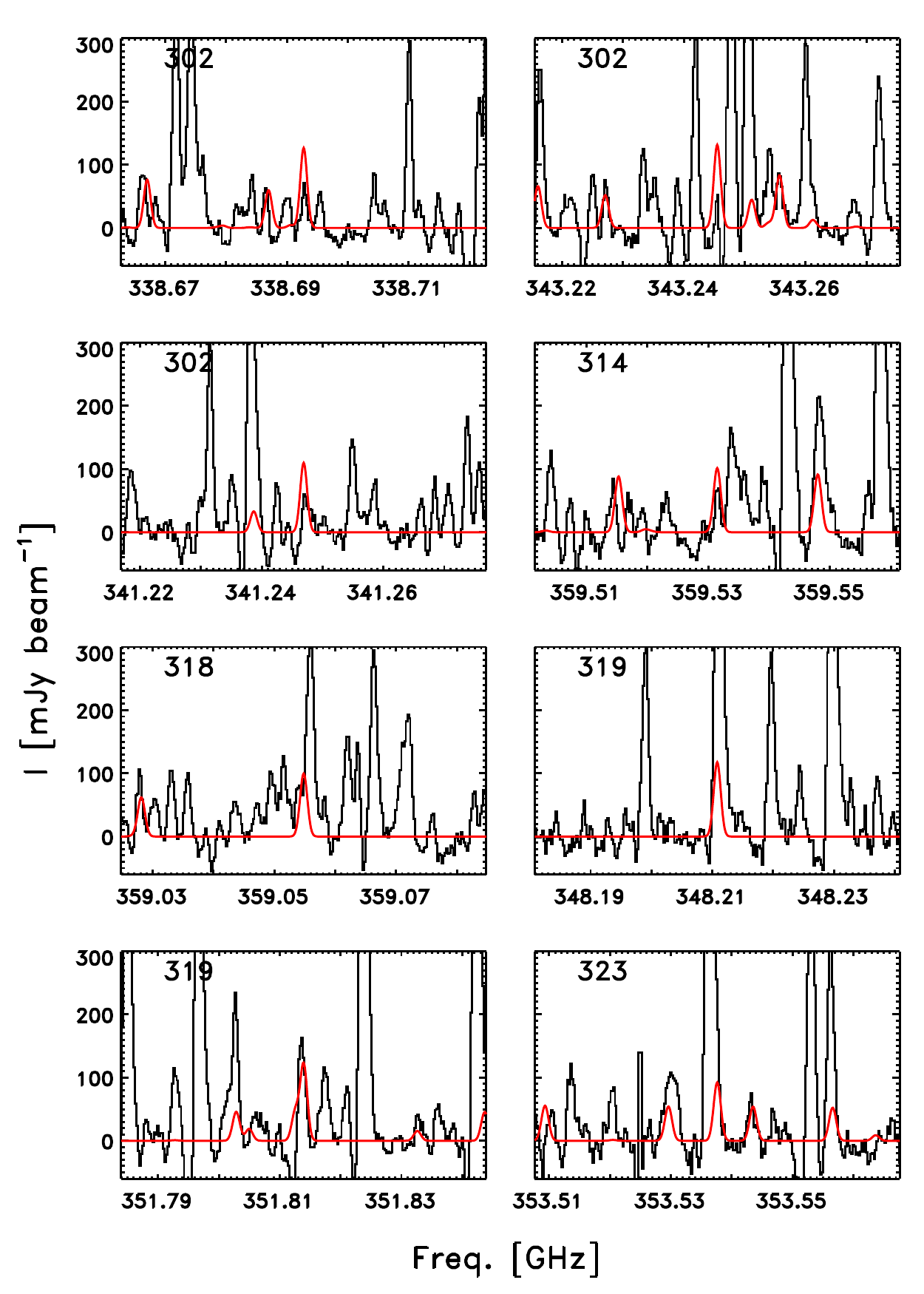}}
\resizebox{0.44\textwidth}{!}{\includegraphics{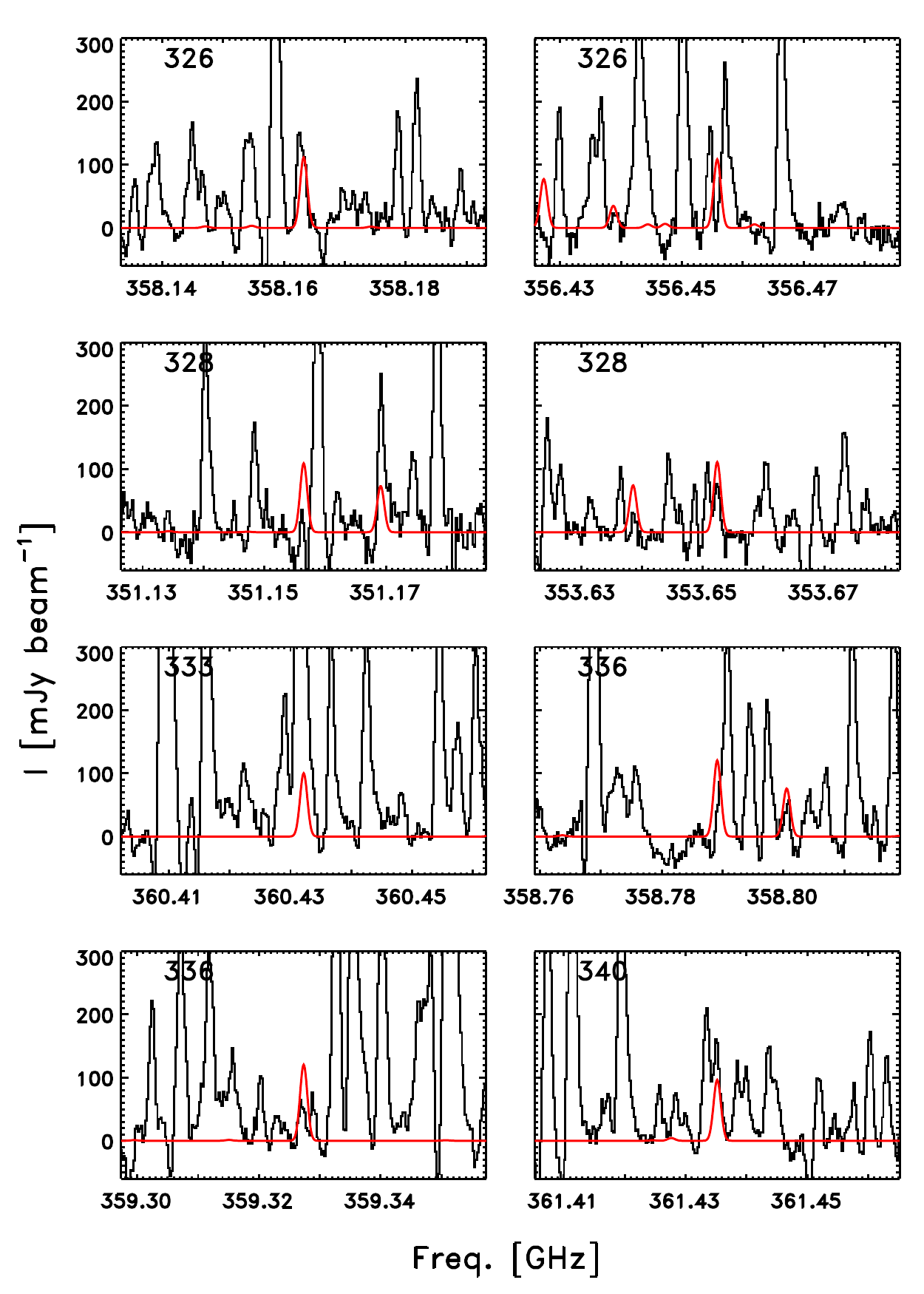}}
\captionof{figure}{As in Fig.~\ref{18methanol_spectra1} for the 24 brightest lines of $gGg'$ ethylene glycol with $\tau < 0.2$ as expected from the synthetic spectrum.}\label{Gg-ethyleneglycol_spectra1}
\end{minipage}

\clearpage

\noindent\begin{minipage}{\textwidth}
\subsection{Acetic acid}
\resizebox{0.88\textwidth}{!}{\includegraphics{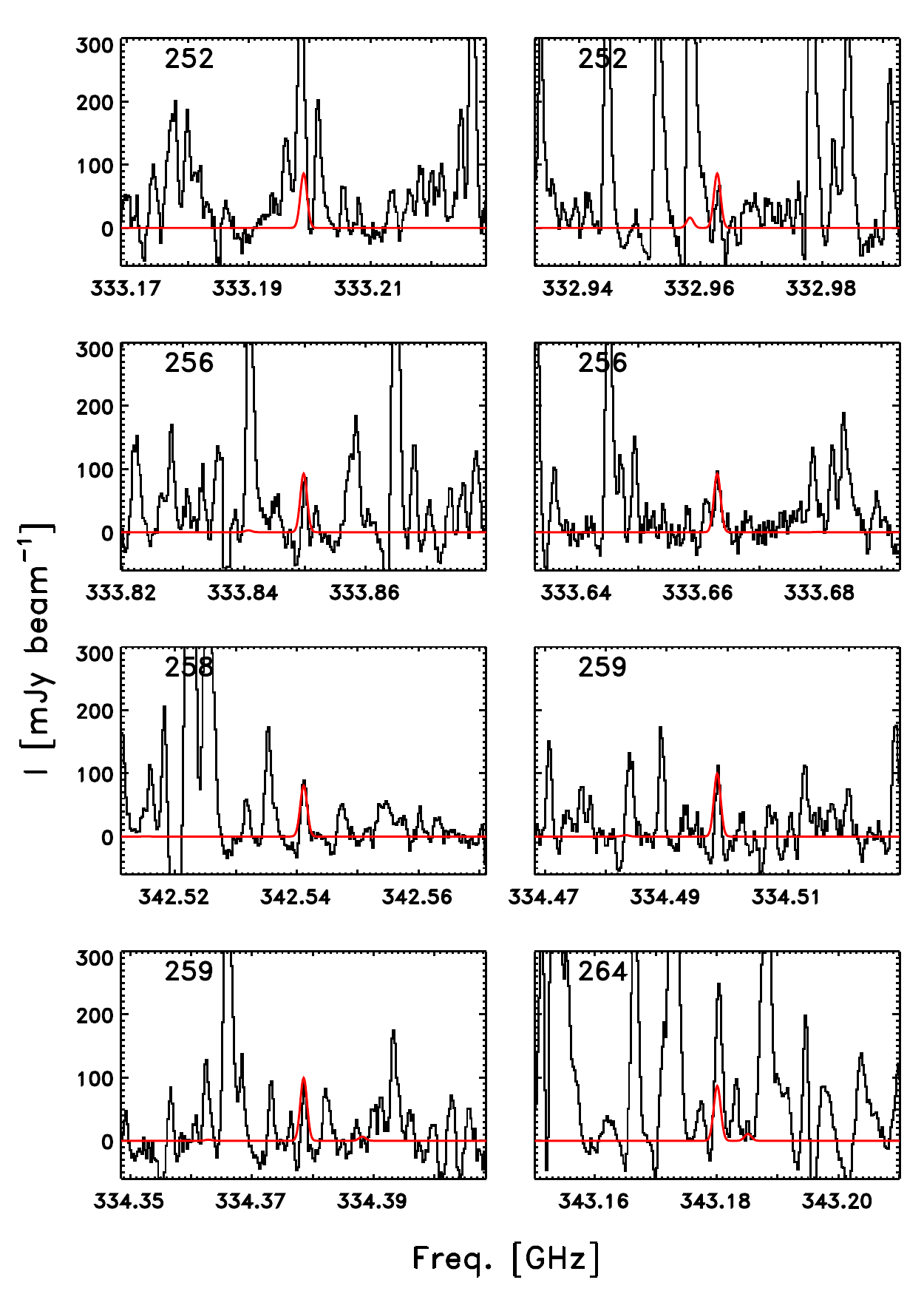}\includegraphics{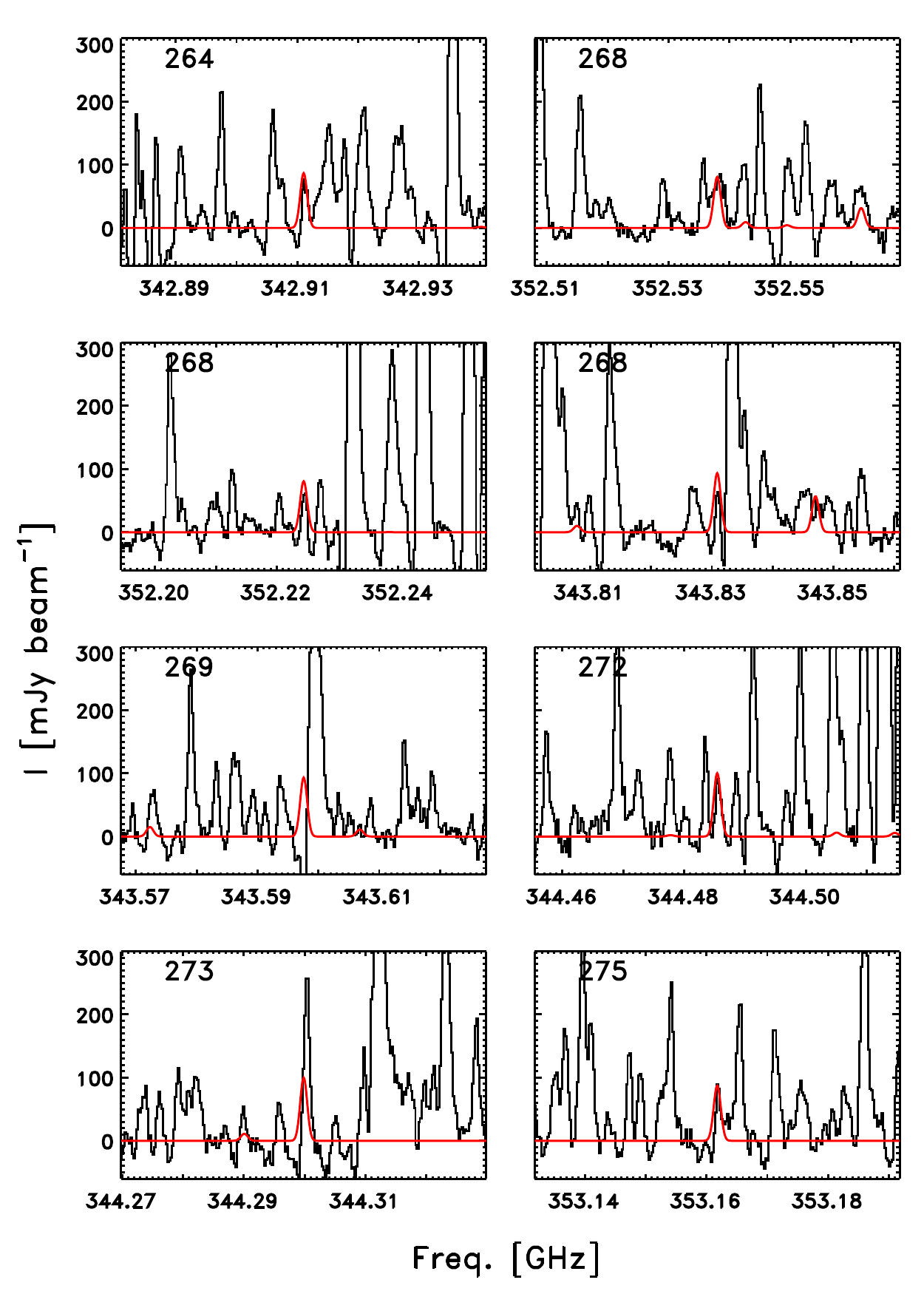}}
\resizebox{0.44\textwidth}{!}{\includegraphics{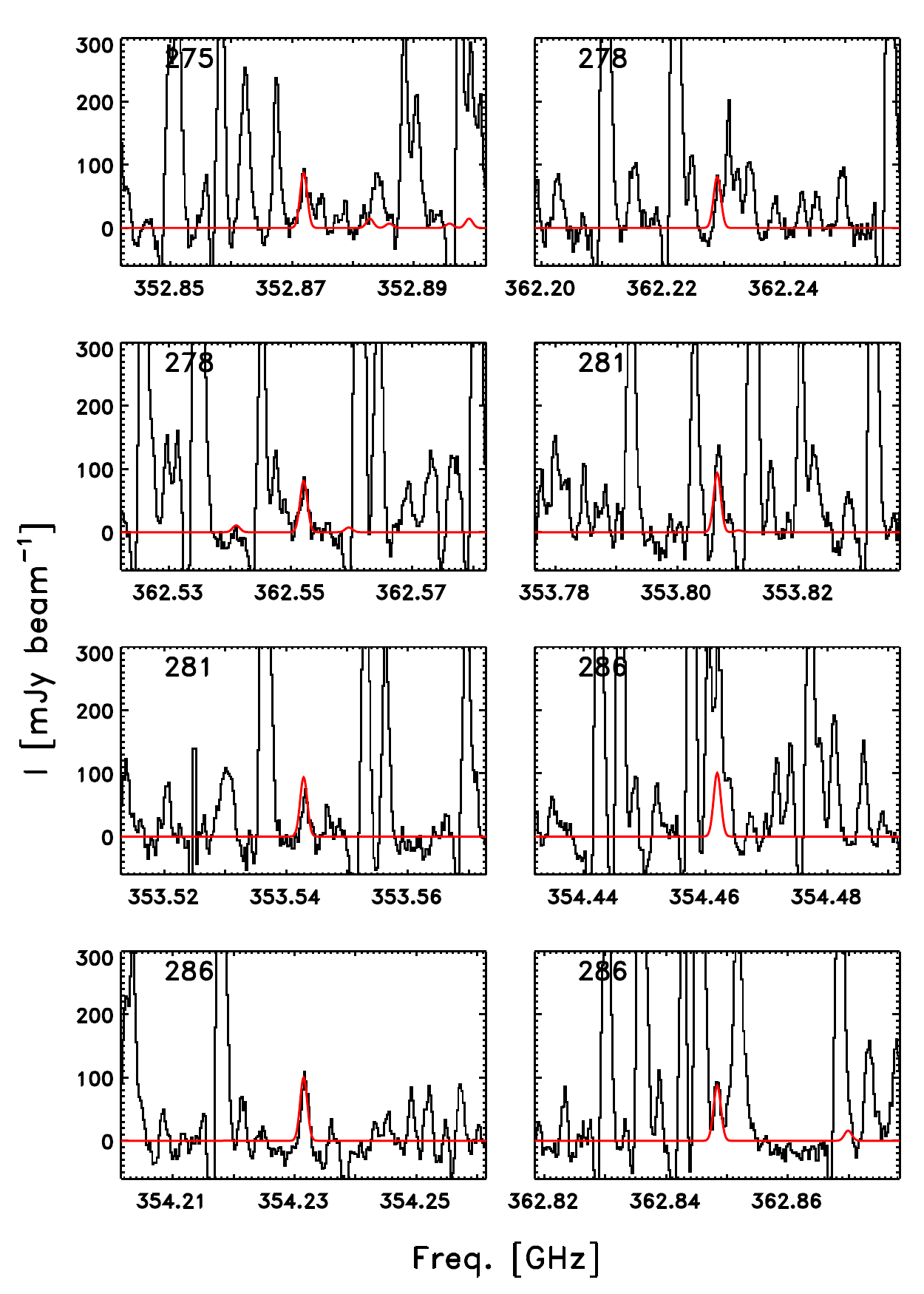}}
\captionof{figure}{The 24 brightest lines of acetic acid with $\tau < 0.2$ as expected from the synthetic spectrum sorted according to $E_{\rm up}$.}\label{aceticacid}
\end{minipage}

\clearpage

\noindent\begin{minipage}{\textwidth}
\subsection{Glycolaldehyde: Deuterated isotopologues}
\resizebox{0.88\textwidth}{!}{\includegraphics{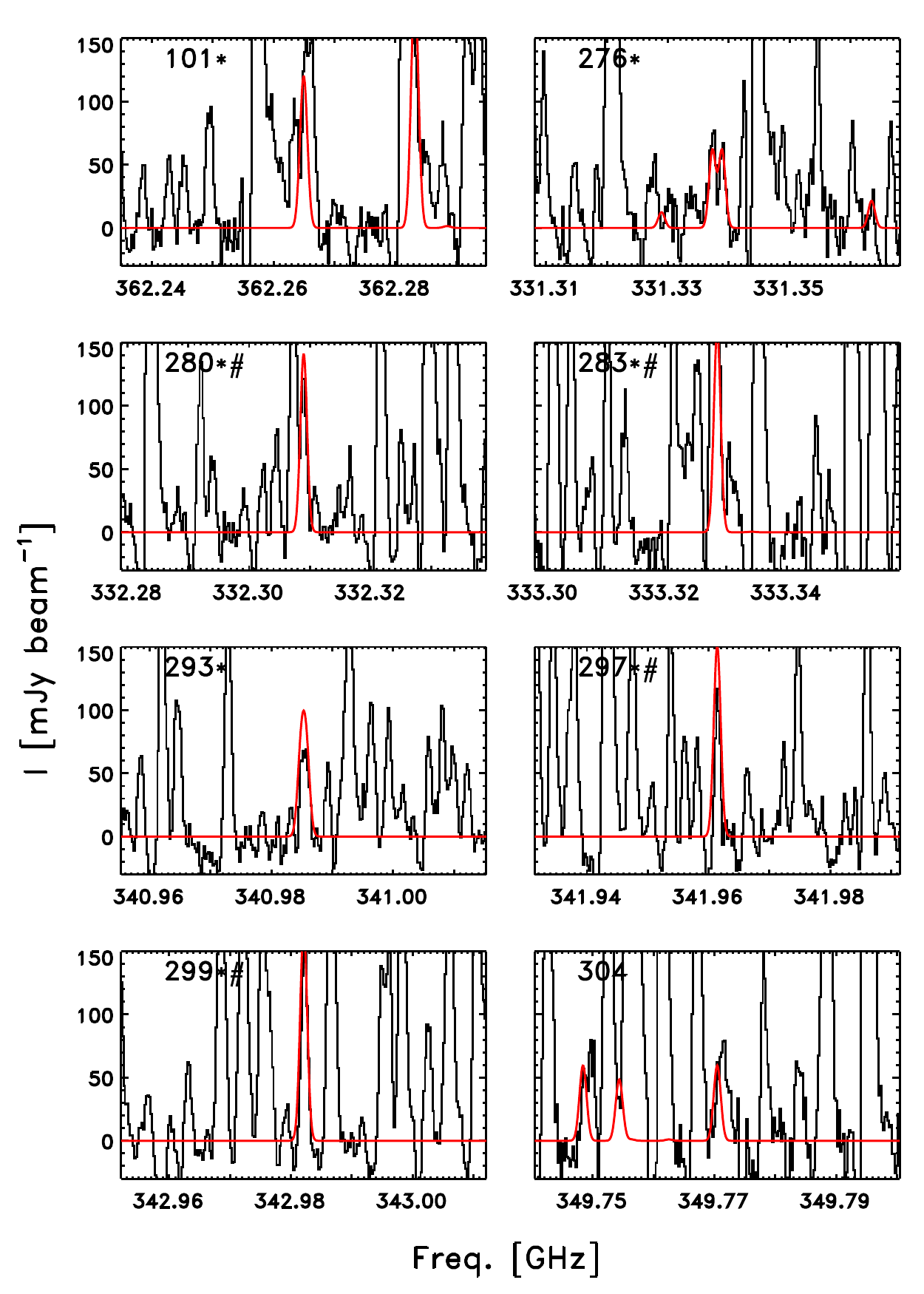}\includegraphics{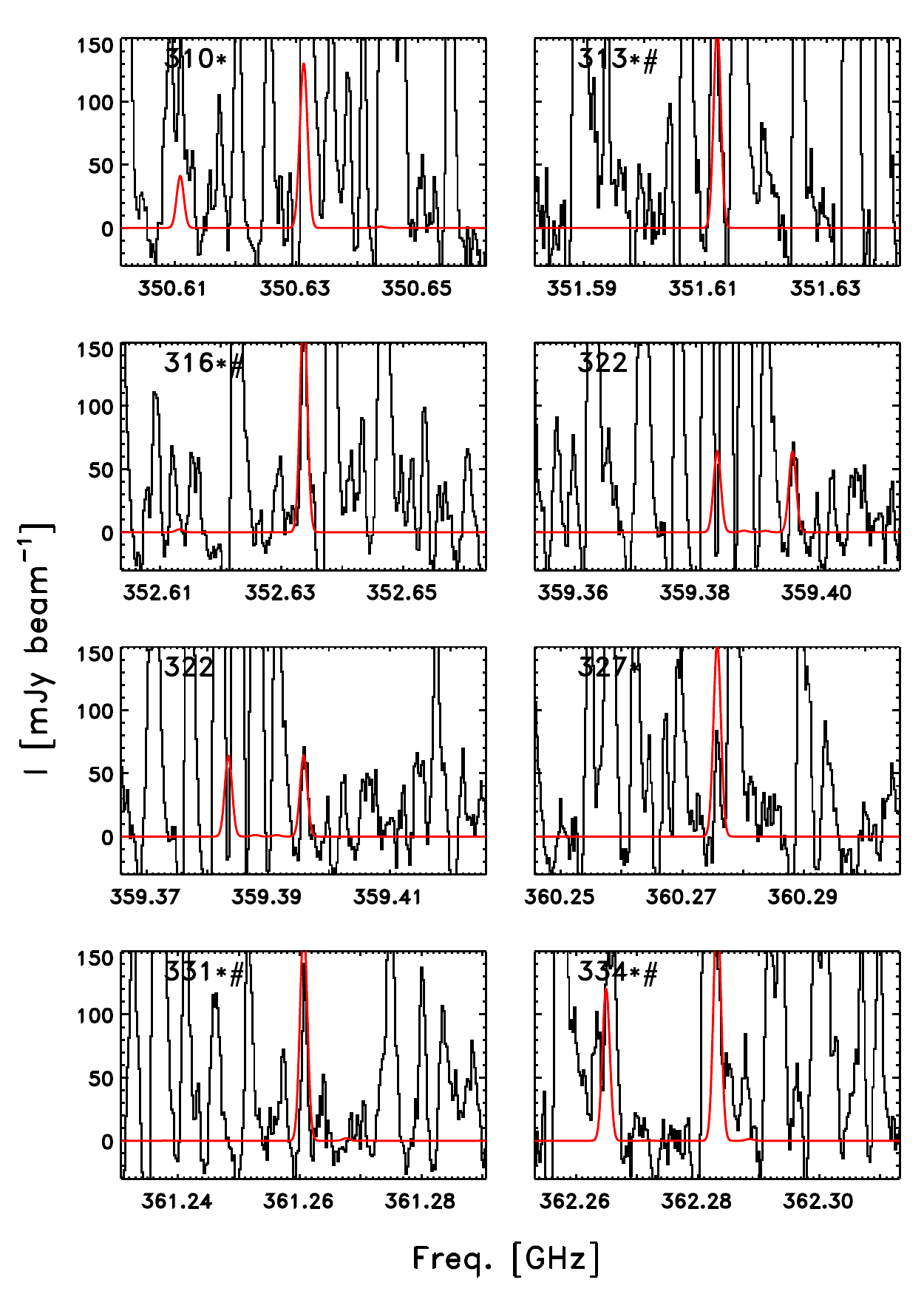}}
\resizebox{0.88\textwidth}{!}{\includegraphics{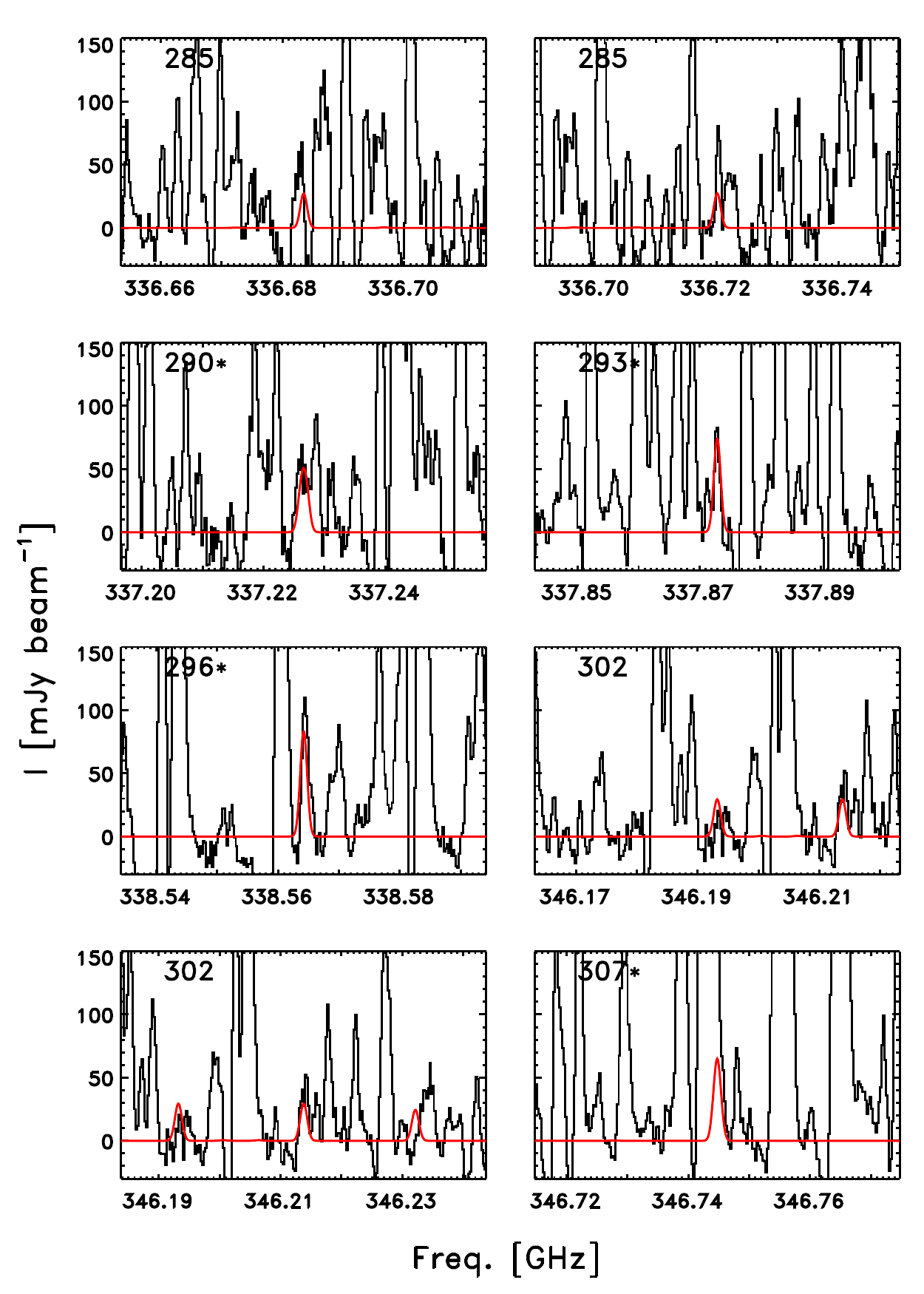}\includegraphics{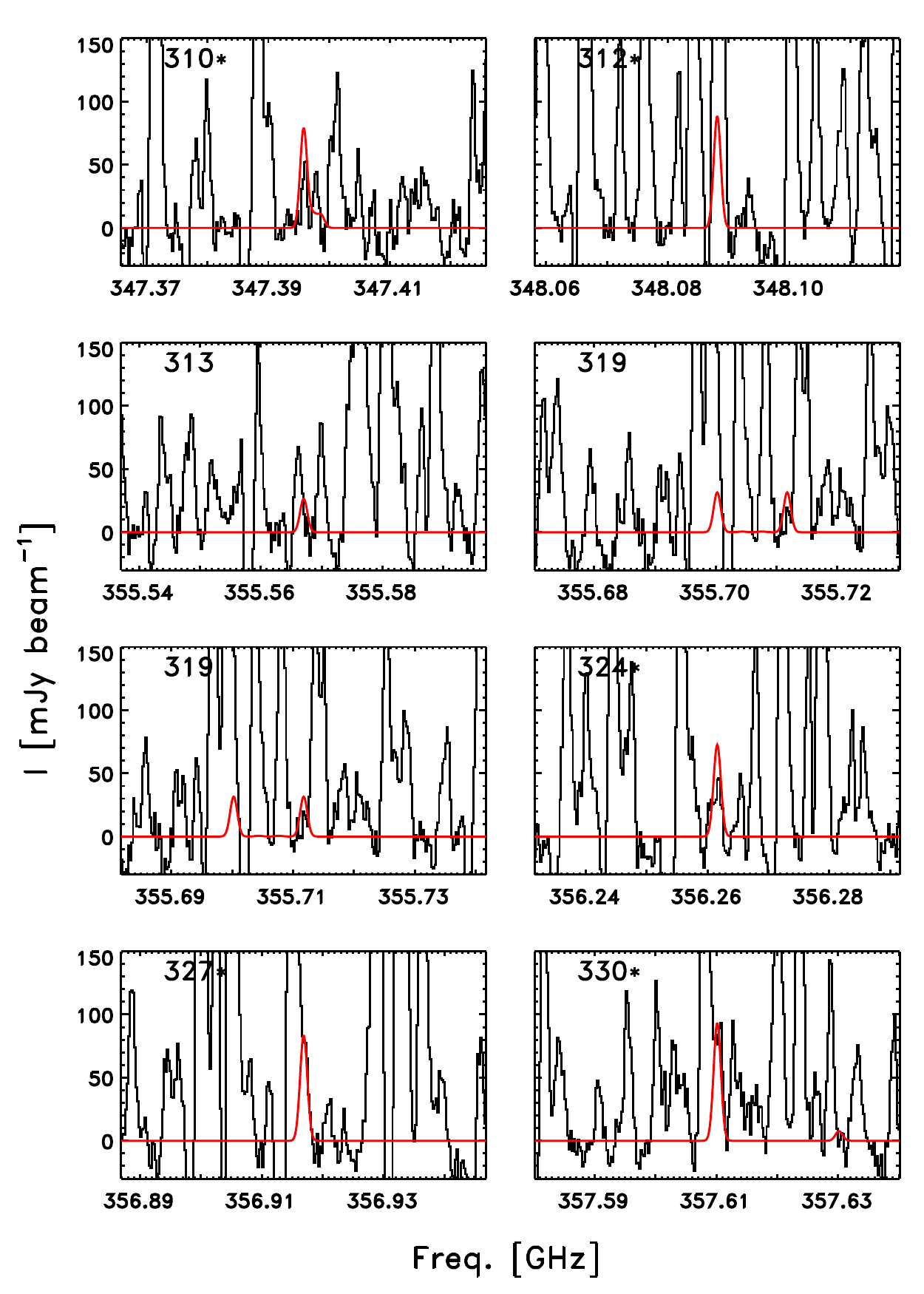}}
\captionof{figure}{Detection CHDOHCHO (upper panels) and CH$_2$OHCDO (lower panels) showing the 16 brightest lines as expected from the synthetic spectrum.}\label{chdohcho_1}\label{ch2ohcdo_1}
\end{minipage}
\clearpage

\begin{figure}[!htb]
\resizebox{\hsize}{!}{\includegraphics{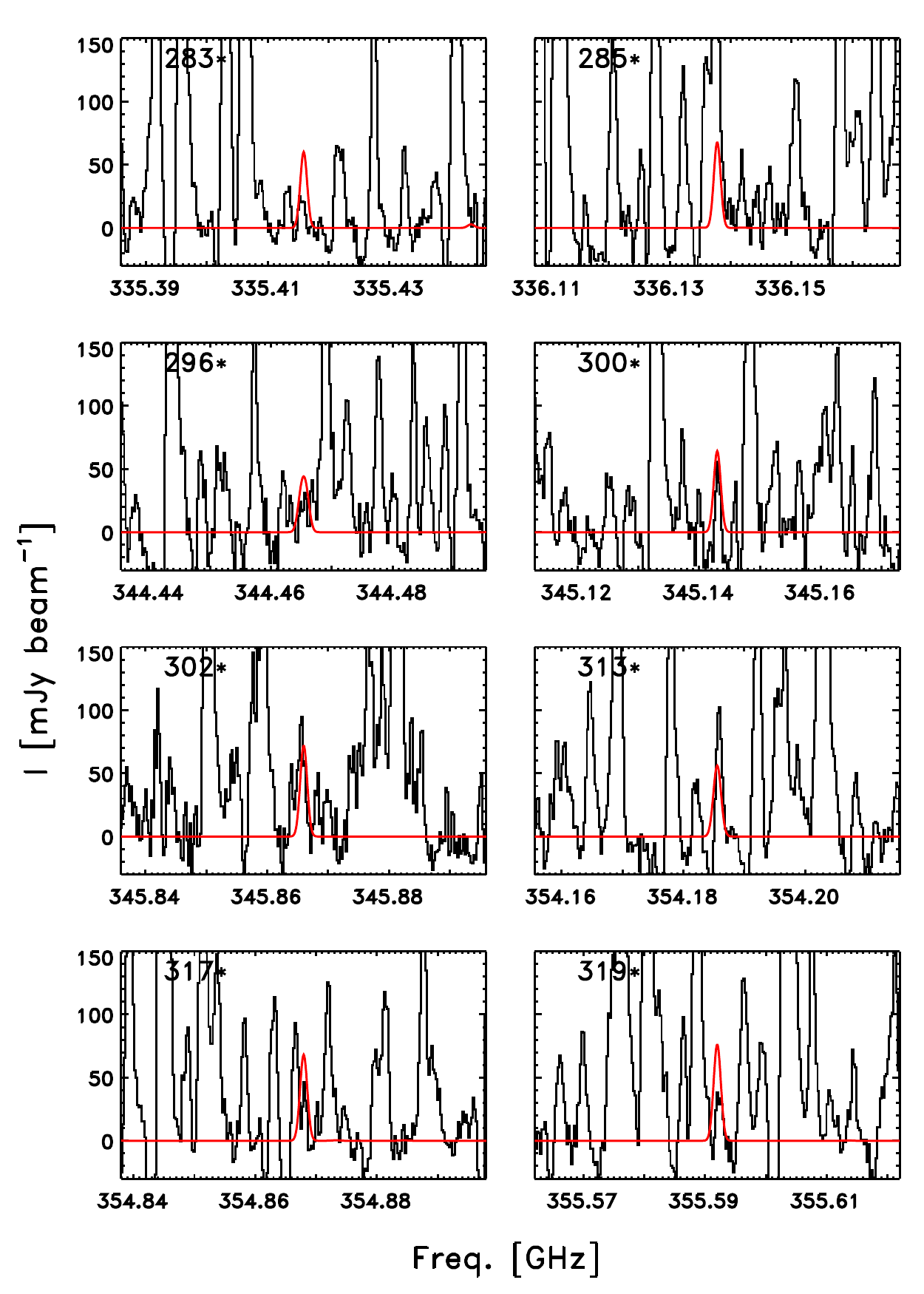}}
\caption{Detection of CH$_2$ODCHO showing the 8 brightest lines as expected from the synthetic spectrum.}\label{ch2odcho_1}
\end{figure}

\clearpage

\noindent\begin{minipage}{\textwidth}
\subsection{Glycolaldehyde: $^{13}$C-isotopologues}
\resizebox{0.88\textwidth}{!}{\includegraphics{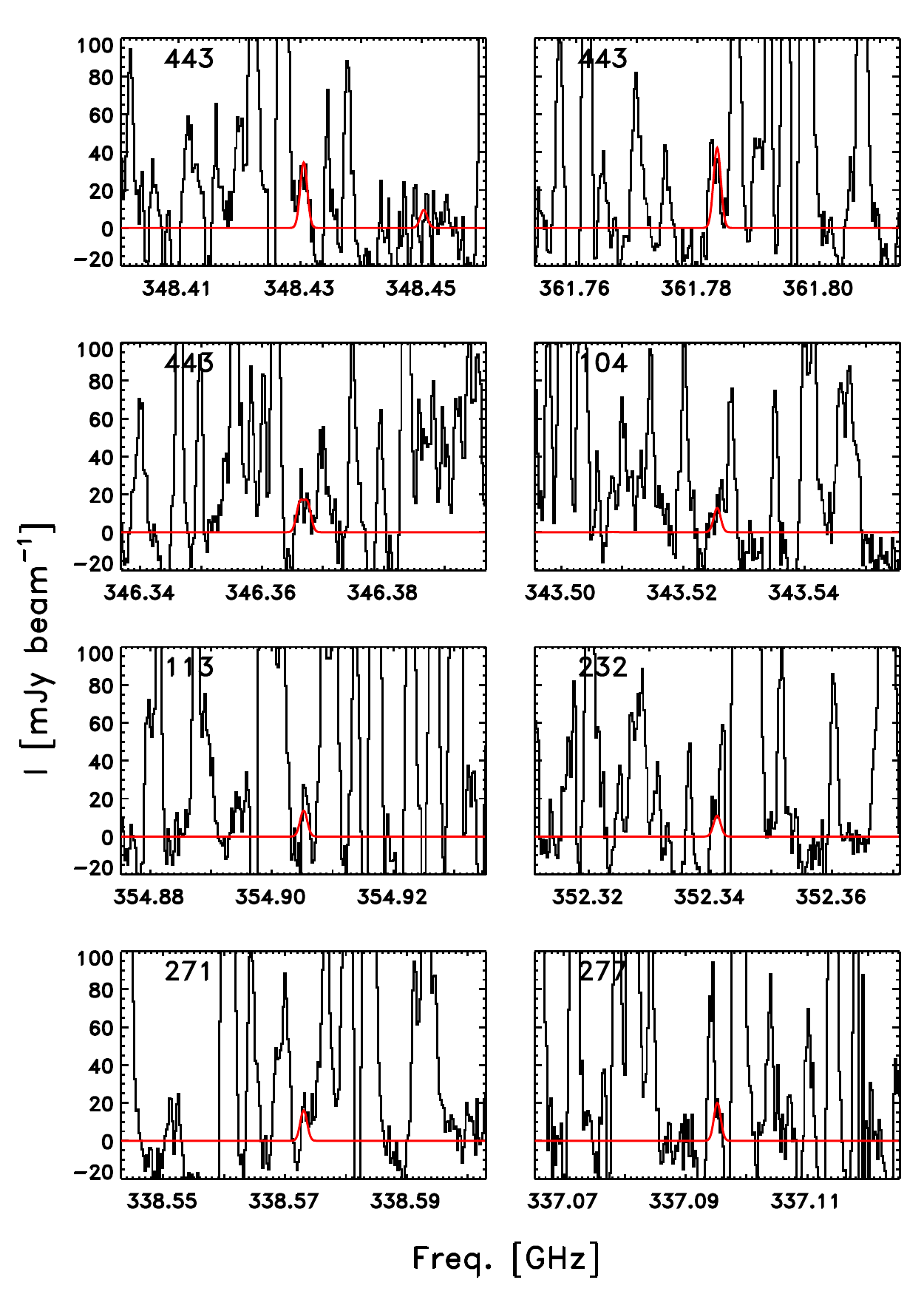}\includegraphics{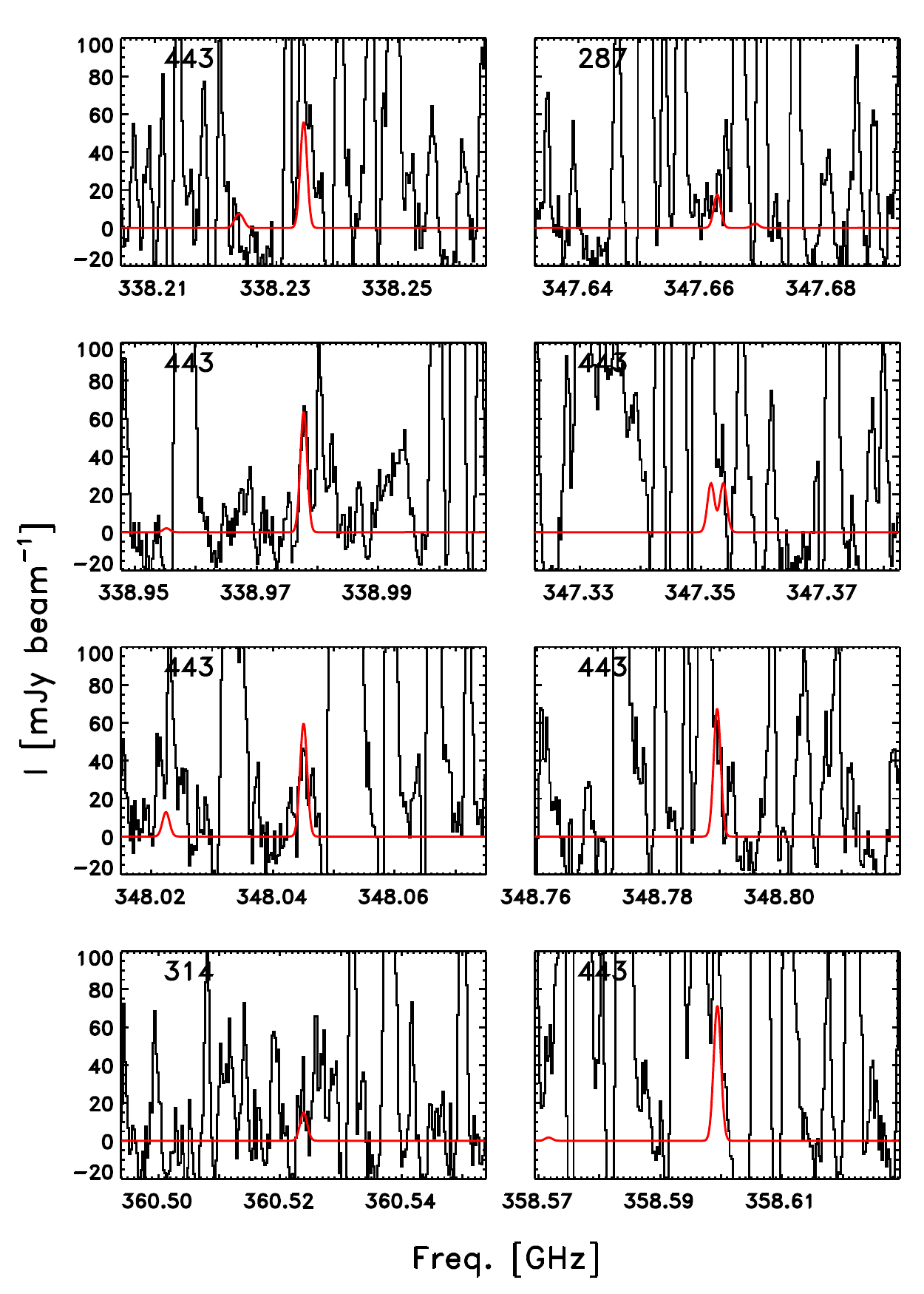}}
\resizebox{0.44\textwidth}{!}{\includegraphics{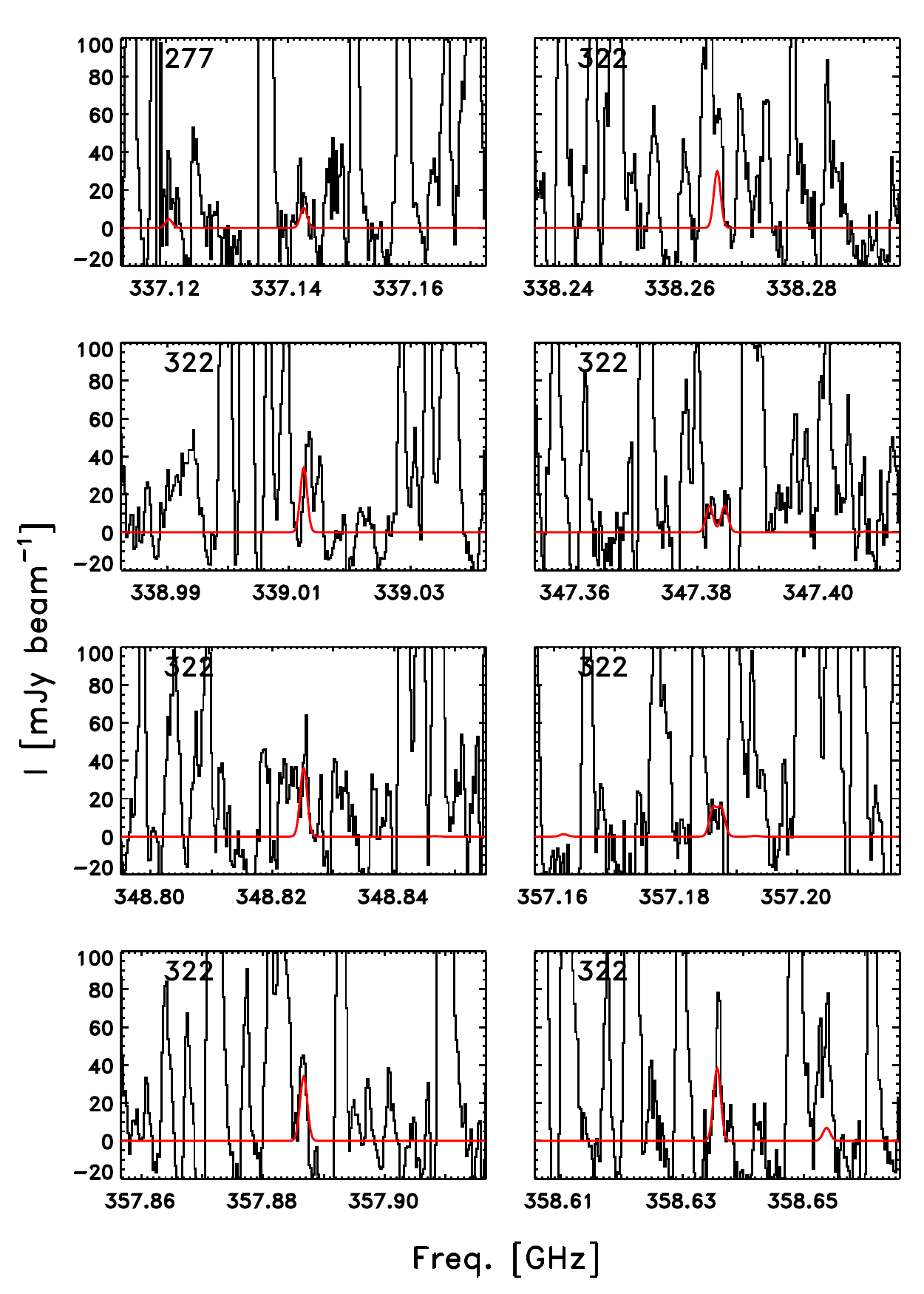}}
\captionof{figure}{Detection of the $^{13}$C isotopologues of glycolaldehyde with the upper panels showing the 16 best (bright/relatively well-isolated) lines of  $^{13}$CH$_2$OHCHO and the lower panel the 8 best lines of CH$_2$OH$^{13}$CHO.}\label{13glycolaldehyde}
\end{minipage}

\end{document}